\documentclass[12pt,psf,epsf]{article}
\usepackage[centertags]{amsmath}
\usepackage{amssymb}
\usepackage{graphicx}
\usepackage{epsfig}
\usepackage{ulem}
\usepackage[english]{babel}
\usepackage{array}
\usepackage{amsthm}
\usepackage{latexsym}
\usepackage[mathcal]{euscript}
\pdfoutput=1
\usepackage{epsfig}
\usepackage{jheppubm}
\usepackage{mathdots}
\usepackage{MnSymbol}
\usepackage{physics}
\usepackage{multirow}
\usepackage{ dsfont }
\usepackage{capt-of}
\usepackage{caption}
\usepackage{subcaption}

\usepackage{tikz}
\usepackage{pgfplots}           
\usepackage{tikz-qtree}

\renewcommand{\emph}{\textit}

\newcommand{\be}{\begin{equation}}
\newcommand{\ee}{\end{equation}}
\newcommand{\bea}{\begin{eqnarray}}
\newcommand{\eea}{\end{eqnarray}}

\newcommand{\e}{\mathrm{e}}

\newcommand{\RR}{\mathbb{R}}

\newcommand{\mO}{\mathcal{O}}

\newcommand{\ie}{\textit{i.e.,} }

\definecolor{dgreen}{rgb}{0.,0.6,0.}

\title{Delicate windows into evaporating black holes}
\author{Ben Craps$^1$, Juan Hernandez$^1$, Mikhail Khramtsov$^{1,2}$ and Maria Knysh$^1$}
\affiliation{$^1$ Theoretische Natuurkunde, Vrije Universiteit Brussel (VUB) and The International Solvay Institutes, Pleinlaan 2, B-1050 Brussels, Belgium \\
$^2$ David Rittenhouse Laboratory, University of Pennsylvania,\\
209 S. 33rd Street, Philadelphia, PA 19104, USA}

\emailAdd{ben.craps@vub.be}
\emailAdd{juan.hernandez@vub.be}
\emailAdd{mikhail.khramtsov@vub.be}
\emailAdd{maria.knysh@vub.be}

\abstract{We revisit the model of an AdS$_2$ black hole in JT gravity evaporating into an external bath. We study when, and how much, information about the black hole interior can be accessed through different portions of the Hawking radiation collected in the bath, and we obtain the corresponding full quantitative Page curves. As a refinement of previous results, we describe the island phase transition for a semi-infinite segment of radiation in the bath, establishing access to the interior for times within the regime of applicability of the model. For finite-size segments in the bath, one needs to include the purifier of the black hole microscopic dual together with the radiation segment in order to access the interior information. We identify four scenarios of the entropy evolution in this case, including a possibility where the interior reconstruction window is temporarily interrupted. Analyzing the phase structure of the Page curve of a finite segment with length comparable to the Page time, we demonstrate that it is very sensitive to changes of the parameters of the model. We also discuss the evolution of the subregion complexity of the radiation during the black hole evaporation. }

\begin{document}
\maketitle
\newpage
\section{Introduction and summary of results}

Recent years have seen significant progress towards the resolution of one of the open puzzles of quantum gravity: the black hole information paradox \cite{Penington191,Almheiri191,Almheiri192,Almheiri193,Penington192,Almheiri194,Hartman20,Goto20} (see also \cite{Review} for a detailed review). The information paradox first appeared in the context of evaporating black holes, in which semi-classical computations seemed to imply that black hole evaporation is non-unitary, in stark contrast with the principles of quantum mechanics \cite{Hawking76}. The first computations of the spectrum of Hawking radiation indicated that it is completely thermal, and therefore an evaporating black hole would eventually leave behind a cloud of thermal radiation, independently of the initial state from which it was formed. However, one could imagine forming a black hole from a pure state with enough energy in a compact region of a quantum gravity system. Such a black hole seems to evolve from a pure state to a mixed thermal state which amounts to a loss of information and thus is incompatible with unitary time evolution. 

This tension between black hole evaporation and unitarity can be quantified through the lens of (entanglement) entropy. For a unitary process, the entanglement entropy of the whole system should remain invariant, and so the entropy of a black hole formed from a pure state should eventually vanish after the evaporation process. However, the initial semi-classical analysis by Hawking \cite{Hawking75} seemed to imply that the entanglement entropy simply keeps increasing without bound. This was the first indication for an information paradox. While unsettling at first, this initial formulation of the information paradox provided a naive natural resolution. The contradiction between expectations from unitarity and the semi-classical result would only be realized at the end of the evaporation process. However, as the size of the black hole keeps decreasing throughout the evaporation, non-perturbative effects from quantum gravity can become more and more important, to the point that they could produce very large corrections towards the late stages of the evaporation. It was natural to expect that quantum gravity effects could restore unitarity towards the very end of the evaporation. 

But further investigations led to a sharper contradiction at earlier times, when the semi-classical results were still expected to be valid. For a unitary quantum theory, the entanglement entropy of a subsystem is always bounded from above by the thermodynamic entropy of its coarse grained description. Specifically, for a black hole, the thermodynamic entropy is identified with the area of its event horizon in Planck units \cite{Bekenstein73,Hawking-entropy} which decreases during the evaporation. Thus, after some time of initial growth, the entanglement entropy of the black hole should start decreasing to eventually vanish after the evaporation \cite{Page93}. However, the first semi-classical computations of entanglement entropy during the evaporation would violate this bound and keep growing even before the black hole would reach a size in which the semi-classical picture was expected to fail. The modern understanding of the information paradox is formulated as a(n apparent) violation of this entropy bound in quantum gravity, and can be considered in much more generality. For example, see e.g.\ \cite{Almheiri193,Penington192,Almheiri194} for the discussion of the information paradox for non-evaporating black holes, and e.g.\ \cite{Krishnan20,Hartman202,Balasubramanian202,Shaghoulian21} for proposals of the information problem in the cosmological context.

In the last few years, newly constructed models have produced the first computations in which the entropy of evaporating black holes satisfies the entropy bound expected by unitarity \cite{Almheiri191,Penington191,Almheiri193,Penington192}. This has been possible thanks to the Engelhardt-Wall prescription for computing holographic entanglement entropy \cite{QES}. To compute the entanglement entropy of a boundary subregion $\mathbf{B}$, one considers all possible codimension two surfaces $\sigma_{\mathbf{B}}$ which are homologous to the boundary subregion $\mathbf{B}$ -- that is, whose boundary is anchored along the boundary of $\mathbf{B}$ and which are smoothly retractable to $\mathbf{B}$. For each $\sigma_\mathbf{B}$ one can define the generalized entropy
\be\label{eq:Sgen}
S_{\rm gen}(\sigma_\mathbf{B}) =   \frac{A(\sigma_\mathbf{B})}{4G_N} + S_{\rm bulk}(\Sigma_{\mathbf{B}})\,,
\ee
where $\Sigma_\mathbf{B}$ is a bulk codimension-one surface bounded by $\sigma_\mathbf{B}$ and $\mathbf{B}$, \ie $\partial\Sigma_{{\mathbf{B}}} = \sigma_{\mathbf{B}} \cup \mathbf{B}$. The first term is the area of $\sigma_\mathbf{B}$, and corresponds to the Bekenstein-Hawking area term in the Hubeny-Rangamani-Ryu-Takayanagi prescription , which constitutes the leading semi-classical contribution \cite{RT,HRT}. The second term incorporates the quantum corrections in the form of the von Neumann entropy of the quantum fields on $\Sigma_{\mathbf{B}}$. To compute the entanglement entropy of a boundary subregion $\mathbf{B}$, one then extemizes the generalized entropy over all surfaces $\sigma_\mathbf{B}$ that are homologous to $\mathbf{B}$. Surfaces $\sigma_\mathbf{B}$ that extremize the generalized entropy are referred to as quantum extremal surfaces (QES). In the event that there are multiple QESs, the one with minimal generalized entropy is referred to as the dominant QES and the holographic entanglement entropy is given by the value of the generalized entropy evaluated on this QES.

The new tractable models consist of a double sided black hole in two dimensional Jackiw-Teitelboim (JT) gravity with conformal matter, which is allowed to evaporate into a non-gravitational reservoir coupled to one side of the black hole \cite{Almheiri191,Penington191,Chen19}. This is done by changing asymptotic boundary conditions with a joining quench in which the asymptotic boundary is glued to a semi-infinite interval with the same conformal matter in the ground state. The conformal matter in the semi-infinite interval acts as a zero temperature bath which absorbs the Hawking radiation emitted by the evaporating black hole. The non-evaporating side of the black hole can be considered as a purification of the evaporating side which is coupled to the bath. This two dimensional black hole + bath model benefits from various simplifications which allows for the precise computation of the generalized entropy, something which in general can be a very complicated task. One important simplification comes from working with a two dimensional theory of gravity. This implies that codimension-two surfaces are simply points on which the generalized entropy can be evaluated. The extremization condition then simply turns into two differential equations to find the QES. Furthermore, working with JT gravity considerably simplifies the computation of the Bekenstein-Hawking area term, which corresponds to the value of the dilaton in Planck units $\frac{A}{4G_N} \sim \frac{\phi}{4G_N} \equiv S_\phi$. Indeed, in these models, the back-reaction of any matter on the geometry can be succinctly incorporated into the dilaton by a simple integral of the matter stress tensor~\cite{Almheiri14}. The computation of the bulk von Neumann entropy of the quantum fields is significantly simplified by working with conformal matter because the matter degrees of freedom are constrained by an infinite dimensional symmetry algebra. This enables the use of powerful tools of 2d CFT to compute the entanglement entropy. More precisely, the quantum state of the conformal matter can be mapped to the ground state on the upper half plane by a coordinate transformation and a local Weyl rescaling, and the entanglement entropy can be computed using the method of images \cite{Cardy06,Calabrese07,Calabrese09,Polyakov,Recknagel13}. With this method, the bulk von Neumann entropies are determined up to a function $\cal F$ of the conformal cross ratio $\eta$ which depends on the precise details of the CFT in question. Further restricting the conformal matter to be holographic fully fixes the von Neumann entropy, and allows for an exact computation of the generalized entropy in this setup.\footnote{For CFTs which satisfy certain sparseness conditions on the spectrum and OPE coefficients of bulk and boundary operators, the entanglement entropy is the same as for a holographic CFT~\cite{Sully20}.} These properties of the model facilitated progress in further generalizations and variations of the model, including \cite{Almheiri193,Almheiri194,Almheiri195,Penington192,Chen19,Chen20,Bra-ket,MarolfMaxfield,Goto20,Balasubramanian20,Balasubramanian202,Balasubramanian21,Balasubramanian212,Hollowood20,Hollowood21,EasyIslands1,EasyIslands2,EasyIslands3,EasyIslands4,Bhattacharya21,Bhattacharya212,Bhattacharya213,Geng20,Geng21,geng2021inconsistency,Ghosh21,DeVuyst22,Gyongyosi22,Basu22,Ghodrati22}.

This model is an exciting playground to study when, and how much, information about the black hole interior can be accessed through different portions of the Hawking radiation collected in the bath. Our goal is to obtain the corresponding full quantitative Page curves for several interesting setups as a refinement of previous results in~\cite{Almheiri191,Chen19} to shed further light on the black hole interior reconstruction properties of this model. In section~\ref{section: prescription for Sgen} we review the black hole evaporation model and provide the necessary ingredients as well as detailed instructions on how to compute the entanglement entropy for any subsystem in the setup. Section~\ref{sec: JT section} summarizes the semi-classical geometry including the back-reaction from joining the JT gravity system with a non-gravitating bath. Section~\ref{sec:vN entropy} provides details on the quantum state and its von Neumann entropy.

In section~\ref{sec:semi-infinite}, we turn our attention to the Page curve of semi-infinite intervals that collect Hawking radiation in the bath. In section~\ref{sec: with QML}, we review the Page curve of a semi-infinite segment in the bath together with the purification of the black hole ($QM_L$) as was considered in~\cite{Almheiri191,Chen19}. We use this as an opportunity to give a comprehensive demonstration of how to compute the Page curve as outlined in section~\ref{section: prescription for Sgen} and to introduce the necessary terminology. The Page curve in this case is divided into four phases in which distinct QESs control the time dependence of entanglement entropy. Before the interval starts collecting Hawking radiation, the entanglement entropy is simply a constant given by the vacuum entanglement entropy of the interval plus the Bekenstein-Hawking entropy of the initial black hole. Once radiation starts being collected by the interval, there is an initial phase in which the entropy rapidly increases following known results for QFTs in the presence of local quenches \cite{Calabrese07,Calabrese09}. In these two initial phases, the QES remains at the bifurcation point and therefore the black hole interior is not accessible. After enough Hawking radiation is collected by the interval, the QES jumps to a nontrivial location. First, it stays very close to the bifurcation surface. During this phase, a small portion of the black hole interior becomes accessible for reconstruction. The entanglement entropy still grows but at a much lower rate than before. Moreover, when transitioning to this phase, there can be a period of decreasing entanglement entropy which reflects the scrambling of the perturbations introduced by the quench. The final phase corresponds to a QES far from the bifurcation point, behind the event horizon of the evaporating black hole. The transition to this phase happens at the so-called Page time after which the entanglement entropy decreases. During this final phase, a large portion of the black hole interior becomes accessible.

As an extension of previous results, we leverage the same method in order to compute the Page curve for a semi-infinite segment in the bath, but without $QM_L$, in section~\ref{sec: without QML}. We find that the Page curve in this case is characterized by three phases. Similarly to section~\ref{sec: with QML}, before the interval starts collecting Hawking radiation, the entanglement entropy is simply a constant, and as soon as the interval starts collecting radiation, the entropy increases rapidly. In both phases the black hole interior is not accessible. Finally, if enough Hawking radiation is collected by the interval before the semi-classical picture breaks down, the entanglement entropy starts to decrease and we establish access to the black hole interior. This final phase is a true island phase where the island is a region in the bulk which is bound by the bifurcation surface and a nontrivial QES which resides behind the horizon of the evaporating black hole. The difference with section~\ref{sec: with QML} lies in the fact that if there is a nontrivial QES phase transition, it happens much later than the Page time. Moreover, for this transition to occur within the regime of applicability of the model, we determine an upper bound on the vacuum entropy in terms of the increase in entropy of the evaporating black hole.

In section~\ref{sec:Finite+QML} we apply the prescription outlined in section~\ref{section: prescription for Sgen} in order to find the Page curve for finite segments of the bath.\footnote{See \cite{Balasubramanian212,Hollowood21} for studies of Page curves of finite segments in the eternal black hole model, and \cite{Ageev22} for the analogous study in the Schwarzschild black hole.} Without $QM_L$, the Page curve is completely described in terms of trivial QES phases and follows known results for QFTs in the presence of local quenches~\cite{Calabrese07,Calabrese09,Calabrese:2016xau}. Hence, the black hole purifier is essential in order to access the interior information. We find that the overall behaviour of the entanglement entropy is robust to small changes in the parameters of the model and the length of the interval. However, the sequence of dominant QES phases controlling the entanglement entropy is not. We identify four scenarios of the entropy evolution in this case. When the segment is too short, we find that it can never capture enough Hawking radiation in order to reconstruct the black hole interior. For large enough segments, there are three possible scenarios in which interior reconstruction is possible. Specifically, for each of the three scenarios there exists a period during which one can access the black hole interior, which we refer to as the reconstruction window. One possibility is to have a continuous reconstruction window during which only a small portion of the black hole is accessible, meaning that there is a nontrivial QES phase dominating the entropy at some point during the evolution but this QES is still very close to the bifurcation surface. Another possibility is for the reconstruction window to be a long, uninterrupted period. Initially a small portion of the interior is accessible after which there is a transition to a phase in which a large portion of the black hole interior is accessible as the nontrivial QES changes location from close to the bifurcation surface to a point close to the event horizon. The final possibility is for the interior reconstruction window to be interrupted by a period of temporary blindness. As in the previous two cases, initially, a small portion of the interior is accessible, however eventually this interior insight is lost. Evolving further, the interior access is regained, with a much bigger portion of the interior becoming accessible for reconstruction. We analyse the occurrence of this phenomenon in parameter space using a phase diagram and find analytical bounds for one of the parameters. Finally, in section~\ref{sec: complex}, we use the holographic subregion complexity=volume proposal to describe the information theoretic structure of the state in the different reconstructing scenarios. Specifically, we find that during trivial QES phases, the subregion complexity remains at its initial value to leading order in the Fefferman-Graham expansion for the volume \cite{Fefferman07,EasyIslands3}. Furthermore, the complexity of subregion $\mathbf{R} \cup QM_L$ increases discontinuously when transitioning to a nontrivial QES. We associate this discontinuity with the complexity of establishing the short-range entanglement structure of the reconstructable region associated to the nontrivial QES.

In appendix~\ref{section: Dilaton Solution}, we review the derivations of a simplified form of the dilaton solution in JT gravity in the presence of matter.
Appendix~\ref{section: QES} contains a detailed calculation on how to find the QES location for the different phases.
In appendix~\ref{section: Bounds on blind spot}, we find the bounds for one of the parameters for which the interruption to the interior reconstruction window occurs. Finally, the accompanying Mathematica file can be found at \href{https://doi.org/10.5281/zenodo.7104771}{https://doi.org/10.5281/zenodo.7104771}. This file contains all the numerical and analytical calculations discussed in this paper.

\section{Background and setup}
\label{section: prescription for Sgen}

\begin{figure}[t]
	\centering
	\includegraphics[scale=0.7]{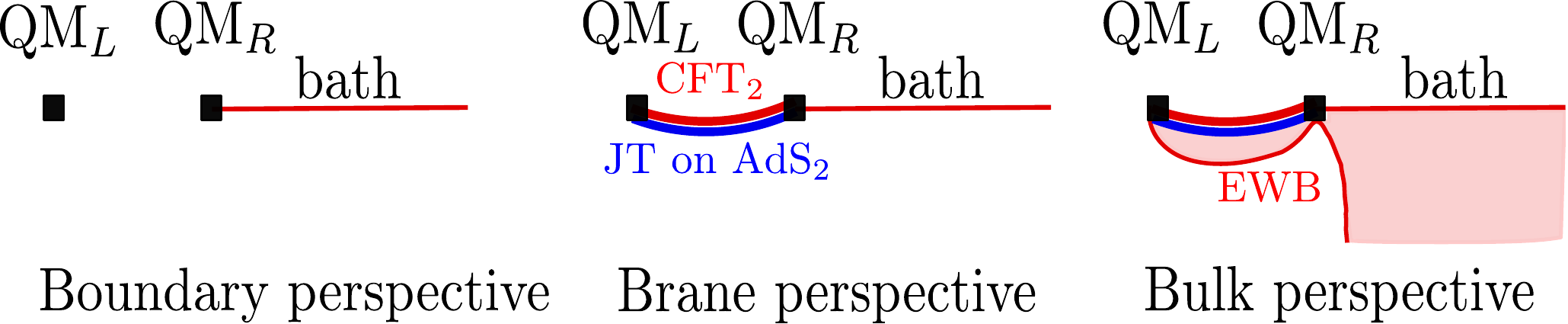}
	\caption{Evaporating black hole coupled to a bath from 3 different perspectives.}
	\label{fig: perspectives of the model}
\end{figure}

\begin{figure}[t]
	\centering
	\includegraphics[scale=0.7]{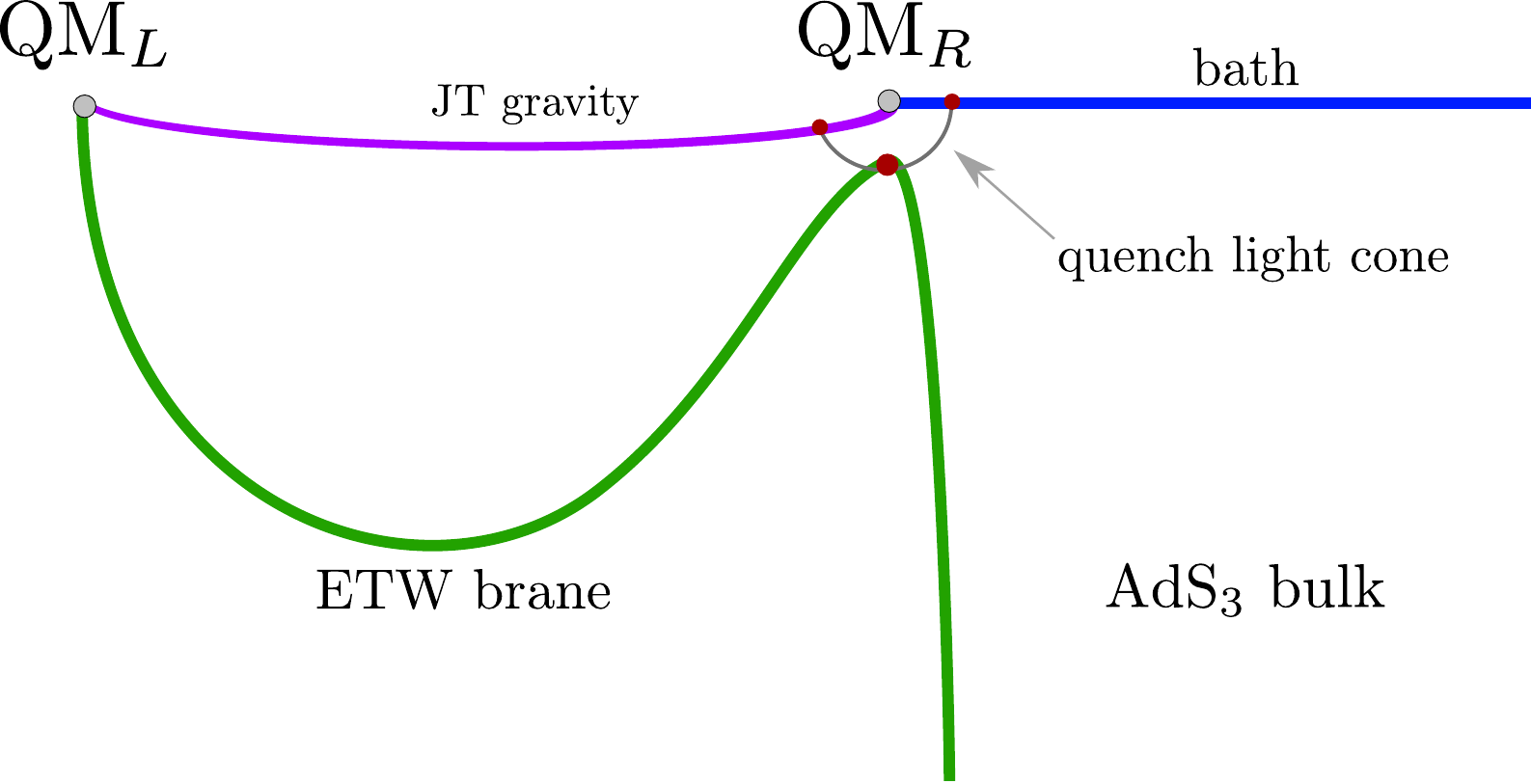}
	\caption{The evaporating black hole model of~\cite{Almheiri191,Almheiri192} summarized in one figure.}
	\label{fig:AEM4Z}
\end{figure}

We consider the black hole evaporation model presented in \cite{Almheiri191,Chen19}. This model can be described from 3 different perspectives as is shown in figure~\ref{fig: perspectives of the model}.
The brane	 perspective considers a double-sided AdS$_2$ black hole in Jackiw-Teitelboim (JT) gravity coupled to holographic CFT$_2$ matter. At some finite time, the system is coupled at one of the asymptotic boundaries with a non-gravitational region which acts as a bath on which the black hole can evaporate. The bath consists of the same holographic CFT$_2$ on a half line, prepared in the ground state. The joining quench creates two shockwaves, one of which propagates into the bath and the other into the black hole such that the black hole temperature changes from some initial value $T_0$ to some final value $T_1$. After the initial heating of the black hole by the quenching process, the black hole slowly evaporates into the non-gravitational region. Correspondingly, the effective temperature of the black hole slowly decays to zero. Since the JT gravity region is in AdS$_2$, it can be interpreted by its holographic dual, which consists of a thermofield double (TFD) of a strongly coupled quantum mechanical system. We denote the two components of the TFD by QM$_L$ and QM$_R$, and each is the holographic dual to the left and right outside region of the black hole, respectively. Only the combined system contains the interior of the black hole in its gravitational dual. This interpretation of the JT gravity region as its quantum mechanical dual is referred to as the boundary perspective and the joining quench couples the right boundary system QM$_R$ to the bath. The bulk perspective emerges by translating the holographic CFT$_2$ matter to its bulk AdS$_3$ dual geometry. From this bulk perspective, there is an end-of-the-world (ETW) brane anchored at the asymptotic boundaries of the AdS$_2$ JT gravity region, and an ETW brane on the boundary of the bath. When the two systems are coupled, the ETW brane detaches from the AdS$_2$ -- bath junction and falls into the bulk. Figure \ref{fig:AEM4Z} provides a detailed illustration of the model in this perspective.

With the context of the information paradox in mind, it is interesting to study the structure of entanglement entropy of various regions in this model. As was shown in \cite{Almheiri191,Chen19}, these models capture the initial entropy growth predicted by Hawking's semi-classical calculation, but after enough time has elapsed, there is a phase transition after which the entropy of the radiation decreases, as expected by unitarity. This transition between the early time increase in entanglement entropy and the late time decrease is referred to as the Page transition and an entanglement curve that follows this behaviour is called the Page curve. The main objective in the rest of the paper is to study the Page curve for different regions of the evaporating black hole model of \cite{Almheiri191,Chen19} in more detail.

The essential ingredient required to compute the Page curve is the generalized entropy $S_{\rm gen}$. There exists a simple prescription to compute the generalized entropy in the evaporating AdS$_2$ black hole in JT gravity, see~\cite{Almheiri191,Almheiri192,Chen19}. The procedure consists of the following three steps:
\begin{enumerate}
    \item Compute the dilaton profile, including its back-reaction due to the quenching of the two systems. This gives the leading contribution to the generalized entropy due to the dilaton $S_\phi$. See section \ref{sec: JT section}.
    \item Compute the von Neumann entropy $S_{\rm vN}$ of the CFT matter. This accounts for the quantum corrections to the generalized entropy. Note that since we are working with a holographic CFT, there are various possible configurations to account for in the RT prescription. See section~\ref{sec:vN entropy}.
    \item Extremize the resulting generalized entropy $S_{\rm gen}=S_\phi+S_{\rm vN}$ for each configuration independently, and pick the one for which the resulting entropy is smallest. See section~\ref{sec:semi-infinite}.
\end{enumerate}

In the remainder of this section, we provide an introduction to the specific model and we compute the two components of the generalized entropy, $S_\phi$ and $S_{\rm vN}$.

\subsection{Gravitational background} \label{sec: JT section}
We begin by reviewing the geometry of the model, including the back-reaction of the gravitational region when coupled to an external bath on which the black hole evaporates. For a more complete introduction to this system, we refer the reader to previous work on this model~\cite{Almheiri191,Almheiri192,Chen19,Chen20}.

The gravitational region of the model is described by a double sided AdS$_2$ black hole in JT gravity coupled to holographic conformal matter, described by the action
\begin{equation}
	I=\frac{1}{16\pi G_N}\qty[\int_{\mathcal{M}}d^2x\sqrt{-g}\qty{\phi_0R+\phi(R+2)}+2\int_{\partial\mathcal{M}}\qty{(\phi_0+\phi_b)K}]+I_{CFT}.\label{eq: JT action}
\end{equation}
Following~\cite{Almheiri191}, we work in the limit in which the JT gravity sector can be treated semi-classically, so that the metric and dilaton obey the following equations of motion
\begin{equation}
	R=-2
	\label{eq: result of dilaton qn of motion}\,,
\end{equation}
\begin{equation}
	(\phi+\phi_0)\qty(R_{\mu\nu}-\frac{1}{2}g_{\mu\nu}R)=8\pi G_N \langle T_{\mu\nu}\rangle+\left(\nabla_\mu\nabla_\nu-g_{\mu\nu}\left(\Box+\frac{R}{2}\right)\right)\phi,\label{eq:dil}
\end{equation}
where  $T_{\mu\nu}=\frac{-2}{\sqrt{-g}}\frac{\delta I_{CFT}}{\delta g^{\mu\nu}}$. The dilaton equation of motion fixes the geometry to be locally AdS$_2$ with cosmological constant $\Lambda = -2$, and the metric equations of motion relate the dilaton profile to the expectation value of the CFT stress tensor $\langle T_{\mu\nu}\rangle$. Note that in the above and for the rest of the paper, we have set the AdS length to one, $L_{\rm AdS}=1$. 

\begin{figure}[t]
	\centering
	\includegraphics[scale=0.7]{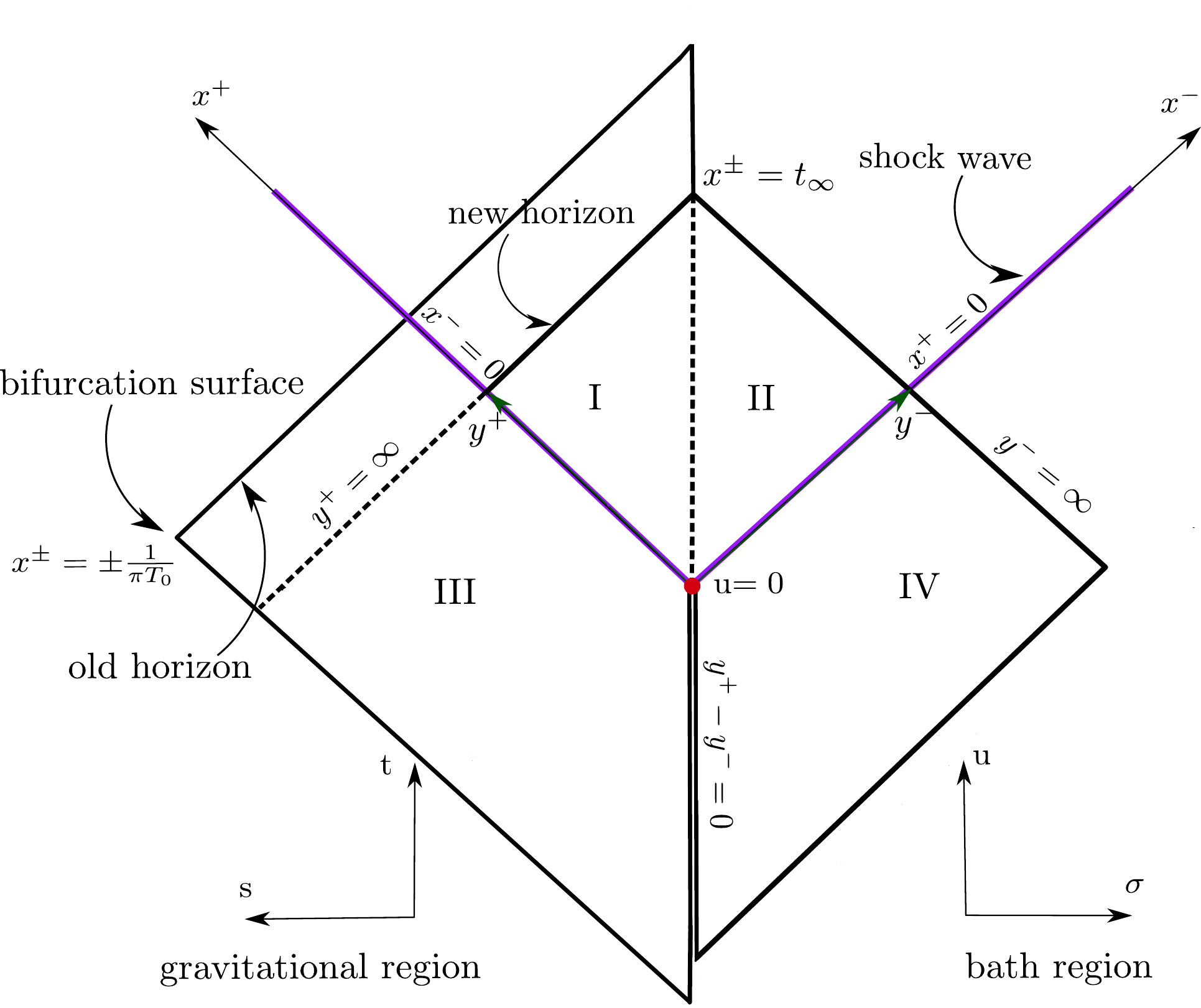} 
	\caption{ This figure shows the Poincare patch of a black hole in AdS$_2$ which is glued to the thermal bath in flat spacetime along the boundary $\sigma=0$ at time $u = 0$. The purple lines are the shock waves that are generated by this quench. This picture also shows the location of the bifurcation surface and the horizon of the initial equilibrium black hole as well as the new horizon which is shifted w.r.t.\ the old one due to the propagating shock wave.}
	\label{figure: spacetime}
\end{figure}
We introduce Poincar\'e coordinates for the gravitational region
\be
ds_{\text{AdS}}^2 = -\frac{4 dx^+ dx^-}{(x^+ - x^-)^2}\,, \label{metricPoincare}
\ee
where 
\be
x^{\pm} = t \pm s\,, \qquad t = \frac{x^+ + x^-}{2} \in \RR\,, \qquad s = \frac{x^+ - x^-}{2} > \epsilon\,.
\ee
We have introduced a radial cutoff $\epsilon$. The bath region consists of a half line with the same conformal CFT$_2$ in the ground state. We use flat coordinates 
\be\label{eq:metric-bath}
ds^2_{\text{bath}} = -\frac{dy^+ dy^-}{\epsilon^2}\,,
\ee
where 
\be
y^{\pm} = u \mp \sigma\,, \qquad u = \frac{y^+ + y^-}{2} \in \RR\,, \qquad \sigma = \frac{y^- - y^+}{2} >-\epsilon\,.
\ee

The coupling of the two systems happens at $t=u=0$, $s =-\sigma= \epsilon$,\footnote{The AdS$_2$ cutoff $s=\epsilon$ needs to be included in the bath $\sigma = -\epsilon$ to be able to smoothly extend the $x$ and $y$ coordinates as in eq.~\eqref{eq:xpm}}. After the coupling, $t,u>0$, the gluing is determined by a map between the worldline of the bath boundary $(u\geq 0,\sigma = -\epsilon)$ and the AdS$_2$ boundary which is parametrized by a coordinate transformation $t = f(u)$. By imposing that the induced metrics from eq.~\eqref{metricPoincare} and eq.~\eqref{eq:metric-bath} match on the common boundary, the boundary of AdS$_2$ follows the trajectory $s=\epsilon f'(u)$.\footnote{The initial conditions $t=u=0$, $s=-\sigma = \epsilon$ therefore fix $f(0)=0$ and $f'(0)=1$.} To determine the function $f$, one can use the energy balance equation as will be explain below. For later use, we will extend the $y^\pm$ coordinates from the bath into the AdS$_2$ region through \cite{Almheiri191}
\be\label{eq:xpm}
x^\pm = f(y^\pm)\,,
\ee
and in these physical coordinates the bulk metric will take the form 
\be
ds^2_{\text{AdS}} = -\frac{4 f'(y^+) f'(y^-) dy^+ dy^-}{(f(y^+) - f(y^-))^2}\,. \label{MetricAdSphys}
\ee

We now determine the function $f$ by considering the energy balance equation. Imposing Dirichlet boundary conditions
\begin{equation}
    g_{uu} = \frac{1}{\epsilon^2}\,, \quad \phi_b = \frac{\phi_r}{\epsilon}\,,
\end{equation}
reduces the JT gravity action in eq.~\eqref{eq: JT action} to the following boundary term
\begin{equation}
    I_{\rm JT}\Big|_{\text{on shell}} = \frac{\phi_r}{8\pi G_N} \int_{\partial {\cal M}} K + \text{topological term}\,,
\end{equation}
where\footnote{The extrinsic curvature $K$ can be computed using Poincar\'e coordinates given in eq.~\eqref{metricPoincare} for the boundary located at $(t=f(u),s=\epsilon f'(u))$, or equivalently in the physical coordinates given in eq.~\eqref{MetricAdSphys} for the boundary $(u,\sigma=-\epsilon)$ \cite{Maldacena16}.} 
\begin{equation}
    \frac{\phi_r}{8\pi G_N} \int_{\partial {\cal M}} K = - \int du E(u)\,,
\end{equation}
with\footnote{The Schwarzian derivative is given by $\{f(u),u\}\equiv \frac{f'''}{f'}- \frac{3}{2} \left(\frac{f''}{f'}\right)^2\,.$}
\be\label{eq:ADM}
E(u) = -\frac{\phi_r}{8 \pi G_N} \{ f(u),u\}
\ee
being the ADM energy of the JT gravity system. 

First, consider a JT gravity system not coupled to an external bath. In this case, the ADM energy is conserved due to reparametrization invariance of the boundary time $u \to u+ \delta u$, 
\be\label{eq:ADM-0}
E(u) = E_0 = \frac{\phi_r \left(\pi T_0\right)^2}{4 \pi G_N} \,.
\ee

Next, consider the combined AdS$_2$ + bath system. The coupling to an external bath breaks the time translation invariance of the JT system, so that the ADM energy is no longer conserved. More generally, when varying the total action in eq.~\eqref{eq: JT action} with respect to the boundary time, we find the energy balance equation~\cite{Engelsoy16}
\be\label{eq:energy-balance}
\partial_u E(u) = \langle T_{u\sigma}\rangle = f'(u)^2\left( \langle T_{x^-x^-}\rangle -\langle T_{x^+ x^+}\rangle \right)\,,
\ee
which implies energy conservation of the combined AdS$_2$ + bath system.

For times before the AdS$_2$ and bath are coupled ($u<0$) there is no energy exchange, which implies that the ADM energy -- and therefore the Schwarzian of $f$ -- is constant, and is given by eq.~\eqref{eq:ADM-0}. A solution for $f$ that satisfies the initial conditions $f(0)=0$, $f'(0)=1$ is given by\footnote{There is an $SL_2(\mathbb{R})$ family of solutions to $\{f,u\} = -2 (\pi T_0)^2$, related by the isometries of AdS$_2$. The choice of initial conditions for the coupling (that is, $u=t=0$ and $s=-\sigma=\epsilon$) restricts the $f$ function to a one dimensional subgroup of the solutions given by
\begin{equation*}
    f_a(u) = \frac{a}{\pi T_0} \frac{e^{\pi T_0 u}-e^{-\pi T_0 u}}{e^{\pi T_0 u}+(2a-1)e^{-\pi T_0 u}}\,,
\end{equation*}
with $a\neq 0$. The conventional solution given in eq.~\eqref{eq: Rindler} corresponds to the choice $a=1$, and is the only solution that is odd under time reversal, which simplifies the analysis below.
} 
\be\label{eq: Rindler}
\begin{aligned}
f(u) &= \frac{1}{\pi T_0} \tanh(\pi T_0 u)\,, \quad u < 0\,.
\end{aligned}
\ee
Therefore, the physical coordinates $y^\pm = f^{-1}(x^\pm)$ in AdS$_2$ before the quench cover the Rindler patch with temperature $T_0$, and the metric reads
\be
ds^2_{\rm AdS}= - \frac{\left(2 \pi T_0\right)^2 dy^+ dy^-}{\left(\sinh(\pi T_0 (y^+-y^-))\right)^2}\,, \quad y^\pm < 0\,.
\ee
The CFT matter in AdS$_2$ is initially in the vacuum of the Poincar\'e coordinates (\ie $\langle T_{x^+x^+} \rangle  =\langle T_{x^-x^-} \rangle =\langle T_{x^+x^-} \rangle =0$ for $s>0$), which corresponds to the TFD in the physical coordinates (\ie $\langle T_{y^\pm y^\pm} \rangle =\frac{c}{12 \pi} (\pi T_0)^2 \sech^2( \pi T_0 y^\pm)$ for $\sigma < 0$). The dilaton profile is therefore
\be\label{eq: dilaton-before}
\phi(x) = 2 \phi_r  \frac{1-(2\pi T_0)^2 x^+ x^-}{x^+-x-} = 2\phi_r \pi T_0 \coth\left( \pi T_0 (y^+-y^-) \right)\,,
\ee
where $\phi_r$ is related to the boundary value of the dilaton by $\phi = \frac{\phi_r}{\epsilon}$ at $\sigma = - \epsilon$. The matter in the bath is initially in the vacuum, so that $\langle T_{y^\pm y^\pm }\rangle = 0$ for $\sigma >0$. 

For times after the bath is initially coupled to the black hole ($u>0$), the system contains a localized shock of positive energy \be
E_S \equiv \frac{\phi_r \pi}{4 G_N} (T_1^2-T_0^2)\,,
\label{eq: formula for energy of the shock}
\ee
which propagates into the bath and into AdS$_2$. The energy shock contributes a delta function to stress energy of the bath and of AdS$_2$, which quickly heats up the black hole to a temperature $T_1>T_0$, and shifts the future event horizon away from its initial position aligned with the bifurcation surface at $x^\pm = \pm\frac{1}{\pi T_0}$. Together with the initial conditions $\langle T_{x^{\pm} x^{\pm}} \rangle = 0$ for $s>0$ and $\langle T_{y^{\pm} y^{\pm}} \rangle=0$ for $\sigma > 0$, we find $\langle T_{x^- x^-} \rangle = E_S \delta(x^-) + E_{\rm flux} \Theta (x^-)$ and $\langle T_{y^+ y^+} \rangle = E_S \delta(y^+)+ E_{\rm flux}' \Theta(y^+)$. After the shock, there are fluxes of energy $E_{\rm flux}$ and $E_{\rm flux}'$ across the now transparent boundary of the gravity region. These energy fluxes can be computed from the anomalous transformation rule of the stress tensor 
\begin{equation}
    \begin{aligned}
        \label{eq:Ttrans}
\langle T_{x^-x^-}\rangle &= \left(\frac{dy^-}{dx^-} \right)^2\langle T_{y^-y^-}\rangle - \frac{c}{24\pi}\{y^-,x^-\}\,,\\
\langle T_{y^+y^+}\rangle &= \left(\frac{dx^+}{dy^+} \right)^2\langle T_{x^+x^+}\rangle - \frac{c}{24\pi}\{x^+,y^+\}\,.
    \end{aligned}
\end{equation}

Finally, by using $f'(0)=1$, the stress tensor in AdS is given by\footnote{The subscript ``AdS'' should be understood as a restriction to the AdS$_2$ region $x^+-x^->0$. Similarly the subscript ``bath'' should be understood as a restriction to the bath region $y^--y^+>0$. Note that the energy shock $\langle T_{x^+x^+} \rangle = E_S \delta (x^+)$ is only in the bath, and not in AdS, which is why it was omitted. A similar argument applies to $\langle T_{y^-y^-} \rangle$ in the bath.}
\be\label{eq:TAdS}
\langle T_{x^- x^-} \rangle_{\rm AdS} = E_S\, \delta (x^-) - \frac{c}{24\pi} \{y^-,x^-\} \, \Theta(x^-)\,, \quad \langle T_{x^+ x^+} \rangle_{\rm AdS} =0\,,
\ee
while the stress tensor in the bath is
\be\label{eq:Tbath}
\langle T_{y^+y^+} \rangle_{\rm bath} = E_S\, \delta (y^+) - \frac{c}{24\pi} \{x^+,y^+\} \, \Theta(y^+)\,, \quad \langle T_{y^-y^-} \rangle_{\rm bath} =0\,.
\ee
After the shock, there is a flux of infalling negative energy $\langle T_{x^-x^-} \rangle_{\rm AdS} = -\frac{c}{24\pi} \{y^-,x^-\}$ which is proportional to the ADM energy given in eq.~\eqref{eq:ADM}.\footnote{The proportionality constant can be found by using the Schwarzian inversion formula $\{f,u\} = - f'(u)^2 \{u,f\}$. It is manifestly negative,  $\langle T_{x^-x^-} \rangle_{\rm AdS} = - \frac{kE(y^-)}{f'(y^-)^2} $, where $k$ is defined in eq.~\eqref{eq: eqn for k related to other parameters}.} This energy flux is due to the Hawking radiation which no longer reflects back from the asymptotic boundary, and instead escapes into the bath. Similarly, the bath experiences a positive energy flux after the shock coming from the AdS$_2$ region, $\langle T_{y^+ y^+} \rangle_{\rm bath} = -\frac{c}{24\pi} \{x^+,y^+\}=k E(y^+)$, which is carried by the Hawking quanta after the shock. We have defined
\be
k = \frac{c G_N}{3\phi_r},
\label{eq: eqn for k related to other parameters}
\ee
which is a parameter of dimension energy that controls the strength of the back-reaction and therefore sets the scale of the evaporation process. We take this parameter to be small compared to other dimensionful scales, such as the temperature. In this limit, the semi-classical picture breaks down for times of order $\frac{1}{k}\log\frac{1}{k}$.\footnote{The semi-classical picture works as long as the parameter $k$ characterizing the rate of evaporation is smaller than the energy of the system. That is $\frac{k}{E(u)}\ll 1$ which in turn implies $u \ll \frac{1}{k} \log \frac{1}{k}$.}

The matter stress tensor in eq.~\eqref{eq:TAdS} back-reacts on the AdS$_2$ geometry, modifying the dilaton above the shock
\be\label{eq: dilaton-integral}
\phi(x) = 2 \phi_r \frac{1-(2\pi T_0)^2x^+x^-+kI}{x^+-x^-}\,,
\ee
where 
\be
I = - \frac{24\pi}{c} \int_{0^-}^{x^-} dt (x^+-t)(x^--t) \langle T_{x^-x^-}(t)\rangle
\ee
accounts for the back-reaction. 

For a stress tensor of the form given in eq.~\eqref{eq:TAdS}, the dilaton above the shock as given in eq.~\eqref{eq: dilaton-integral} can be re-expressed in the following simplified form \cite{Moitra:2019xoj,Hollowood20}
\be\label{eq: dilaton-closed}
\phi(x) = \phi_r \left(\frac{f''(y^-)}{f'(y^-)} + 2 \frac{f'(y^-)}{x^+-f(y^-)}\right)\,.
\ee
We refer the reader to appendix~\ref{section: Dilaton Solution} for details on this derivation. 

Integrating the $E_S \delta(x^-)$ contribution of $\langle T_{x^-x^-}\rangle$ in the energy balance equation in eq.~\eqref{eq:energy-balance} gives the ADM energy immediately after the quench
\be\label{eq:E1}
E(u) = E_1 =  \frac{\phi_r \left(\pi T_1\right)^2}{4 \pi G_N} \,, \quad u \to 0^+\,.
\ee
Writing the ADM energy and matter stress tensor in eq.~\eqref{eq:ADM} and eq.~\eqref{eq:TAdS} explicitly, we find a differential equation for the Schwarzian derivative of $t=f(u)$\footnote{For the following, we use the Schwarzian inversion formula 
\be
\{f,u\} = - f'(u)^2 \{u,f\}\,. 
\ee}
\be
\partial_u \{f(u),u\} = - k \{f(u),u\}\,,  \quad u>0\,, \label{eq:SchwEq}
\ee
which, together with the boundary condition in eq.~\eqref{eq:E1} fixes the Schwarzian
\be
\{f(u),u\}= 2\left( \pi T_1 \right)^2 e^{-ku}\,.
\ee
Matching the solution to the $u<0$ coordinate reparametrization in eq.~\eqref{eq: Rindler} requires $f(0)=f''(0)=0$ and $f'(0)=1$, and the unique solution is then \cite{Engelsoy16,Almheiri191}
\be
f(u) = \frac{1}{\pi T_1} \frac{I_0 (\frac{2\pi T_1}{k})K_0 (\frac{2\pi T_1}{k}\e^{-\frac{k u}{2}}) -K_0 (\frac{2\pi T_1}{k})I_0 (\frac{2\pi T_1}{k}\e^{-\frac{k u}{2}}) }{I_1 (\frac{2\pi T_1}{k})K_0 (\frac{2\pi T_1}{k}\e^{-\frac{k u}{2}})+K_1 (\frac{2\pi T_1}{k})I_0 (\frac{2\pi T_1}{k}\e^{-\frac{k u}{2}})}\,,
\label{f(u)}
\ee
where $I_n(x)$ and $K_n(x)$ are the modified Bessel functions of the first and second kind.

We conclude this section with the expression of the dilaton contribution to the generalized entropy
\begin{equation}
    S_\phi=\frac{\phi_0+\phi(x)}{4 G_N}\,,
    \label{eq: dilaton contribution to generelized entropy}
\end{equation}
where the dilaton profile is given by eq.~\eqref{eq: dilaton-before} and eq.~\eqref{eq: dilaton-closed} which can be summarized as follows
\begin{equation}
    \phi(x)=\left\{
                \begin{array}{ll}
               
                2 \phi_r  \frac{1-(2\pi T_0)^2 x^+ x^-}{x^+-x-} & \qquad \text{below the shock}\, (x^-<0),\\
                
                \phi_r \left(\frac{f''(y^-)}{f'(y^-)} + 2 \frac{f'(y^-)}{x^+-f(y^-)} \right) & \qquad  \text{above the shock}\, (x^->0).
                \end{array}
              \right.
    \label{eq: dilaton profile before and after quench}
\end{equation}

\subsection{Von Neumann entropy of the radiation}
\label{sec:vN entropy}

With the leading dilaton contribution to the generalized entropy under control, we compute the quantum corrections, which are given by the von Neumann entropy of the CFT matter. This can be done by mapping the quantum state to the vacuum of the upper half plane through a coordinate transformation and a local Weyl rescaling \cite{Calabrese04,Calabrese09,Almheiri191}.

We begin by performing a coordinate transformation
\bea \label{eq:Euclid}
&& x^-=-x \,, \quad x^+= \bar{x} \,, \\ 
&& y^-= -y \,, \quad  y^+=\bar{y}\,.
\eea
Next, with a careful choice of the coordinate transformation $(y, \bar{y}) \to (z, \bar{z})$  (and similarly $(x, \bar{x}) \to (z, \bar{z})$), the CFT matter can be mapped to the ground state of the upper half plane \cite{Almheiri191} (\ie $\langle T_{zz} \rangle =  \langle T_{\bar{z}\bar{z}} \rangle = 0$). The details of this coordinate transformation will be described below, see eqs.~\eqref{eq:z}, \eqref{map y to z} and \eqref{map x to z}. The metric with these new coordinates is given by
\bea\label{eq:metric-UHP}
ds^2_{\text{AdS}} &=& \Omega_{\text{AdS}}^{-2} (x, \bar{x}) dz d\bar{z}\,,\\
ds^2_{\text{bath}} &=& \Omega_{\rm bath}^{-2} (y, \bar{y}) dz d\bar{z}\,,
\eea
where the Weyl factors are given by
\bea\label{eq:Weyl}
\Omega_{\text{AdS}} (x, \bar{x}) &=&  \frac{x + \bar{x}}{2} \sqrt{z'(x) \bar{z}'(\bar{x})}\,,\\
\Omega_{\rm bath} (y, \bar{y}) &=& \epsilon \sqrt{z'(y) \bar{z}'(\bar{y})}\,.
\eea
Finally, we can perform a local Weyl rescaling which transforms the piece-wise defined upper half plane metric in eq.~\eqref{eq:metric-UHP} to the flat metric
\be
ds^2_{\rm AdS} \to dz d\bar{z}\,, \quad ds^2_{\rm bath} \to dz d\bar{z}\,,
\ee             
while keeping the CFT matter in the ground state.

With this relation between the vacuum state in the upper half plane and the CFT matter state in the evaporation model, it is then straightforward to compute the entanglement entropy of the CFT matter using the well known vacuum results~\cite{Cardy06,Calabrese07,Calabrese09}. That is, the task of computing the bulk entanglement entropy of an interval or union of intervals with endpoints $\{x_i,\bar{x}_i\}$ reduces to writing down the entanglement entropy of the vacuum state in the upper half plane $S_{\rm vac} (\{z_i,\bar{z}_i\})$. After that, one needs to convert this expression to the bath or AdS$_2$ coordinates and include the effect of the local Weyl rescalings $g_{\mu\nu} \to \Omega^{-2}g_{\mu\nu}$,\footnote{The following transformation may be interpreted as resulting from the rescaling of UV cutoffs with respect to which the entropy is defined.} which leads to
\be\label{eq: Weyl-S}
S_{\rm vN} = S_{\rm vac} - \frac{c}{6} \sum_{\rm endpoints} \log \Omega({\rm endpoint})\,,
\ee
where the Weyl factors are given by eq.~\eqref{eq:Weyl}.

We now turn our attention to computing the vacuum entanglement entropy on the upper half plane. For the purposes of our analysis, we only need two results: the entanglement entropy of a semi-infinite interval with a single endpoint at $(z, \Bar{z})$, and the entanglement entropy of a finite interval with two endpoints at $(z_1, \Bar{z}_1)$ and $(z_2,\Bar{z}_2)$.

For a semi-infinite interval with only one endpoint, conformal symmetry constrains the entropy up to a non-negative constant\footnote{The standard derivation of this result involves using the replica trick to relate entanglement entropy to twist operator one-point functions in the upper half plane, which in turn can be computed via the method of images, see for example \cite{Recknagel13}.}
\begin{equation}
    S_{\rm vac}(z,\bar{z})=\log g+\frac{c}{6}\log \left(\frac{-i\left(z-\Bar{z}\right)}{\delta}\right),
    \label{eq: holographic onepoint function}
\end{equation}
where $c$ is the central charge, $\log g\geq 0$ is the Affleck-Ludwig boundary entropy and $\delta$ is the UV cutoff, see \cite{Calabrese04,Cardy06,Calabrese07,Calabrese09,Recknagel13} for more details.

For the case of a finite interval with two endpoints at $(z_1, \Bar{z}_1)$ and $(z_2,\Bar{z}_2)$, the entanglement entropy is given by
\begin{equation}
    S_{\rm vac}(z_1,\bar{z}_1,z_2,\bar{z}_2)=\log G(\eta)+\frac{c}{6}\log\left(\frac{ |z_1-z_2|^2}{\delta^2} \eta\right), \qquad \eta=\frac{\left(z_1-\Bar{z}_1\right)\left(z_2-\Bar{z}_2\right)}{\left(z_1-\Bar{z}_2\right)\left(z_2-\Bar{z}_1\right)}\,,
    \label{eq: holographic twopoint function}
\end{equation}
where $\eta$ is the conformal cross ratio. The unspecified function $G(\eta)$ depends on the theory and boundary conditions and is related to the chiral four point function of twist operators \cite{Recknagel13,Sully20}.

Furthermore, by considering a bulk OPE limit ($\eta \to 1$) or an operator boundary expansion ($\eta \to 0$), the $G(\eta)$ function can be shown to satisfy the following two boundary conditions
\begin{equation}
    \lim_{\eta\to 1}G(\eta)=1, \qquad  \lim_{\eta\to 0}G(\eta)=g^2.
\end{equation}

These expressions hold true for any two-dimensional BCFT with a conformal boundary  at $z-\bar{z}=0$ but these expressions are not very convenient to work with. From here on, we can assume that the matter theory is a holographic BCFT, in which case the function $G(\eta)$ is explicitly known \cite{Takayanagi11,Sully20}, and is given by 
\be
G\left(\eta\right)=\Theta\left(\eta-\eta^*\right)\eta^{-c/6}+\Theta\left(\eta^*-\eta\right)\frac{g^2}{\left(1-\eta\right)^{c/6}}.
\label{eq: explicit form of Geta for hologrphic BCFT}
\ee

Using this expression, the entanglement entropy reduces to
\begin{equation}
    S_{\rm vac}(z_1,\bar{z}_1,z_2,\bar{z}_2)=\left\{
                \begin{array}{ll}
               
                \frac{c}{6}\log \left(\frac{|z_1-\Bar{z}_1||z_2-\Bar{z}_2|}{\delta^2} \right)+2\log g& \qquad \eta<\eta^*,\\
                
                 \frac{c}{6}\log \left(\frac{|z_1-z_2||\Bar{z}_1-\Bar{z}_2|}{\delta^2} \right)& \qquad \eta>\eta^*\,,
                \end{array}
              \right.
    \label{eq: holographic BCFT twopoint function}
\end{equation}
where $\eta^*=\frac{1}{1+g^{12/c}}$. Holographic BCFTs have bulk duals which terminate at an end-of-the-world (ETW) brane anchored at the boundary of the BCFT, and the Affleck-Ludwig boundary entropy $\log g$ is related to the tension of the ETW brane \cite{Affleck91,Takayanagi11}. For computational simplicity we will take the limit of no boundary entropy $\log g =0$, which in the gravitational picture corresponds to a tensionless ETW brane.\footnote{For positive tension branes ($\log g >0$) the entropy of disconnected configurations receives an increase relative to connected configurations, favoring configurations that can reconstruct the black hole interior. This causes, for example, the Quench to Scrambling phase transition to occur faster in section~\ref{sec:semi-infinite} but will not affect the Page transition, which is a transition between two reconstructing phases~\cite{Chen19}.} The first expression ($\eta < \eta^*$) in eq.~\eqref{eq: holographic BCFT twopoint function} corresponds to the disconnected configuration in which the candidate RT surfaces connect the endpoints of the interval to the ETW brane. The second one ($\eta > \eta^*$) is given by the connected configuration with one candidate RT surface connecting both endpoints of the interval. 

Now, let us construct the explicit coordinate transformation that maps the state of the matter CFT to the vacuum. To find such a transformation, we consider the form of the stress tensor in Poincar\'e and physical coordinates, given by eq.~\eqref{eq:TAdS} and eq.~\eqref{eq:Tbath}, as well as the transformation rule of the stress tensor under coordinate transformations given in eq.~\eqref{eq:Ttrans}, and demand that the stress tensor in the upper half plane coordinates $z$ vanish everywhere. Because the stress tensor vanishes in Poincar\'e coordinates for $x>0$,\footnote{Recall that $x^-<0$ corresponds to $x>0$.} the coordinate transformation $z(x)$ must be a M\"{o}bius transformation for $x>0$. Similarly, because the stress tensor vanishes in physical coordinates for $y<0$,\footnote{That is, for $y^->0$.} $z(y)$ must be a M\"{o}bius transformation for $y<0$. Following the convention established in~\cite{Almheiri191}, we map the AdS$_2$ space to the region $(0,z_0)$ and the bath to $(z_0,i\infty)$, which fixes the map to be
\bea
	\label{eq:z} z &=& \left\{ \begin{array}{cc}  \frac{-iz_0^2}{x-iz_0}\,, \quad \,\,\,\,  x >0\,, \\ 
		z_0-iy\,, \quad y < 0\,,  \end{array}\right.
\eea
where $-iz_0>0$. The second derivative of this transformation is discontinuous, and therefore the Schwarzian is proportional to a delta distribution. This induces a localized shock of energy in the Poincar\'e and bath coordinates
\be
E_S = \frac{ic}{12 \pi z_0}\,,
\ee
as was described around eq.~\eqref{eq: formula for energy of the shock}. We take the limit of high energy $E_S \gg T_1$ for which the map in eq.~\eqref{eq:z} becomes
\bea
\begin{aligned}
	\label{map y to z} z(y) &=& \left\{ \begin{array}{cc} \left(\frac{c}{12 \pi E_s}\right)^2 \frac{i}{f(y)}\,, \quad \quad & y > 0\,, \\ 
		-iy\,, \quad & y < 0\,,  \end{array}\right.
\end{aligned}
\eea
\bea
\begin{aligned}
	\label{map x to z} z(x) &=& \left\{ \begin{array}{cc} \left(\frac{c}{12 \pi E_s}\right)^2 \frac{i}{x}\,, \quad \quad  & x > 0\,, \\ 
		-if^{-1}(x)\,, \quad  & x < 0\,. \end{array}\right.
\end{aligned}
\eea
A similar argument gives the same result for the coordinate transformation of the barred coordinates $\bar{z}(\bar{x})$ and $\bar{z}(\bar{y})$.\footnote{Note that the relation with the original coordinates $x^\pm$ and $y^\pm$ is slightly less symmetric due to the minus signs in eq.~\eqref{eq:Euclid}.}

Putting together the entropy formulas in the upper half plane given in eq.~\eqref{eq: holographic onepoint function} and eq.~\eqref{eq: holographic BCFT twopoint function}, the relation between the UHP coordinates and the Poincar\'e/physical coordinates given in eq.~\eqref{map y to z} and eq.~\eqref{map x to z} and the relation in eq.~\eqref{eq: Weyl-S}, we can write the bulk von Neumann entropy of the CFT matter. Because the coordinate transformation in eq.~\eqref{map y to z} and eq.~\eqref{map x to z} are piece-wise defined, we will need to split the expressions into multiple cases. Therefore, we divide the spacetime described in figure \ref{figure: spacetime} into four distinct regions
\begin{equation}
        x^{\pm}=\left\{
                \begin{array}{lll}
                 \text{I} & \text{Post-shock in AdS} & x^+ \geq x^- \geq 0,\\
                 
                 \text{II} & \text{Post-shock in bath} & x^- \geq x^+ \geq 0,\\
                 
                 \text{III} & \text{Pre-shock in AdS} & x^+ \geq 0 \geq x^-,\\
                 
                 \text{IV} & \text{Pre-shock in bath} & x^- \geq 0 \geq x^+.
                \end{array}
              \right.
    \label{eq: regions in spacetime}
\end{equation}
The von Neumann entropy of semi-infinite intervals with one endpoint are given by
\begin{equation}
    S_{\rm vN}(x_1)=\frac{c}{6}\log\left\{
                \begin{array}{ll}
                \Big|\frac{24 \pi  E_S y^-x^+}{c \left(x^+-x^-\right)\delta \sqrt{ f'(y^-)}}\Big|+\mathcal{O}\left(\frac{T_1}{Es}\right) & x^{\pm} \in \text{I},\\
                 
                \Big|\frac{12 \pi E_S}{c \delta \epsilon } y^- \frac{f\left(y^+\right)}{\sqrt{f'\left(y^+\right)}}\Big|  +\mathcal{O}\left(\frac{T_1}{Es}\right) & x^{\pm} \in \text{II},\\
                 
                \left|\frac{2}{\delta}\right| & x^{\pm}\in \text{III},\\
                 
                \left|\frac{y^- - y^+} {\epsilon\delta}\right|& x^{\pm} \in \text{IV}.
                \end{array}
              \right.
    \label{eq: vonNeumannn entropy via onepoint function}
\end{equation}

These expressions will be relevant in section~\ref{sec:semi-infinite}. Notice that the von Neumann entropy for semi-infinite intervals with one endpoint in AdS below the shock ($x^\pm \in \text{III}$) is independent of the exact position of the endpoint. This is due to a cancellation between the position dependence of the vacuum entropy in BCFT given in eq.~\eqref{eq: holographic onepoint function} and that of the Weyl factor contribution given in eq.~\eqref{eq:Weyl} in eq.~\eqref{eq: Weyl-S}. Furthermore, for semi-infinite intervals with an endpoint in the bath below the shock ($x^\pm \in \text{IV}$), the entropy corresponds to the vacuum entropy in a BCFT and is therefore time independent. 

We now proceed to outline the von Neumann entropy of finite intervals, which will be useful for section~\ref{sec:Finite+QML}. There are in principle ten different expressions for the von Neumann entropy of finite intervals with two endpoints depending on where each boundary is situated in the classification given by eq.~\eqref{eq: regions in spacetime}. However, only four of them will be relevant for our calculation. We begin with the entropy of intervals that are entirely above the shock -- that is, with endpoints in regions I and/or II. Since we compute the entanglement entropy of intervals of the bath, at least one of the endpoints will be in region II, the other endpoint can be a late time QES in region I, or another bath interval endpoint in region II. These are given by
\begin{align}
    S_{\rm vN}(x_{1}\in \text{II}, x_2 \in \text{I})&=\frac{c}{6}\log\left(\frac{2 (y^-_1-y^-_2) (x^+_2-f(y^+_1)) }{\delta^2  \epsilon  (x^+_2-x^-_2)}\sqrt{\frac{f'(y^-_2)}{f'(y^+_1)}}\right)\,,
    \label{eq: vonNeumann entropy via twopoint function for x1 above the shock in bath and x2 above the shock in AdS}\\
    S_{\rm vN}\left(x_1\in \text{II}, x_2 \in \text{II}\right)&=\frac{c}{6}\log\left(\frac{ (y^-_1-y^-_2) (f(y^+_1)-f(y^+_2))}{\delta^2  \epsilon^2   \sqrt{f'(y^+_1)} \sqrt{f'(y^+_2)}}\right)\,.
    \label{eq: vonNeumann entropy via twopoint function for x1 above the shock in bath and x2 above the shock in bath}
\end{align}
Note that the entropy of intervals which are completely above the shock is always dominated by the connected configuration, since the coordinate transformation in eq.~\eqref{map y to z} implies $\eta=1+\mathcal{O}\left(\frac{T_1}{E_S}\right)$ in this region. For the cases we will consider below, both the connected and disconnected configurations can be important. However, note that in the disconnected configurations ($\eta < \eta^*$) the expressions factorize into a sum of the one point functions given by eq.~\eqref{eq: vonNeumannn entropy via onepoint function}
\be
S_{\rm vN}(x_1,x_2) = S_{\rm vN}(x_1) + S_{\rm vN}(x_2)\,, \quad \eta < \eta^*\,.
\ee
Therefore, we only specify the entropy of the connected phase ($\eta > \eta^*$) in what follows.

For intervals which are crossing the shock, we consider two possibilities. First, the interval can join a point in the bath above the shock to an early time QES in region III, the entropy is given by
\begin{equation}
    S_{\rm vN}\left(x_1\in \text{II},x_2 \in \text{III}\right)=\frac{c}{6}\log\left(\frac{24\pi E_S y^-_1 f(y^+_1)}{ c \delta ^2 \epsilon  \sqrt{f'(y^+_1)}}\right)+\mathcal{O}\left(\frac{T_1}{E_S}\right) \,, \quad \eta>\eta^*,
    \label{eq: vonNeumann entropy via twopoint function for x1 above the shock in bath and x2 is below the shock in AdS}
\end{equation}
where $\eta=\frac{f(y^+_1) (x^-_2-x^+_2)}{x^+_2 (x^-_2-f(y^+_1))}+\mathcal{O}\left(\frac{T_1}{Es}\right)$. The second option is for the interval to be completely in the bath and cross the shock, the entropy is given by
\begin{equation}
    S_{\rm vN}\left(x_1\in \text{II},x_2 \in \text{IV}\right)=\frac{c}{6}\log\left(\frac{12  \pi E_S y^-_1 f(y^+_1) (y^-_2-y^+_2)}{c \delta ^2 \epsilon ^2 \sqrt{f'(y^+_1)}}\right)+\mathcal{O}\left(\frac{T_1}{E_S}\right) \,, \quad \eta>\eta^*\,,
    \label{eq: vonNeumann entropy via twopoint function for x1 above the shock in bath and x2 in bath below the shock}
\end{equation}
where $\eta=\frac{y^-_1 (y^-_2-y^+_2)}{y^-_2 (y^-_1-y^+_2)}+\mathcal{O}\left(\frac{T_1}{E_S}\right)$.

Lastly, for intervals that do not cross the shock, we only need the entropy of bath intervals entirely below the shock which is given by
\begin{equation}
    S_{\rm vN}\left(x_1\in \text{IV},x_2 \in \text{IV}\right)=\frac{c}{6}\log\left(\frac{(y^-_1-y^-_2) (y^+_1-y^+_2)}{\delta ^2 \epsilon ^2}\right)  \,, \quad  \eta>\eta^*\,,
    \label{eq: vonNeumann entropy via twopoint function for x1 and x2 below the shock in bath}
\end{equation}
where $\eta=\frac{(y^-_1-y^+_1) (y^-_2-y^+_2)}{(y^-_1-y^+_2) (y^-_2-y^+_1)}$.  As mentioned above, the disconnected configuration dominates when $\eta<\eta^*$, in which case the bulk entropy is simply given by the sum of the disconnected contributions $S _{\rm vN}\left(x_1\in\text{II}\right)+S _{\rm vN}\left(x_2\in\text{IV}\right)$ in eq.~\eqref{eq: vonNeumannn entropy via onepoint function}. The von Neummann entropy of intervals below the shock agrees with its vacuum value for holographic CFTs~\cite{Cardy06}.

We end this section with some comments on the subtleties of including or excluding QM$_L$ in the computation of entanglement entropy. That is, for a given bath interval $\mathbf{R}$, we can compute the entropy of $\mathbf{R}$ or of $\mathbf{R}  \cup QM_L$. The computations of the generalized entropy for $\mathbf{R}$ and $\mathbf{R}  \cup QM_L$ are very similar. The only difference is due to the homology condition of the RT prescription, which will introduce/remove an extra RT surface, which is anchored at the bifurcation point and ends on the ETW brane, to each of the configurations to be considered.\footnote{To be precise, the homology condition requires the candidate RT surfaces corresponding to the entanglement entropy of $\mathbf{R} \cup QM_L$ ($\mathbf{R}$) to have an odd (even) number of anchor points at the Planck brane where the JT gravity theory is located. A simple example where this is illustrated is shown in figure \ref{figure: entanglement wedges of semi-infinite segment at later times+QML} (\ref{figure: entanglement wedges of semi-infinite segment at later times}).} These extra RT surfaces contribute to the generalized entropy by adding a Bekenstein-Hawking entropy term.\footnote{This definition of the Bekenstein-Hawking entropy includes quantum corrections, given by the bulk von Neumann entropy of the CFT matter in the black hole exterior.}
\be
 S_{\rm BH}(T_0)=\frac{c}{6}\log \frac{2}{\delta}+ \frac{2\pi T_0 \phi_r+\phi_0}{4G_N},
\label{eq: definition of SQML}
\ee
which corresponds to the length of the RT surface connecting the ETW brane and the bifurcation surface, and equals the entanglement entropy of a single side of the black hole before coupling to the bath. This prescription leads to the following relation
\be\label{eq: r vs nr}
S^R_{\rm gen}(\mathbf{R}) = S^R_{\rm gen}(\mathbf{R} \cup QM_L) + S_{BH}(T_0)\,, \quad S^{NR}_{\rm gen}(\mathbf{R}) = S^{NR}_{\rm gen}(\mathbf{R} \cup QM_L) - S_{BH}(T_0)\,,
\ee
where $S^R_{\rm gen}$ ($S^{NR}_{\rm gen}$) refers to the generalized entropy of RT saddle points which (do not) allow for reconstruction of the black hole interior. 

Considering the dominant RT configurations and their respective entanglement wedges, it is possible to determine when enough information is encoded on $\mathbf{R}$ and/or \\ $\mathbf{R}  \cup QM_L$ to reconstruct a portion of the black hole interior. More precisely, reconstruction is possible when $S_{\rm gen}^{NR} > S_{\rm gen}^R$. Therefore, the relation in eq.~\eqref{eq: r vs nr} implies that it is much easier to reconstruct a portion of the black hole interior if one has access not only to the radiation $\mathbf{R}$, but also of the purification of the black hole QM$_L$. Indeed, eq.~\eqref{eq: r vs nr} indicates that the radiation $\mathbf{R}$ alone can reconstruct a portion of the black hole interior only when there is a reconstructing configuration for $\mathbf{R} \cup QM_L$ which has at least $2S_{BH}(T_0)$ less generalized entropy than the non-reconstructing configurations, since
\be\label{eq: r vs nr 2}
S^{NR}_{\rm gen}(\mathbf{R}) - S^{R}_{\rm gen}(\mathbf{R})   =  S^{NR}_{\rm gen}(\mathbf{R}\cup QM_L) - S^{R}_{\rm gen}(\mathbf{R} \cup QM_L) - 2 S_{BH}(T_0)\,.
\ee

\section{Page curve of the semi-infinite radiation segments}
\label{sec:semi-infinite}

In this section, we compute the Page curve for a semi-infinite segment $\mathbf{R}$ of the bath. We review the results of~\cite{Almheiri191,Chen19} concerning the Page curve when the black hole purifier $QM_L$ is part of the subsystem, that is, the Page curve of $\mathbf{R} \cup {QM_L}$. We then compute the Page curve of the bath segment $\mathbf{R}$ without the black hole purifier and describe the island phase transition. Along the way, we introduce the necessary ingredients to compute the Page curve for finite segments of radiation in the bath. 
\begin{figure}[h]
	\centering
	\includegraphics[scale=0.7]{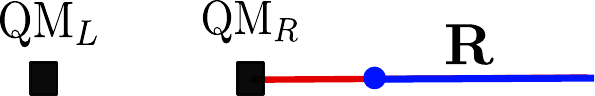}
	\caption{A semi-infinite radiation segment from the boundary perspective.}
	\label{fig infinite R}
\end{figure}

Let $\mathbf{R}$ be an evolving semi-infinite segment of the bath, extending from a finite spatial location to infinity. Figure~\ref{fig infinite R} illustrates the segment in the bath from the boundary perspective. Throughout the evaporation, $\mathbf{R}$ collects more and more Hawking radiation and slowly gains information about the black hole. Eventually, $\mathbf{R}$ will contain enough information so that $\mathbf{R} \cup QM_L$ can reconstruct a large portion of the black hole interior. This occurs at the Page transition,\footnote{Note that in the present model, there is an earlier transition after which a very small portion of the black hole interior becomes reconstructable by $\mathbf{R} \cup QM_L$. This small reconstruction window is not universal for generic CFTs, while the Page transition is expected to be quite general.} in which the dominance between two different nontrivial QESs of $\mathbf{R} \cup QM_L$ is exchanged. As the evaporation process continues, even more information is collected by $\mathbf{R}$ until eventually the black hole purifier $QM_L$ is no longer needed for $\mathbf{R}$ to reconstruct a portion of the black hole interior. This occurs at very late times when there is an island transition in the generalized entropy of $\mathbf{R}$. However, for large $\phi_0$, this transition would occur at times beyond the regime of applicability of the semi-classical model. As we will see in section~\ref{sec: without QML}, by relaxing the large extremal entropy restriction ($\phi_0 \gg \phi$) required for top-down constructions of JT gravity from compactifications of higher dimensional black holes, we can find a range of $\phi_0$ for which the island transition remains in the regime of validity of the model.

\begin{figure}[t]
     \centering
          \begin{subfigure}[b]{0.45\textwidth}
         \centering
         \includegraphics[scale=1.2]{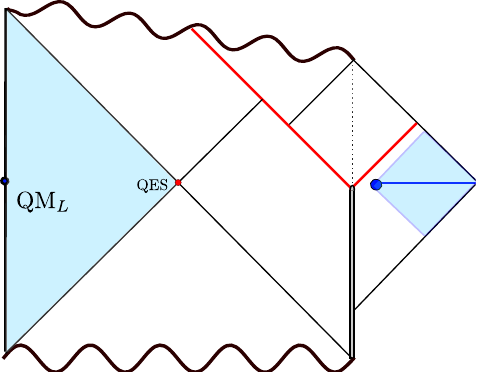}
         \caption{Trivial phase}
         \label{figure: EW trivial+QML}
     \end{subfigure}
     \begin{subfigure}[b]{0.45\textwidth}
         \centering
         \includegraphics[scale=0.55]{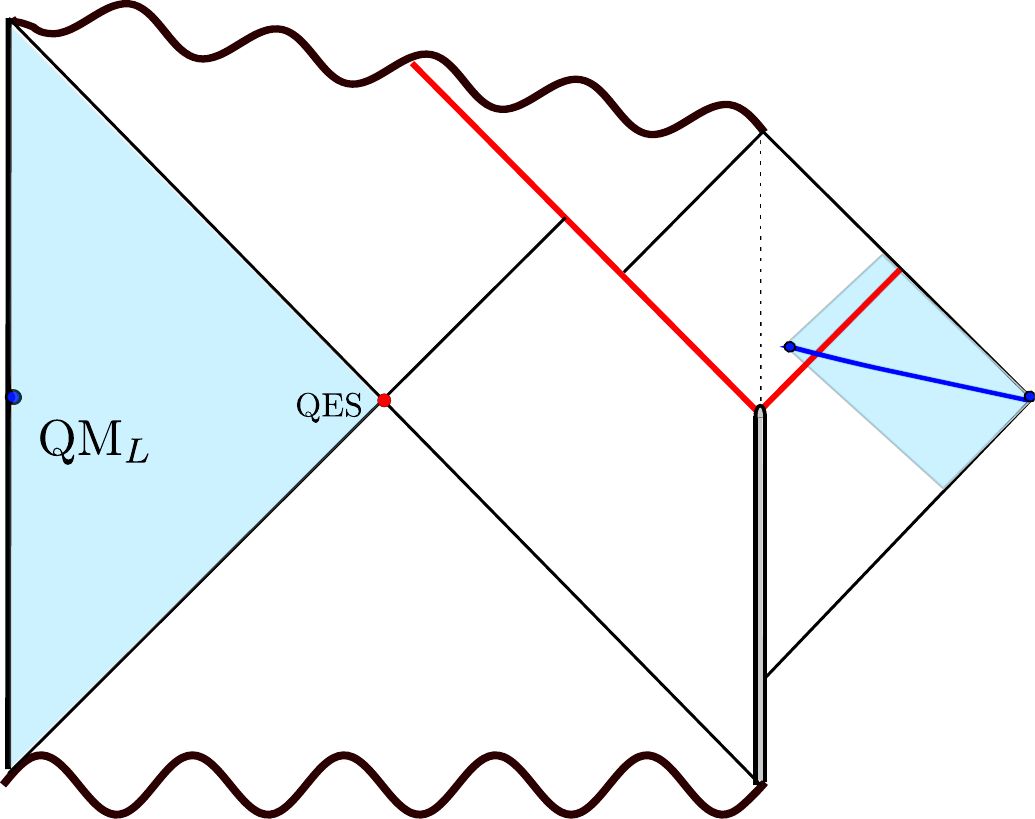}
         \caption{Quench phase}
         \label{figure: EW quench+QML}
     \end{subfigure}\\
     \hfill
     \begin{subfigure}[b]{0.45\textwidth}
         \centering
         \includegraphics[scale=0.5]{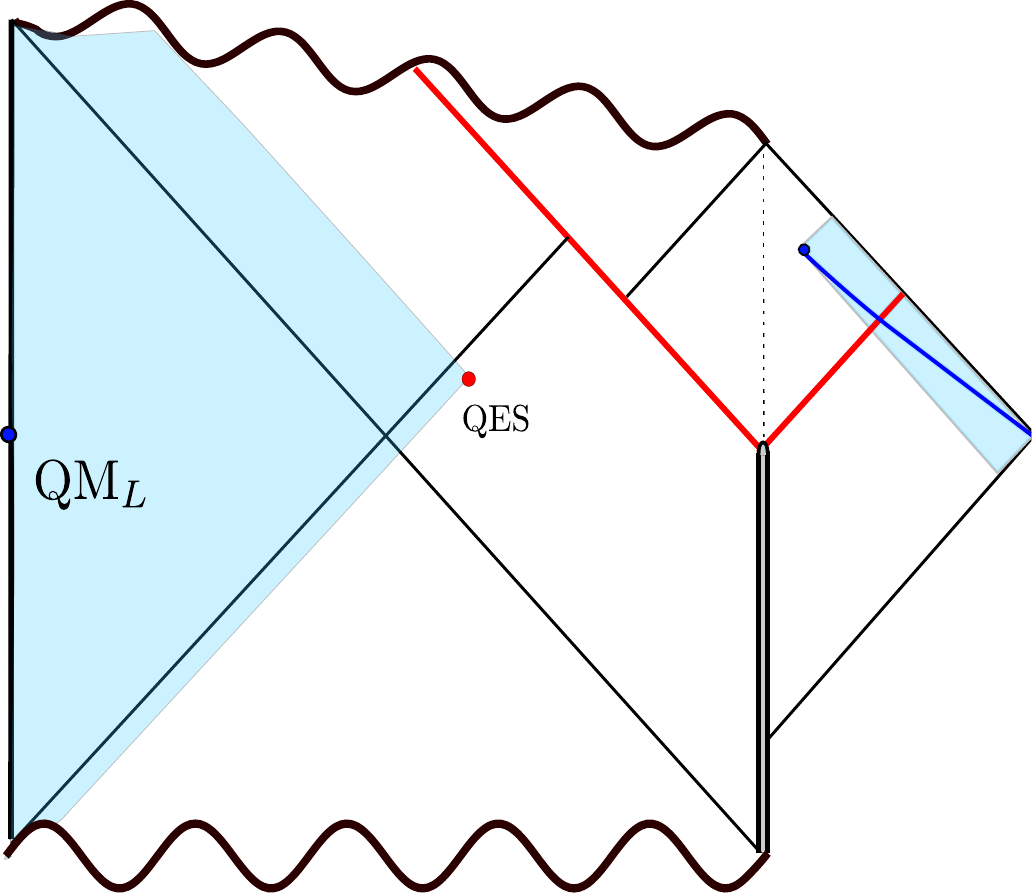}
         \caption{Scrambling phase}
         \label{figure: EW scrambling+QML}
     \end{subfigure}
     \hfill
     \begin{subfigure}[b]{0.45\textwidth}
         \centering
         \includegraphics[scale=0.5]{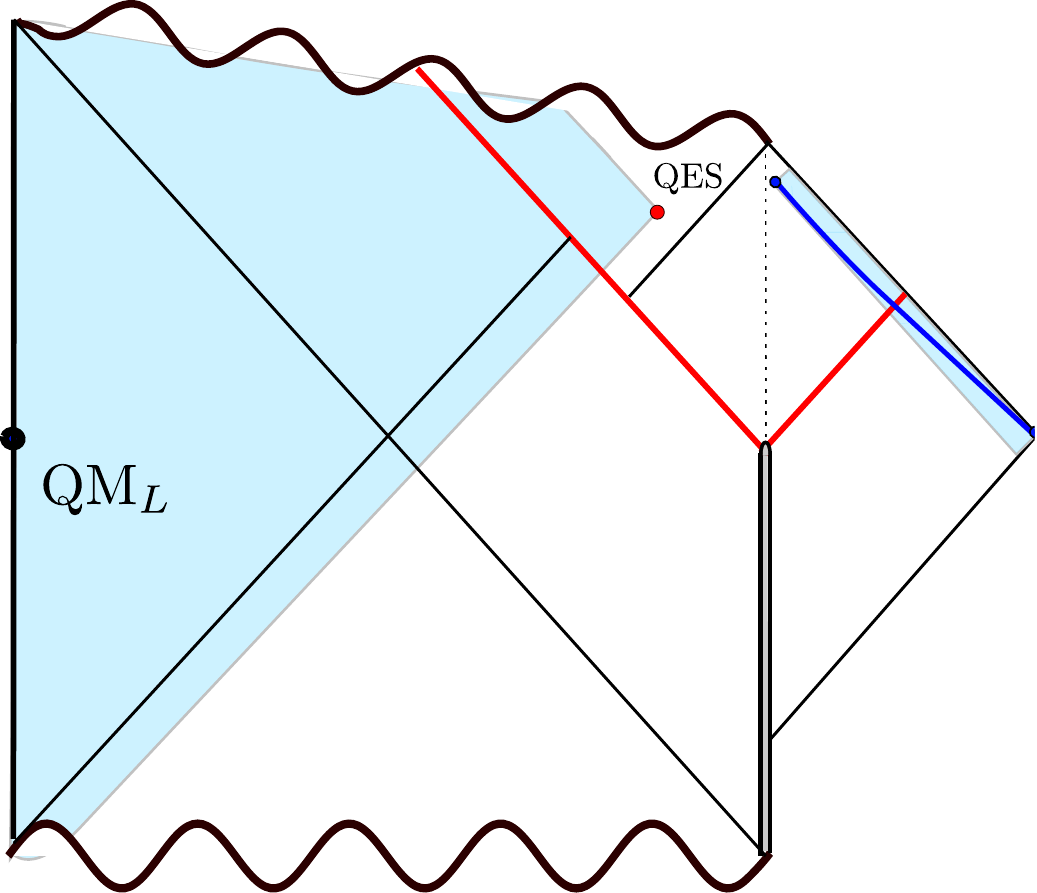}
         \caption{Late phase}
         \label{figure: EW late+QML}
     \end{subfigure}
     \caption{A sketch of the entanglement wedges for the different phases in the Page curve of semi-infinite intervals plus QM$_L$.}
    \label{figure: entanglement wedges of semi-infinite segment at later times+QML}
\end{figure}

Following the prescription outlined in section \ref{section: prescription for Sgen}, we compute the generalized entropy of the radiation segment and the black hole purifier $\mathbf{R} \cup QM_L$. The corresponding Page curve will evolve through four phases with the following dominant saddles and their corresponding generalized entropies\footnote{\label{foot:quench}Notice that the von Neumann entropy of the Trivial and Quench saddles is independent of the position of the QES. Therefore, the location of the QES will extremize the dilaton contribution and thus will corresponds to the bifurcation surface. The contribution to the generalized entropy from the RT surface anchored at the bifurcation surface corresponds to the Bekenstein-Hawking entropy of the initial black hole $S_{\rm BH}(T_0)=S_{\rm vN}\left(x^\pm=\pm \frac{1}{\pi T_0}\right)+ S_\phi\left(x^\pm=\pm \frac{1}{\pi T_0}\right)$.}
\begin{equation}
\begin{aligned}
   &\text{Trivial saddle}\qquad & S^T_{\rm gen}&=S_{\rm vN} \left(x\in \text{IV}\right)+S_{\rm BH},\\
    &\text{Quench saddle}\qquad & S^Q_{\rm gen}&=S_{\rm vN} \left(x\in \text{II}\right)+S_{\rm BH},\\
    &\text{Scrambling saddle}\qquad & S^S_{\rm gen}&=S_{\rm vN} \left(x\in \text{II},x_{QS}\in\text{III}\right)+S_{\phi_{QS}},\\
   &\text{Late (island) saddle}\qquad & S^L_{\rm gen}&=S_{\rm vN} \left(x\in \text{II},x_{QL}\in\text{I}\right)+S_{\phi_{QL}}.
    \label{eqn: schematic Sgen semi-inf+QML}
\end{aligned}
\end{equation}
The entanglement wedges of the corresponding phases are illustrated in figure \ref{figure: entanglement wedges of semi-infinite segment at later times+QML} and are described as follows. 

\paragraph{Trivial saddle} This saddle dominates until the endpoint of the semi-infinite interval has crossed the shock as would be expected by causality.\footnote{A semi-infinite interval of the bath ending below the shock is causally disconnected from the AdS$_2$ space. Since the bath is initially not entangled with the black hole, the entanglement entropy therefore cannot receive contributions from nontrivial QES which depend on the state of the evaporating black hole. This can also be understood from a geometric point of view in the bulk perspective. An HRT geodesic anchored at a nontrivial QES would have to reach deeper into the bulk than the available space opened up by the ETW brane, whose tip falls into the bulk no faster than the speed of light and is therefore within the lightcone produced by the shock.} The three-dimensional description involves an RT surface consisting of two disconnected segments. One of the RT segments connects the endpoint of the bath interval $\mathbf{R}$ to the ETW brane, and another RT segment connects the bifurcation point to the ETW brane. Since the endpoint of the interval is causally disconnected from the quench, the RT surface remains in a static subregion of the bulk geometry. Thus, the length of the RT surface (and therefore the entropy) is constant and given by the Bekenstein-Hawking entropy of the initial black hole with temperature $T_0$ plus the vacuum entanglement entropy of $\mathbf{R}$. The mutual information between $QM_L$ and $\mathbf{R}$ vanishes because there has been no entanglement transferred by Hawking radiation yet.

\paragraph{Quench saddle} The entanglement entropy transitions to a phase dominated by this saddle immediately after the endpoint of the semi-infinite interval crosses the shock wave. The three-dimensional description involves an HRT surface with two disconnected segments. One HRT segment connects the endpoint of the bath interval to the ETW brane, while the other connects the bifurcation point to the ETW brane. Since the ETW brane falls into the bulk, the HRT segment connecting the bath interval to the ETW brane stretches as time evolves leading to a characteristic rapid increase in the entanglement entropy. The generalized entropy remains factorized between a $QM_L$ contribution and an $\mathbf{R}$ contribution at leading order in the central charge of the holographic CFT, implying that the entanglement transferred by the initial Hawking radiation is negligible. Most of the time dependence of the generalized entropy in this phase is due to the quench between the black hole ($QM_R$) and the bath itself, rather than the black hole evaporation, and reproduces known results about entanglement entropy of intervals in quenched systems \cite{Calabrese07,Calabrese09}. 

\paragraph{Scrambling saddle} During the corresponding phase, the entanglement entropy may at first show some transient behaviour characterized by an initial decrease due to scrambling of the shock wave in the black hole. This is the only phase in which scrambling physics can dominate the evolution of entanglement entropy for some time. After the initial perturbation is scrambled, the entropy begins to grow linearly as the segment absorbs more and more Hawking radiation. The three-dimensional description involves an HRT surface connecting the endpoint of the segment to a point $\left(x^+_{QS},x^-_{QS}\right)$ very close to the bifurcation surface. From the brane perspective, the anchor point $\left(x^+_{QS},x^-_{QS}\right)$ constitutes a nontrivial QES on the Planck brane. Correspondingly, there is a small portion of the black hole interior which is reconstructable by $\mathbf{R} \cup QM_L$ as pictured in figure~\ref{figure: EW scrambling+QML}. 

\paragraph{Late saddle} Once transitioned to the phase dominated by this saddle, the segment collects Hawking quanta emitted both at early and at late times and the correlations between those are manifested by the characteristic decrease in the generalized entropy after the Page time. The three-dimensional description involves an HRT surface connecting the endpoint of the segment to a point behind the event horizon of the evaporating black hole and after the shock, $\left(x^+_{QL},x^-_{QL}\right)$. There is a large portion of the black hole interior encoded in $\mathbf{R} \cup QM_L$, as can be seen in figure~\ref{figure: EW late+QML}. The size of the interior portion that is encoded in $\mathbf{R} \cup QM_L$ increases with the growth of the Einstein-Rosen bridge.

When excluding the black hole purifier, the Page curve of the bath interval $\mathbf{R}$ evolves through three phases whose respective entropy saddles are given by
\begin{equation}
\begin{aligned}
   &\text{Trivial saddle}\qquad & S^T_{\rm gen}&=S_{\rm vN} \left(x\in \text{IV}\right),\\
    &\text{Quench saddle}\qquad & S^Q_{\rm gen}&=S_{\rm vN} \left(x\in \text{II}\right),\\
    &\text{Late (island) saddle}\qquad & S^L_{\rm gen}&=S_{\rm vN} \left(x\in \text{II},x_{QL}\in\text{I}\right)+S_{\phi_{QL}}+S_{\rm BH}.
    \label{eqn: schematic Sgen semi-inf}
\end{aligned}
\end{equation}
The Trivial to Quench phase transition occurs at as the endpoint of $\mathbf{R}$ crosses the shock, similarly to the case for $\mathbf{R} \cup QM_L$. The following transition, from the Quench phase to the Late phase, is a true island transition as can be seen in figure \ref{figure: EW late}. Specifically, the Late saddle QES ($x_{QL}$) corresponds to one of the boundaries of the island, the other being the bifurcation surface. This island transition for $\mathbf{R}$ occurs at much later times than the Page transition for $\mathbf{R} \cup QM_L$, in fact, in some cases it may occur beyond the regime of applicability of the semi-classical model. In addition to the three saddles in eq.~\eqref{eqn: schematic Sgen semi-inf}, there is a QES analogous to the one in the Scrambling phase for $\mathbf{R}\cup QM_L$. This saddle would correspond to a phase with a smaller island than the Late phase. However, this saddle always has much greater generalized entropy than at least one of the other three QESs and so it will never correspond to a dominant contribution to the entanglement entropy of $\mathbf{R}$. 

We will use the parameters specified in table~\ref{table section3} as numerical input for all numerical calculations in this section.

\begin{table}[h]
	\centering
	\begin{tabular}{l|c|c|c|c|c|c|c|c| c}
		\hline
		\hline
		\textbf{Parameter} & $L_{\rm AdS}$ & $k$ & $T_1$ & $T_0$ & $c$ & $\Tilde{\phi}_0$ & $\phi_r$ & $\sigma_1$ & $\sigma_2$ \\ \hline
	   Value  & $1$ & $\frac{1}{1024}$ & $\frac{1}{\pi}$ & $\frac{509}{512 \pi}$ & $1024$ & 0 & $\frac{1}{1024^2}$ & $3$ & $\infty$ \\
		\hline
		\hline
	\end{tabular}
	\caption{The numerical values of the various parameters that are used to produce the plots in figures~\ref{figure: semi-infinte interval with QML} and ~\ref{figure: semi-infinte interval without QML late times}.}
	\label{table section3}
\end{table}

\subsection{Reviewing the entanglement entropy evolution with QM$_L$}\label{sec: with QML}
We are most interested in the nontrivial behaviour that occurs after the first endpoint crossed the shockwave in the bath. The three competing saddles during this stage are the Quench saddle, Scrambling saddle and Late saddle of eq.~\eqref{eqn: schematic Sgen semi-inf+QML}, which we can write explicitly as 
\begin{equation}
\begin{aligned}
    & \text{Quench saddle} & S^{\rm Q}_{\rm  gen}&=\frac{c}{6}\log\left(\frac{12 \pi \text{Es}}{c \delta \epsilon } y^- \frac{f(y^+)}{\sqrt{f'\left(y^+\right)}} \right)+\frac{c}{6}\log \frac{2}{\delta}+\frac{2\pi T_0\phi_r+\phi_0}{4 G_N},\\
    & \text{Scrambling saddle} & S^{\rm S}_{\rm gen}&=\frac{c}{6}\log\left(\frac{24 \pi  \text{Es} x^-_{QS} y^- (x^+_{QS}-f(y^+))}{c \delta^2  \epsilon  (x^-_{QS}-x^+_{QS}) \sqrt{f'(y^+)}}\right)\\
    &&&\qquad+\frac{\phi_0+\phi(x^+_{QS},x^-_{QS})}{4 G_N},\\
    & \text{Late saddle} & S^{\rm L}_{\rm gen}&=\frac{c}{6}\log\left( \frac{2 (y^--y^-_{QL}) (x^+_{QL}-f(y^+)) }{\delta^2  \epsilon  (x^-_{QL}-x^+_{QL})}\sqrt{\frac{f'(y^-_{QL})}{f'(y^+)}}\right)\\
    & & &\qquad+\frac{\phi_0+\phi(x^+_{QL},y^-_{QL})}{4 G_N}.
\end{aligned}
\end{equation}
One can absorb the UV cutoff into $\phi_0$ using the redefinition $\Tilde{\phi}_{0}=\phi_0+2k\phi_r \log \frac{1}{\delta}$ and renormalize the generalized entropy $S^{\rm ren}= S^{\rm bare}-\frac{c}{6} \log \frac{1}{\delta\epsilon}$
such that the renormalized entropy is given by
\begin{equation}\label{eq:entropies}
\begin{aligned}
    & \text{Quench saddle} &  S^{\rm Q}_{\rm gen}&=\frac{c}{6}\log\left(\frac{12 \pi \text{Es}}{c  } y^- \frac{f(y^+)}{\sqrt{f'\left(y^+\right)}} \right)+\frac{c}{6}\log 2+\frac{2\pi T_0\phi_r+\Tilde{\phi}_0}{4 G_N},\\
    & \text{Scrambling saddle} &  S^{\rm S}_{\rm gen}&=\frac{c}{6}\log\left(\frac{24 \pi  \text{Es} x^-_{QS} y^- (x^+_{QS}-f(y^+))}{c    (x^-_{QS}-x^+_{QS}) \sqrt{f'(y^+)}}\right)\\
    &&&\qquad+\frac{\Tilde{\phi}_0+\phi(x^+_{QS},x^-_{QS})}{4 G_N},\\
    & \text{Late saddle} &  S^{\rm L}_{\rm gen}&=\frac{c}{6}\log\left( \frac{2 (y^--y^-_{QL}) (x^+_{QL}-f(y^+)) }{   (x^-_{QL}-x^+_{QL})}\sqrt{\frac{f'(y^-_{QL})}{f'(y^+)}}\right)\\
    & & &\qquad+\frac{\Tilde{\phi}_0+\phi(x^+_{QL},y^-_{QL})}{4 G_N}.
\end{aligned}
\end{equation}
The last step towards the Page curve is to extremize the generalized entropy of each saddle independently. As mentioned in footnote~\ref{foot:quench}, the QES in the Quench saddle is at the bifurcation surface, so the only nontrivial QES phases are the Scrambling and Late phase.

Let us first focus on the analytic derivation of the QES solutions for the Scrambling saddle. During the corresponding phase, the QES is located below the shock, $x_{QS}\in$ III, and is a solution to the following two equations 
\begin{align}
    0=\frac{\partial S^{S.}_{gen}\left(x^+_{QS},x^-_{QS}\right)}{\partial x^+_{QS}}&\propto k x^-_{QS} \left(x^-_{QS} - 
     x^+_{QS}\right)  + \left(-1 + \pi^2 T_0^2 x^{2-}_{QS}\right)x^+_{QS}\\
     &\quad + \left(1 - \pi^2 T_0^2 x^{2-}_{QS} + 
     k (-x^-_{QS} + x^+_{QS})\right) f(y^+),\nonumber\\
    0=\frac{\partial S^{S.}_{gen}\left(x^+_{QS},x^-_{QS}\right)}{\partial x^-_{QS}}&\propto\left(k x^{2+}_{QS} - x^-_{QS} \left(-1 + k x^+_{QS} + \pi^2 T_0^2 x^{2+}_{QS}\right)\right).
\end{align}
There is an exact and unique solution to these equations.\footnote{There are several solutions but only one that satisfies  $x_{QS}\in$ III} However, for our purposes it is sufficient to find an approximate solution using the small $k$ expansion
\begin{subequations}\label{eq: scrambling QES}
\begin{align}
    x^-_{QS}&= \frac{-1}{\pi  T_0}+ \frac{k (\pi  T_0 f(u-\sigma)+1)}{\pi^2  T_0^2 (1-\pi  T_0 f(u-\sigma))}+\mathcal{O}\left(k^2\right)\,,
    \label{eq: solution of x^-_Q for scrambing phases text}\\
    x^+_{QS}&= \frac{1}{\pi  T_0}-\frac{k}{\pi^2  T_0^2}+\mathcal{O}\left(k^2\right)\,.
    \label{eq: solution of x^+_Q for scrambing phases text}
\end{align}
\end{subequations}

Now, we turn to the analytic derivation of the QES solutions for the Late phase. During this phase, the QES is located above the shock, $x_{QS}\in$ I, and is a solution to the following two equations 
\begin{equation}
    \partial_{x^+_{QL}} S^{\rm L}_{\rm gen}=0\,, \qquad \partial_{x^-_{QL}} S^{\rm L}_{\rm gen}=0\,,
    \label{eqn: eq for QES late}
\end{equation}
where $ S^L_{\rm gen}$ is given in eq.~\eqref{eq:entropies}. Our aim is to approximate the equations in order to find the approximate solutions for $x^+_{QL}$ and $y^-_{QL}$ that determine the location of the QES. To do this, we use the asymptotic expansion of the reparametrization function in eq.~\eqref{f(u)} for small $k$ and finite $k u$. Writing 
\begin{equation}\label{eq:fapprox}
    x^+ = f(y^+) = t_\infty(1-2\lambda)\,,
\end{equation}
we find that the parameter $\lambda$ is exponentially small for finite $k y^+$\footnote{In fact, for times of the order of the Hayden-Preskill time $u_{HP} = \frac{1}{2\pi T_1} \log\left(\frac{8\pi T_1}{3k} \right)$ \cite{Hayden07}, the parameter $\lambda$ is already much smaller than $\frac{k}{T_1}$.}  
\begin{equation}
    \lambda=\exp\left[-\frac{4 \pi  \left(1-e^{-\frac{k y^+}{2}}\right) T_1}{k}+\frac{k}{4 \pi  T_1}+\mathcal{O}\left(k^2\right)\right]\ll \frac{k}{T_1}.
    \label{eq: lambda def}
\end{equation}
Expanding the equations, first in small $\lambda$ and then in small $k$, leads to the following solution
\begin{subequations}\label{eq: late QES}
\begin{align}
    x^+_{QL}& =t_\infty+\frac{2}{3}t_\infty \lambda+\mathcal{O}\left(k\lambda\right)
    \label{eq: solution of x^+_Q for late phases text}\,,\\
    y^-_{QL}& =y^+-\frac{\log \left(\frac{8 \pi  T_1 }{3 k}\right)}{2 \pi T_1}+\mathcal{O}\left(k\left(\log\left(\frac{8\pi T_1}{3k}\right)\right)^2\right)
    \label{eq: solution of y^-_Q for late phases text}.
\end{align}
\end{subequations}
Note that these solutions are obtained under assumption of small $k$ while $k u$ and $k y^+$ are fixed and finite. The QES solutions in this regime will be valid until and including times of order $u = \mO\left(\frac{1}{k}\right)$. For a more detailed explanation of the derivation of these solutions, we refer the interested reader to appendix \ref{section: QES solution for Late Phases}. 

\begin{figure}[t!]
	\centering
	\includegraphics[scale=0.5]{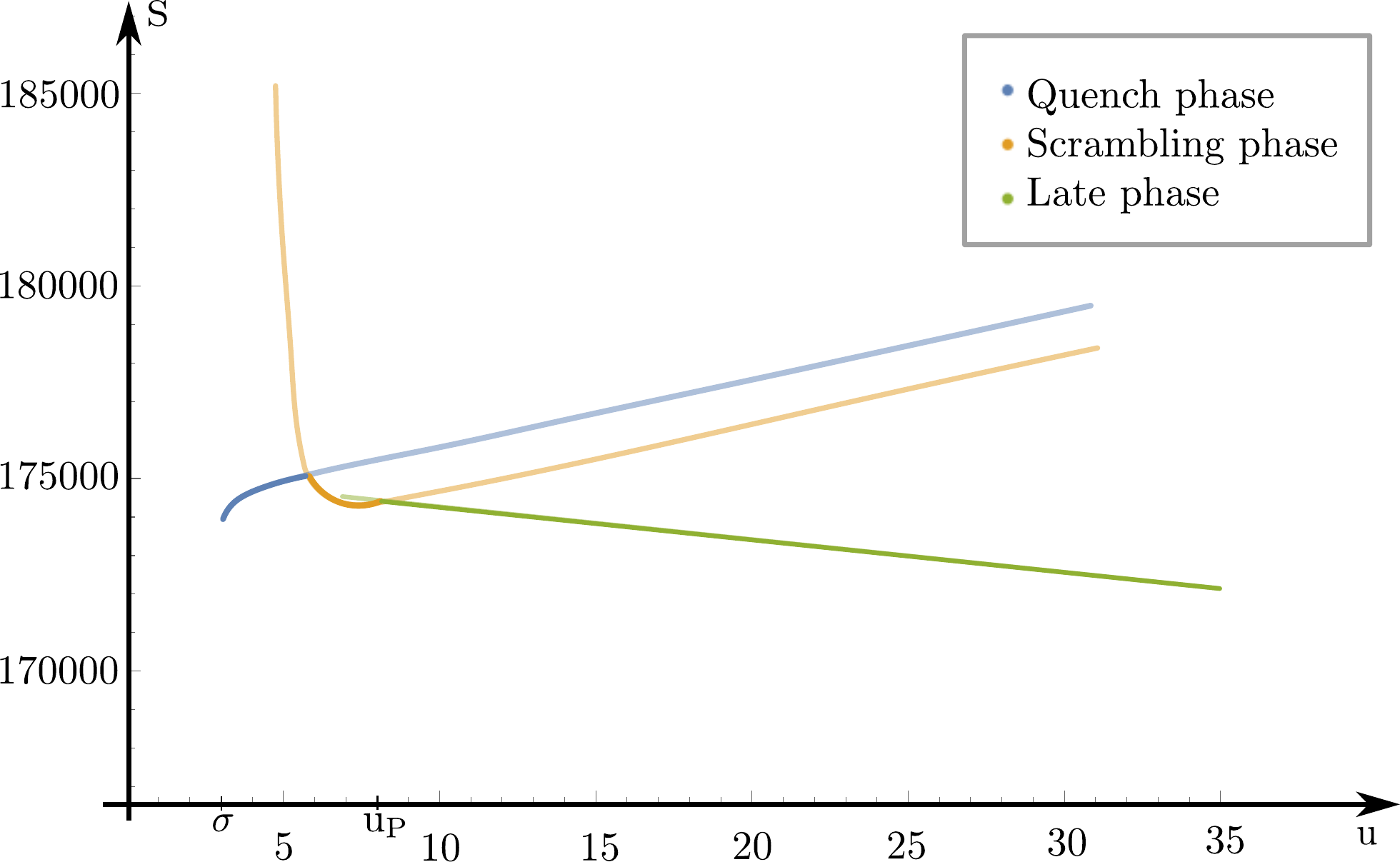} 
	\caption{The generalized entropy of the three saddles of QM$_L$+semi-infinite segment of radiation in the bath. The entanglement entropy is obtained by minimizing between the generalized entropy of these saddles. In this plot, this is reflected in the use of a transparent color whenever the saddle is not dominant. The value of the parameters used in this plot can be found in table~\ref{table section3}.}
	\label{figure: semi-infinte interval with QML}
\end{figure}

The entanglement entropy of $\mathbf{R}\cup QM_L$ after the first endpoint crosses the shock is given by minimizing over the competing saddles in eq.~\eqref{eq:entropies}. An example of the Page curve is plotted in figure~\ref{figure: semi-infinte interval with QML}. To produce the plots, we used the analytic approximations of the QES locations in eqs.~\eqref{eq: scrambling QES} and eqs.~\eqref{eq: late QES} as seeds to numerically minimize the generalized entropies in eq.~\eqref{eq:entropies}. 

After the endpoint crosses the shock, $u=\sigma_1$, the generalized entropy evolves through the Quench phase and subsequently the Scrambling and Late phase. The generalized entropy of the Quench saddle initially increases rapidly, while the generalized entropy of the Scrambling Saddle initially decreases, and so there is a transition from the Quench to the Scrambling phase at~\cite{Chen19}
\begin{equation}
    x^+_1 = \frac{1}{3\pi T_0}-\frac{4k}{9\pi^2T_0^2} + {\cal O}(k^2)\,,
\end{equation}
when $S^{\rm Q}_{\rm gen}=S^{\rm S}_{\rm gen}$. Furthermore, the generalized entropy of the Scrambling saddle transitions to a linearly increasing regime, while the generalized entropy of Late saddle decreases linearly with time. When the two generalized entropies intersect, there is a Page transition between the Scrambling phase and the Late phase at the so-called Page time $u_P=y^+_P+\sigma$ with
\begin{equation}
    \label{eqn: page time with uhp}
\begin{split}
    y^+_P&=\frac{2}{3}\frac{T_1-T_0}{k T_1}+u_{HP}-\frac{1}{3\pi T_1}\\
    &\quad-\frac{2}{3\pi T_1}\log \left(\frac{\pi^2 \left(T_1-T_0\right)^2 (T_1 +T_0) (\sigma_1+u^0_P)}{2k^2 T_0  (u_{HP}+2  \sigma_1 )}\right)+\mathcal{O}\left(k\right)\,.
    \end{split}
\end{equation}
In the above we have used the Hayden-Preskill time $u_{HP} = \frac{1}{2\pi T_1} \log\left(\frac{8\pi T_1}{3k} \right)$. The calculation can be found in appendix~\ref{section: Bounds on blind spot}, specifically in the section leading to eq.~\eqref{eq:Pagetime}.  

The leading term in eq.~\eqref{eqn: page time with uhp} can be understood by considering the main features of the generalized entropies of the Scrambling QES and the Late QES. The initial value of the Scrambling generalized entropy is approximately the entropy of the initial black hole with temperature $T_0$, and increases with a slope proportional to the temperature $T_1$, that is, $S_{\rm gen}^{S} = \frac{\phi_r}{4G_N} \left( 2\pi T_0+ 2\pi T_1 k u \right)+ \cdots$. On the other hand the generalized entropy of the Late QES begins approximately at the entropy of the perturbed black hole with temperature $T_1$ and decreases with a slope proportional to $T_1$, but at half the rate compared to the Scrambling generalized entropy, \ie $S_{\rm gen}^{L} = \frac{\phi_r}{4G_N} \left( 2\pi T_1- \pi T_1 k u \right)+ \cdots$. The leading term in the Page time given in eq.~\eqref{eqn: page time with uhp} is the time required to close the gap between the entropy of the initial and the perturbed black hole. The second term is a delay from subleading corrections to the generalized entropies and coincides with the Hayden-Preskill scrambling time. We have organized the terms in eq.~\eqref{eqn: page time with uhp} by assuming the following scaling $\frac{T_1}{k} \gg \frac{T_1-T_0}{k} > \log\frac{T_1}{k} >1$.

\subsection{Entanglement entropy evolution without QM$_L$}
\label{sec: without QML}

\begin{figure}[t!]
     \centering
     \begin{subfigure}[b]{0.3\textwidth}
         \centering
         \includegraphics[scale=0.8]{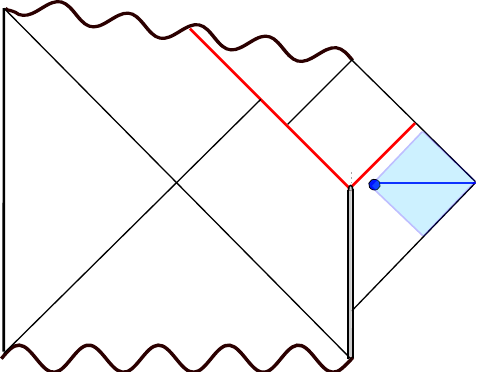}
         \caption{Trivial phase}
         \label{figure: EW trivial}
     \end{subfigure}
     \hfill
     \begin{subfigure}[b]{0.3\textwidth}     
         \centering
         \includegraphics[scale=0.33]{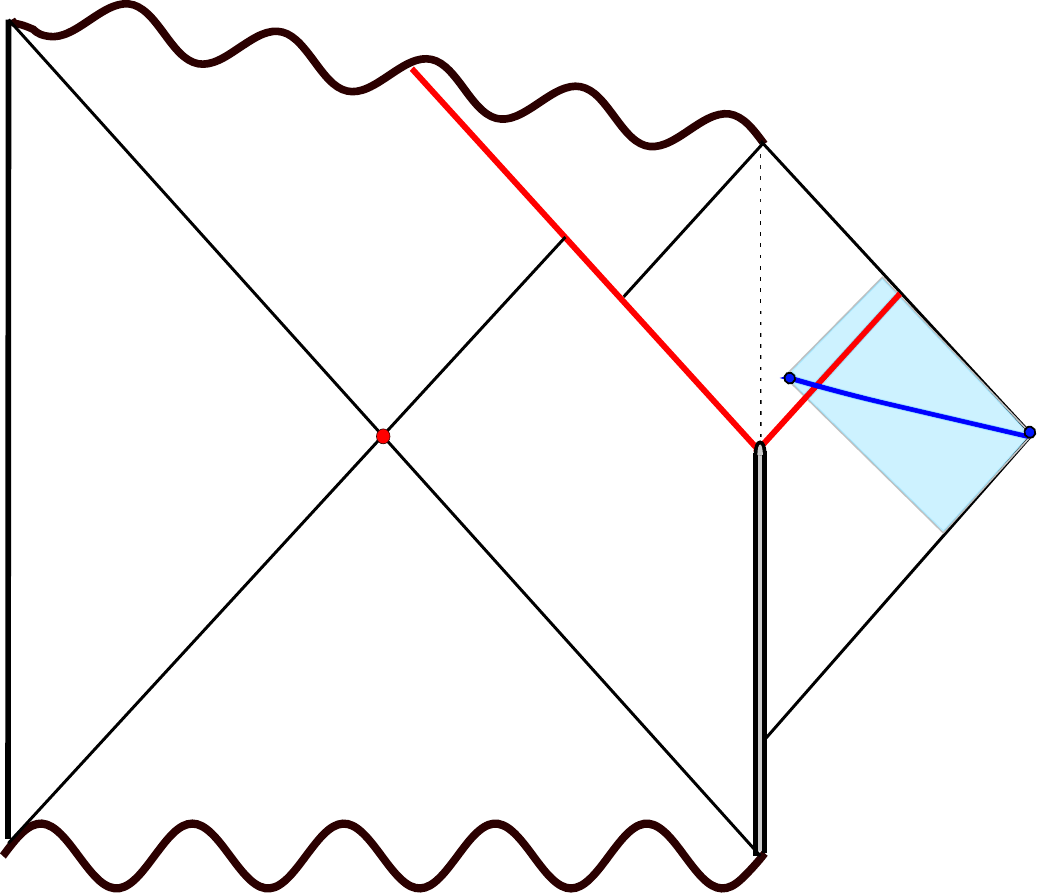}
         \caption{Quench phase}
         \label{figure: EW quench}
     \end{subfigure}
     \hfill
     \begin{subfigure}[b]{0.3\textwidth}
         \centering
         \includegraphics[scale=0.34]{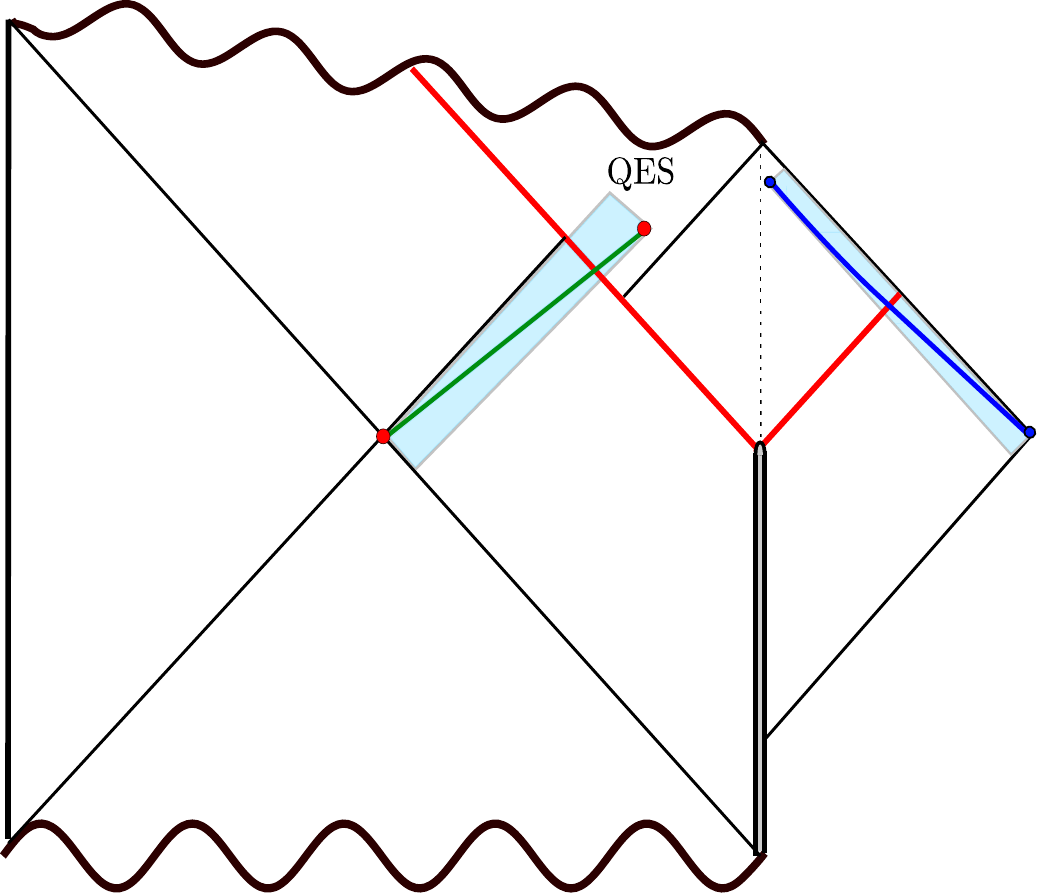}
         \caption{Late phase}
         \label{figure: EW late}
     \end{subfigure}
     \caption{A sketch of the entanglement wedges for the different phases of the entanglement entropy for semi-infinite intervals without QM$_L$.}
    \label{figure: entanglement wedges of semi-infinite segment at later times}
\end{figure}

Now, let us explore what happens when QM$_L$ is not included. Following the prescription outlined in section \ref{section: prescription for Sgen} and performing the redefinition $\Tilde{\phi}_{0}=\phi_0+2k\phi_r \log \frac{1}{\delta}$ and renormalization $S^{\rm ren}= S^{\rm bare}-\frac{c}{6} \log \frac{1}{\delta\epsilon}$, one finds that the relevant expressions of eq.~\eqref{eqn: schematic Sgen semi-inf} after the first endpoint crossed the shock reduce to
\begin{equation}\label{eq:withoutQML}
    \begin{aligned}
        &\text{Quench saddle} &  S^{\rm Q}_{\rm gen}&= \frac{c}{6}\log\left(\frac{12 \pi E_S}{c  } y^- \frac{f(y^+)}{\sqrt{f'\left(y^+\right)}} \right)\\
    &\text{Scrambling saddle} &  S^{\rm S}_{\rm gen}&=\frac{c}{6}\log\left(\frac{24 \pi  E_S x^-_{QS} y^- (x^+_{QS}-f(y^+))}{c   (x^-_{QS}-x^+_{QS}) \sqrt{f'(y^+)}}\right)\\
    &&&\qquad+\frac{\Tilde{\phi}_0+\phi(x^+_{QS},x^-_{QS})}{4 G_N}+\frac{c}{6}\log 2+\frac{2\pi T_0\phi_r+\Tilde{\phi}_0}{4 G_N},\\
    &\text{Late saddle} &  S^{\rm L}_{\rm gen}&=\frac{c}{6}\log\left( \frac{2 (y^--y^-_{QL}) (x^+_{QL}-f(y^+)) }{ (x^-_{QL}-x^+_{QL})}\sqrt{\frac{f'(y^-_{QL})}{f'(y^+)}}\right)\\&&&\qquad+\frac{\Tilde{\phi}_0+\phi(x^+_{QL},y^-_{QL})}{4 G_N}
    +\frac{c}{6}\log 2+\frac{2\pi T_0\phi_r+\Tilde{\phi}_0}{4 G_N}.
    \end{aligned}
\end{equation}
The entanglement wedge of each phase is illustrated in figure \ref{figure: entanglement wedges of semi-infinite segment at later times}. The Late phase is a true island phase, in contrast to the Late phase of section~\ref{sec: with QML} as is shown in figure \ref{figure: EW late+QML}. As was mentioned towards the end of section~\ref{sec:vN entropy}, around eq.~\eqref{eq: r vs nr 2}, the island saddles for $\mathbf{R}$, which can reconstruct the black hole interior, have much larger generalized entropy than the Quench saddle at times comparable to the Page time of $\mathbf{R}\cup QM_L$, since
\begin{equation}
    S^{\rm NR}_{\rm gen}(\mathbf{R}\cup QM_L) - S^{\rm R}_{\rm gen}(\mathbf{R} \cup QM_L) >0\quad \Rightarrow\,\quad S^{\rm NR}_{\rm gen}(\mathbf{R}) - S^{\rm R}_{\rm gen}(\mathbf{R} ) >-2S_{\rm BH}(T_0).
\end{equation}
Hence, the island transition occurs at much later times than the transitions into nontrivial QES phases of section~\ref{sec: with QML}. Moreover, the only transition that may occur, is an island transition from the Quench phase to the Late phase. In what follows, we will determine when (and under which conditions) this island transition will occur.

In order to find the island transition, we look at the asymptotic behaviour of the Quench and Late saddle. The late-time behaviour of the Quench saddle entropy in eq.~\eqref{eq:withoutQML} is\footnote{Recall that the semi-classical picture breaks down for times of order $\frac{1}{k}\log\frac{1}{k}$, so here we take very late times to be ${\cal O}(\frac{1}{k})$.}
\begin{equation}\label{eq:SQ-no-QML}
     S^{\rm Q}_{\rm gen} \approx \frac{c}{6} \left(\frac{2\pi T_1}{k} (1-e^{-k y^+/2}) + \log \left(\frac{6\pi E_S}{c}\frac{y^-}{\pi T_1} \right) + \frac{k}{4} \left( y^+ + \frac{1}{2\pi T_1}\right) + {\cal O}(k^2) \right) 
\end{equation}
which asymptotes to $\frac{c}{6} \left(\frac{2\pi T_1}{k} (1-e^{-\alpha/2}) +  \log \left(\frac{6\pi E_S}{c k}\frac{\alpha}{\pi T_1} \right)+\frac{\alpha}{4} + {\cal O}(k)\right)$ from below for $y^+\to \frac{\alpha}{k}$ for some finite $\alpha$.

\begin{figure}[t!]
	\centering
	\includegraphics[scale=0.6]{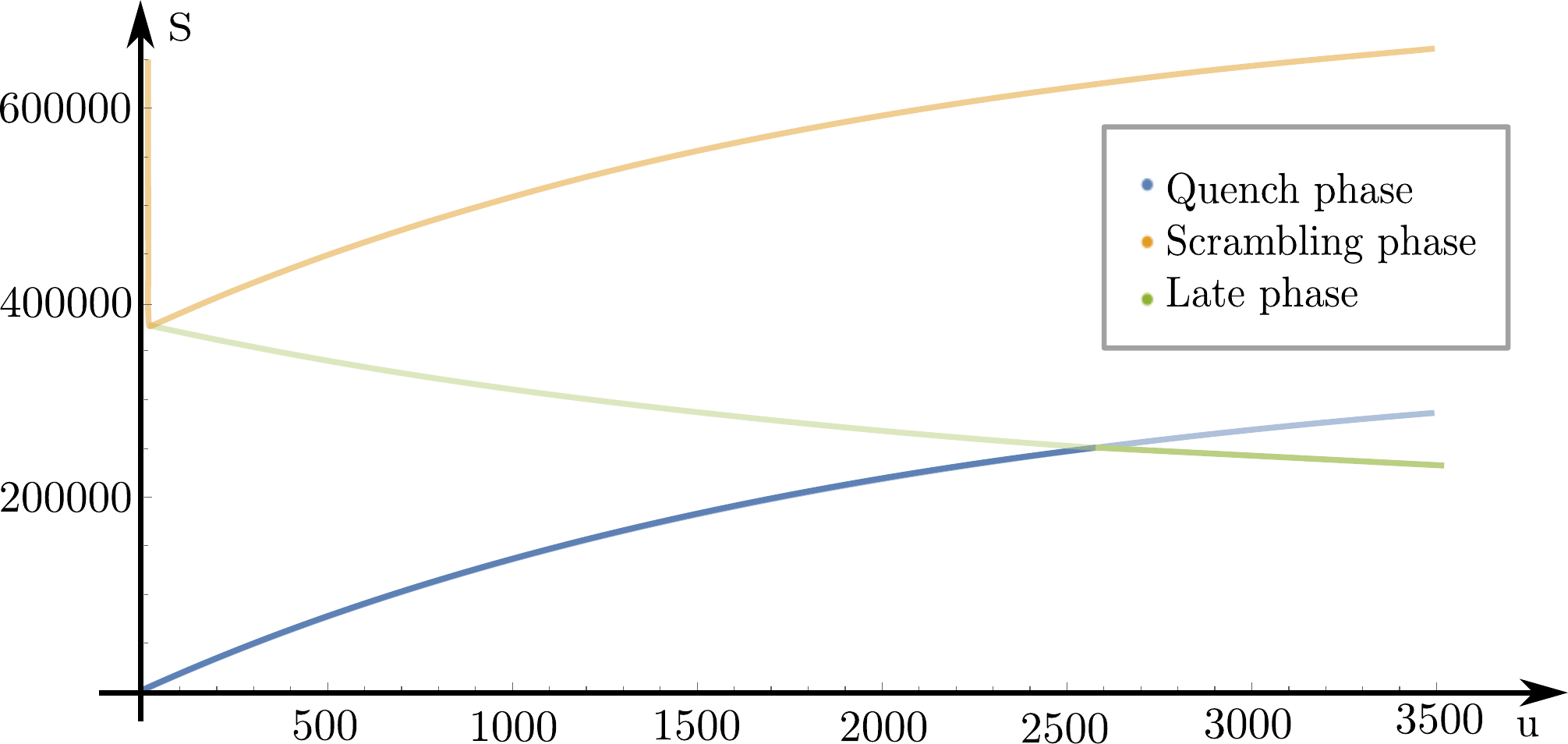} 
	\caption{The generalized entropy of the three saddles for a semi-infinite segment of radiation in the bath. Under the condition that eq.~\eqref{eq:phi0-condition} holds, the transition from the Quench phase to the Late phase happens around $u_{island}$ specified in eq.~\eqref{eqn: very late times}. This transition is a true island transition where an island forms in the Late phase. The parameters used for this figure can be found in table~\ref{table section3}.}
	\label{figure: semi-infinte interval without QML late times}
\end{figure}

The late time behaviour of the Late saddle entropy can be expanded by plugging the position of the late time QES given in eqs.~\eqref{eq: late QES} into the corresponding generalized entropy in eq.~\eqref{eq:withoutQML}
\begin{equation}\label{eq:SL-no-QML}
\begin{split}
     S^{\rm L}_{\rm gen} &\approx \frac{c}{6} \left[ \frac{\Tilde{\phi}_0 + \phi_r(\pi T_0 + \pi T_1 e^{-k y^+/2})}{k \phi_r} \right.\\
     &\left.+ \log\left(\sqrt{\frac{8k}{3\pi^3 T_1^3}}\log\left(\frac{8\pi T_1}{3k}\right) \right) + \frac{e^{-k y^+/2}}{4}  \log\left(\frac{8\pi T_1}{3k}\right) + {\cal O} (k^2)\right]\,.
     \end{split}
\end{equation}
This entropy decreases towards 
\begin{equation}
\begin{split}
    &\frac{c}{6} \left[ \frac{\Tilde{\phi}_0  + \phi_r(\pi T_0 + \pi T_1 e^{- \alpha/2})}{k \phi_r} +\log\left(\sqrt{\frac{8k}{3\pi^3 T_1^3}}\log\left(\frac{8\pi T_1}{3k}\right) \right)  \right.\\
    &\left.\qquad + \frac{e^{-\alpha/2}}{4}  \log\left(\frac{8\pi T_1}{3k}\right) + {\cal O} (k^2)\right]
    \end{split}
\end{equation} 
from above for $y^+ \to \frac{\alpha}{k}$.

Comparing eq.~\eqref{eq:SQ-no-QML} and eq.~ \eqref{eq:SL-no-QML}, it is evident that for the two saddles to intersect in the regime of applicability of the semi-classical model, we need
\begin{equation}\label{eq:phi0-condition}
    \Tilde{\phi}_0  < (2\pi T_1-\pi T_0) \phi_r\,.
\end{equation}
At leading order, this bound constrains the vacuum entropy $S_{\rm vac} = \frac{\Tilde{\phi}_0}{4G_N}+\frac{c}{6}\log2$ to be lower than the increase in entropy $\Delta S=S_{\rm BH}-S_{\rm vac}$ of a black hole with temperature $T_1$ minus half the increase in entropy of a black hole with temperature $T_0$. In this case, the island transition occurs at
\begin{equation}
    y^+_{\rm island}= \frac{2}{k} \log\left(\frac{3\pi T_1 \phi_r}{(2\pi T_1 - \pi T_0)\phi_r -\Tilde{\phi}_0 } \right) + {\cal O}\left(\frac{1}{T_1}\right)\,.
    \label{eqn: very late times}
\end{equation}
Assuming that eq.~\eqref{eq:phi0-condition} holds by more than ${\cal O}(k)$, \textit{i.e.}, $(2\pi T_1-\pi T_0) \phi_r -\Tilde{\phi}_0  > {\cal O}(1)$, one can estimate the earliest time for which an island transition can occur in this model. Using $y^+ = u-\sigma_1$ and taking $\sigma_1 \to 0$, the earliest island forms at times of ${\cal O}\left(\frac{1}{k} \right)$.

We use numerics to plot the time evolution of $S_{\rm gen}$ for each saddle as shown in figure \ref{figure: semi-infinte interval without QML late times}.\footnote{In doing so, we use the analytic approximations in eqs.~\eqref{eq: scrambling QES} and eqs.~\eqref{eq: late QES} as seeds for our numerical computations.} We find that after the endpoint crosses the shock, $u=\sigma$, the generalized entropy remains in the Quench phase for very long times, until eventually it transitions to the Late phase, in which there is an island.

\section{Page curve of finite radiation segments}
\label{sec:Finite+QML}

\begin{figure}[h!]
	\centering
	\includegraphics[scale=0.7]{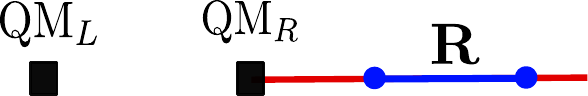}
	\caption{A finite radiation segment from the boundary perspective.}
	\label{fig finite R}
\end{figure}

In this section, we will explore the Page curve of finite radiation segments $\mathbf{R}$ in the bath and consider the black hole purifier $QM_L$ to be part of the subsystem. That is, we compute the Page curve of $\mathbf{R} \cup {QM_L}$.\footnote{The Page curve of finite radiation segments in the bath without the black hole purifier generically does not evolve through nontrivial QES phases.} 
Let us assume that the interval $\mathbf{R}$ spans the space between two endpoints in the bath, both at a fixed spatial location finite location, as shown in figure~\ref{fig finite R}.

As before, at the early stages of the evaporation (more specifically, while the second endpoint of the interval has not crossed the shock), $\mathbf{R}$ collects more and more Hawking radiation. Eventually, $\mathbf{R}$ may contain enough information so that $\mathbf{R} \cup QM_L$ can reconstruct a portion of the black hole interior. However, in contrast to the semi-infinite interval case of section \ref{sec:semi-infinite}, this is not guaranteed in this case. The reason is that when the second endpoint crosses the shock, further evolving in time no longer amounts to capturing more Hawking radiation overall. Instead, as the interval moves forward in time, later Hawking radiation is captured at the expense of early Hawking radiation which is no longer being captured by the interval $\mathbf{R}$. For this reason, in order for interior reconstruction to be possible for the subsystem $\mathbf{R}\cup QM_L$, the transition to a reconstructing phase should happen before the second endpoint crosses the shock. If the transition occurs, the reconstruction is only possible for a finite amount of time - eventually the interior reconstruction window ends. This happens because 1) the flux of Hawking radiation decreases as the temperature of the black hole decreases throughout the evaporation, and 2) the quantum information about the interior becomes more and more diluted in the radiation, so one would need more radiation to reconstruct the same amount of interior data \cite{Chen19}. 

Following the prescription outlined in section \ref{section: prescription for Sgen}, we compute the entanglement entropy of the radiation segment and the black hole purifier $\mathbf{R} \cup QM_L$. We discuss all the possibilities for the corresponding Page curve. For this, we divide the time evolution into $3$ distinct time regimes, as discussed in section \ref{ssection: time windows}. We discuss the possible HRT saddles that can dominate in each time regime and give the corresponding generalized entropy. Similarly to section~\ref{sec:semi-infinite}, we use the redefinition $\Tilde{\phi}_{0}=\phi_0+2k\phi_r \log \frac{1}{\delta}$ and the following renormalization $S^{\rm ren}= S^{\rm bare}-\frac{c}{6} \log \frac{1}{\delta^2\epsilon^2}$. In section~\ref{ssection: page curve}, we discuss the Page curve. The Page curve itself is largely robust to changes of parameters, such as  $\left(l,T_0,T_1,k\right)$, where $l$ is length of the segment, $T_0$ and $T_1$ are the initial and final temperature of the black hole and $k$ is the back-reaction parameter. However, we show that there is an important difference in the sequence of dominant QESs which strongly depend on the chosen parameters. Importantly, the dominance of QES determines when a portion of the black hole interior is reconstructable by $\mathbf{R} \cup QM_L$. We resort to numerics in order to plot different examples in which the Page curve evolves through different QES paths. The main results of this section can be summarized by figures~\ref{figure: finite segments intermediate times}, \ref{fig:flowchart} and \ref{figure: phase diagram for blind spot}, which provide examples of the possible Page curves, the sequence of phases that can dominate the Page curve and a phase diagram highlighting the possible temporary interruption of the interior reconstruction window, respectively. 

For the numerical plots in this section, we use the sets of parameters given in table~\ref{table section4}. 
\begin{table}[h]
	\centering
	\begin{tabular}{l|c|c|c|c|c|c|c|c| c}
		\hline
		\hline
		\textbf{Parameter} & $L_{\rm AdS}$ & $k$ & $T_1$ & $c$ & $\Tilde{\phi}_0$ & $\phi_r$ & $\sigma_1$ & $\sigma_2$ \\ \hline
		 Value & $1$ & $ 0.0025995745070$ & $\frac{1}{\pi}$  & $1024$ & 0 & $\frac{1}{1024^2}$ & $1$ & $30$ \\ 
		\hline
		\hline
	\end{tabular}
	\caption{The numerical values of the various parameters that are used to produce the plots in figures~\ref{figure: finite segments intermediate times}, \ref{figure: phase diagram for blind spot} and \ref{figure: subregion complexity finte inteval with QML intermediate times}. The value of $T_0$ was varied to realize the three reconstructing scenarios described in section~\ref{ssection: page curve}. Notice that this set of parameters is carefully chosen to satisfy eq.~\eqref{eqn:first bound} with $T_0=\frac{459}{512\pi}$.}
	\label{table section4}
\end{table}

\subsection{Catalogue of QESs}\label{ssection: time windows}

\begin{figure}
  \centering
  \includegraphics[width=.4\linewidth]{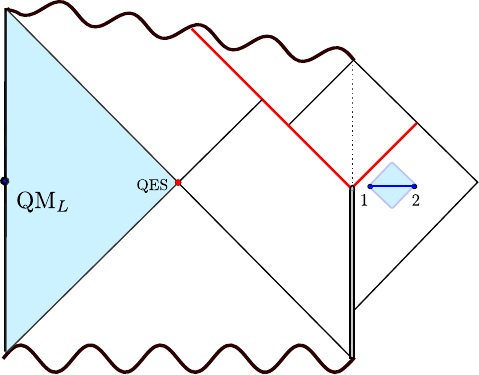}
\caption{ A sketch of the entanglement wedges for the Early phases of the entanglement entropy during early times}
\label{fig: entanglementwedge of Early phase}
\end{figure}

\begin{figure}[t]
     \centering
     \begin{subfigure}{0.3\textwidth}
         \centering
         \includegraphics[width=\textwidth]{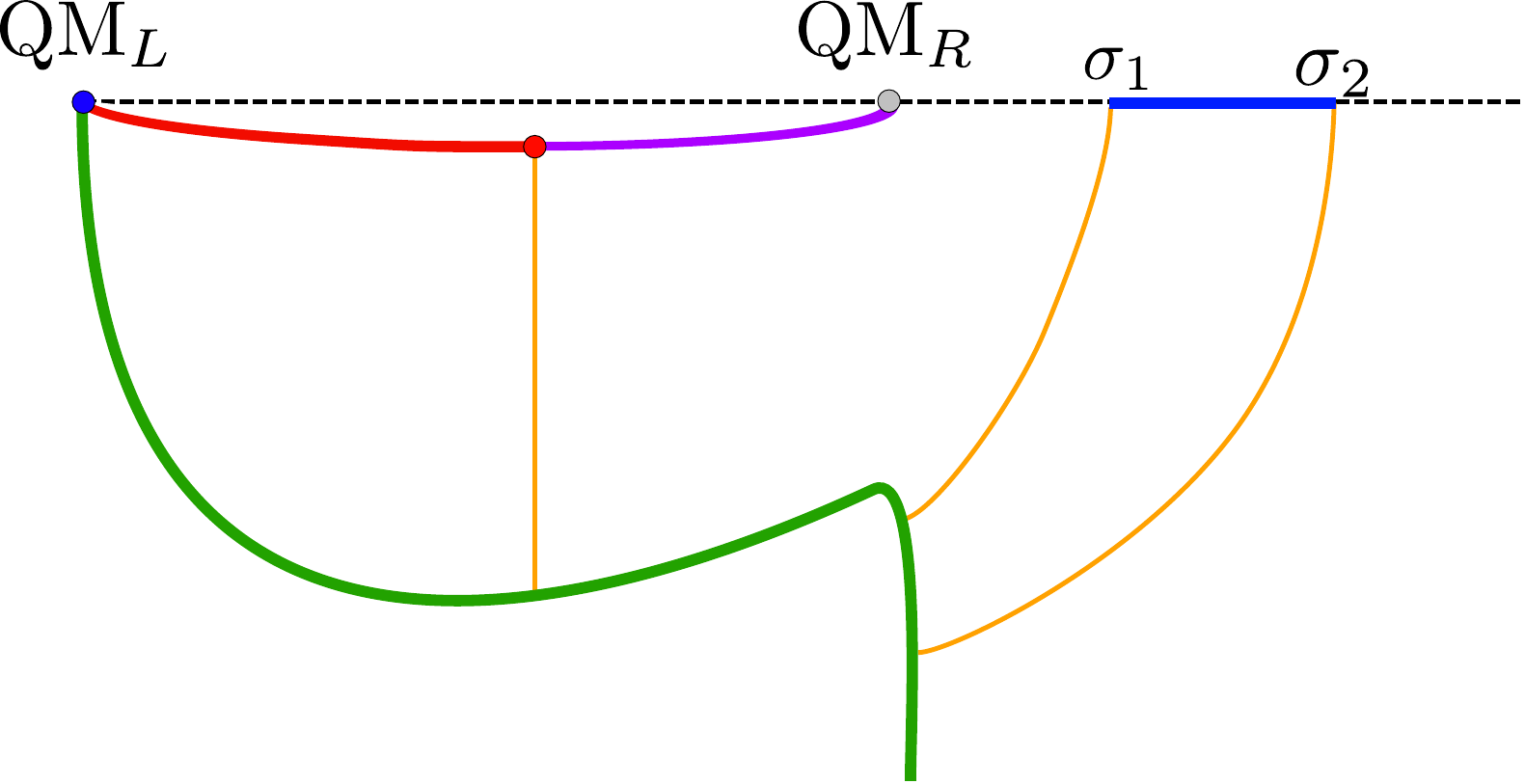}
         \caption{}
         \label{figure: HRT-disc}
     \end{subfigure}
     \hfill
     \begin{subfigure}{0.3\textwidth}
        \centering
         \includegraphics[width=\textwidth]{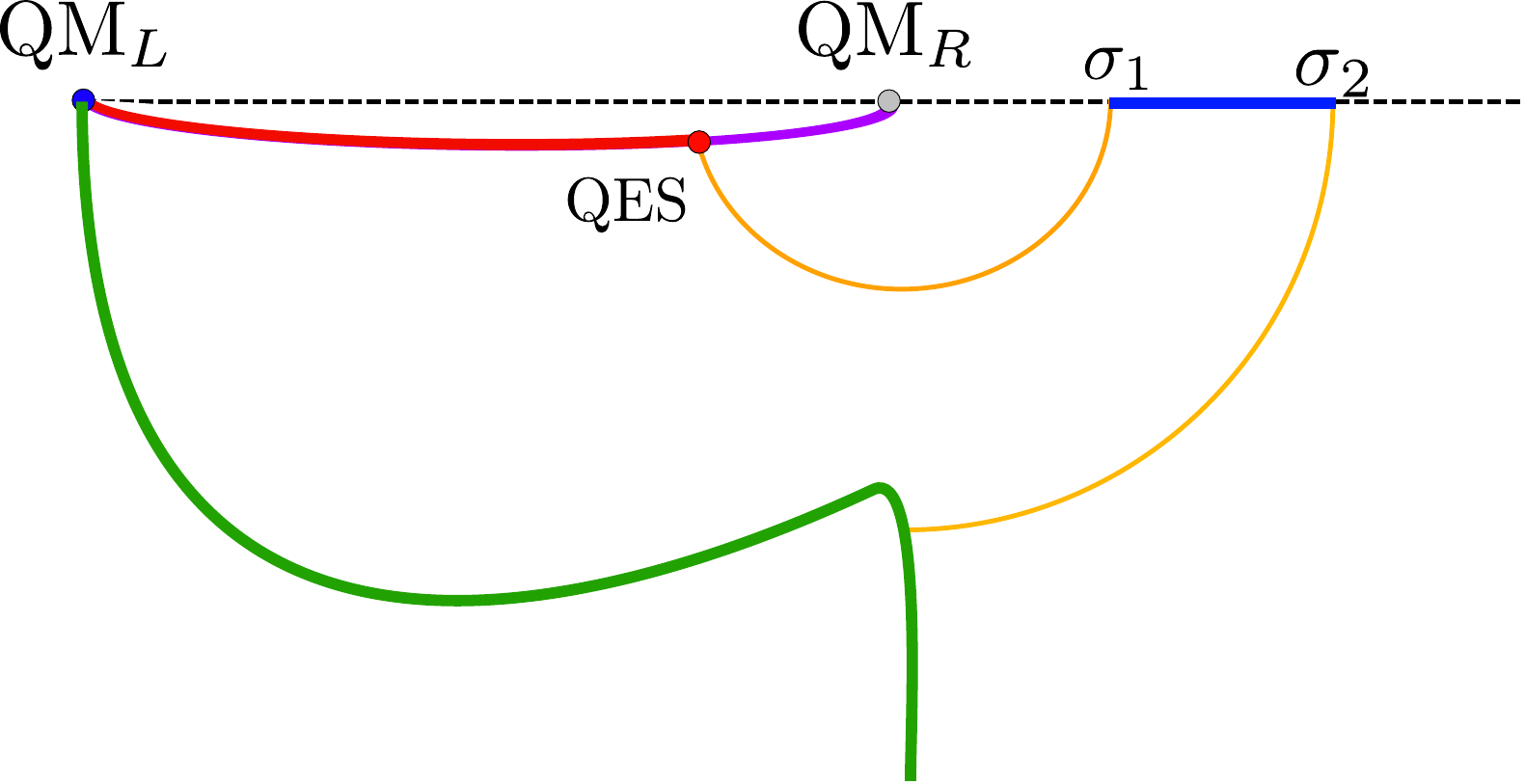}
        \caption{}
         \label{figure: HRT-qes}
     \end{subfigure}
     \hfill
     \begin{subfigure}{0.3\textwidth}
         \centering
         \includegraphics[width=\textwidth]{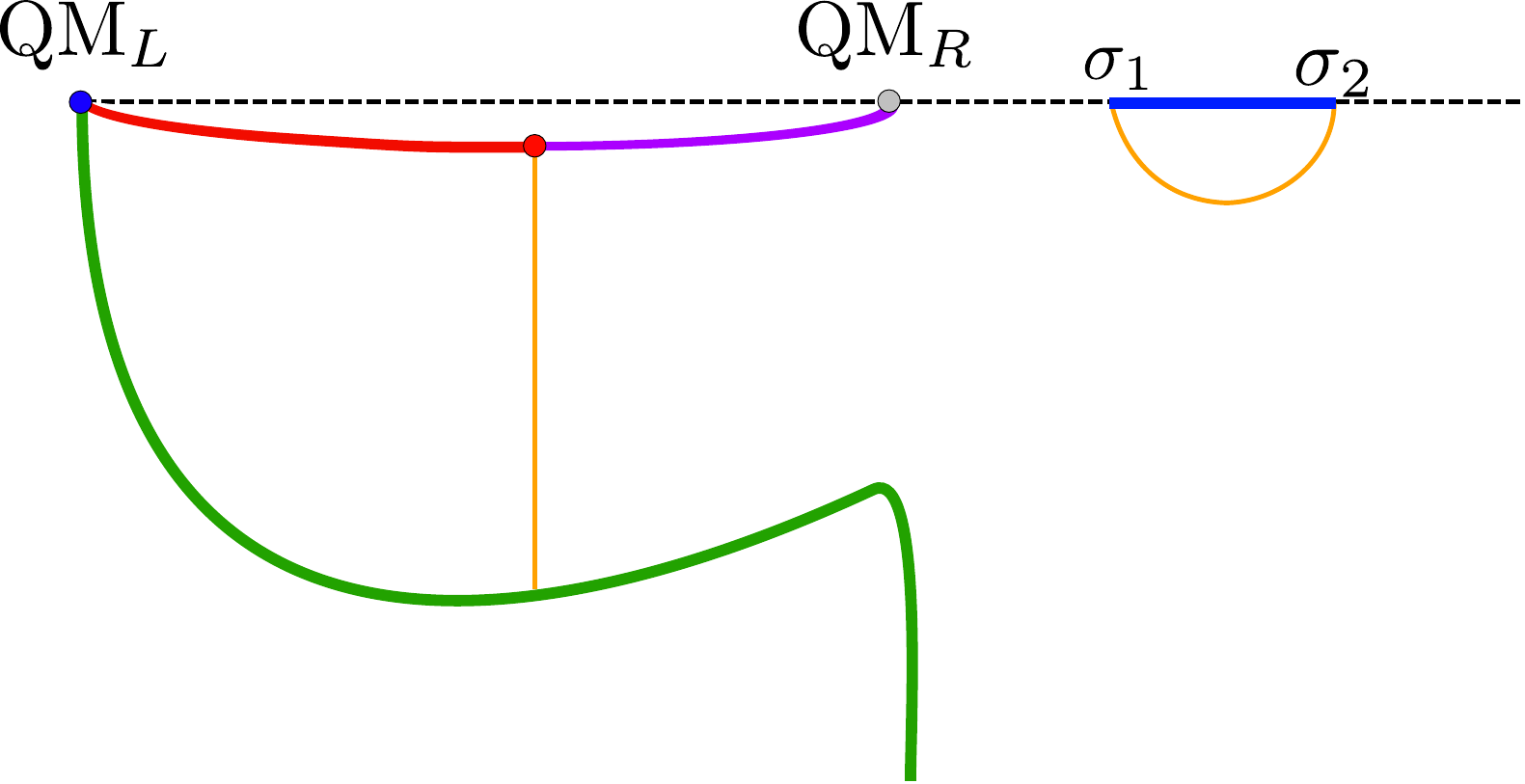}
        \caption{}
         \label{figure: HRT-conn}
     \end{subfigure}
     \caption{HRT geodesic configurations that can compute the matter part of the generalized entropy at intermediate times. (a) CrossDisconnected saddle, (b) ScramblingTrivial and LateTrivial saddles (differing by the position of QES with respect to the shock, not shown), (c) CrossConnected saddle. The red region of the Planck brane is reconstructable by $\mathbf{R}\cup QM_L$ when the corresponding saddle dominates.}
    \label{figure: HRT at intermediate times}
\end{figure}

\begin{figure}[t!]
     \centering
     \begin{subfigure}[b]{0.3\textwidth}
         \centering
         \includegraphics[width=\textwidth]{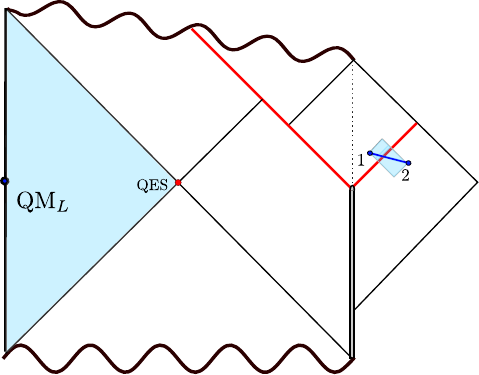}
         \caption{Cross phases}
         \label{figure: EW cross}
     \end{subfigure}
     \hfill
     \begin{subfigure}[b]{0.3\textwidth}
         \centering
         \includegraphics[width=\textwidth]{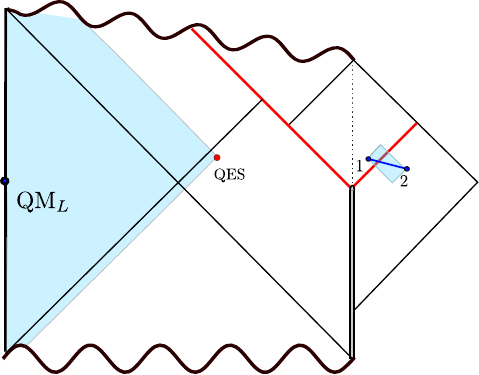}
         \caption{ScramblingTrivial phase}
         \label{figure: EW scramblingtrivial}
     \end{subfigure}
     \hfill
     \begin{subfigure}[b]{0.3\textwidth}
         \centering
         \includegraphics[width=\textwidth]{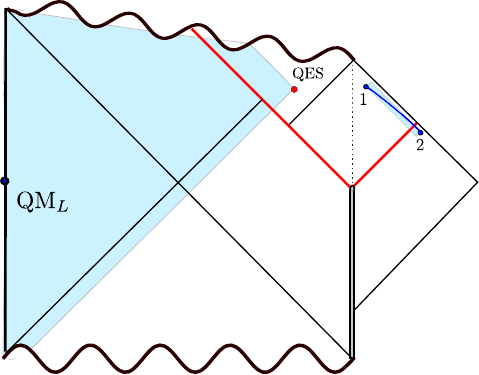}
         \caption{LateTrivial phase}
         \label{figure: EW latetrivial}
     \end{subfigure}
     \caption{ A sketch of the entanglement wedges for the different phases of the entanglement entropy for QM$_L$+finite segment during intermediate times}
    \label{figure: entanglement wedges of QML+finite segment at intermediate times}
\end{figure}

\begin{figure}[t]
      \centering
      \begin{subfigure}[b]{0.4\textwidth}
          \centering
          \includegraphics[scale=1]{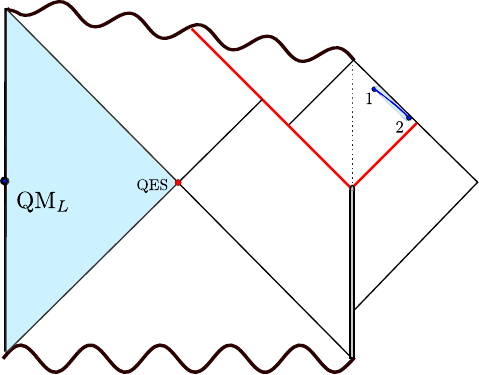}
          \caption{FutureConnected phase}
          \label{figure: EW future}
      \end{subfigure}
      \hfill
      \begin{subfigure}[b]{0.4\textwidth}
          \centering
          \includegraphics[scale=0.30]{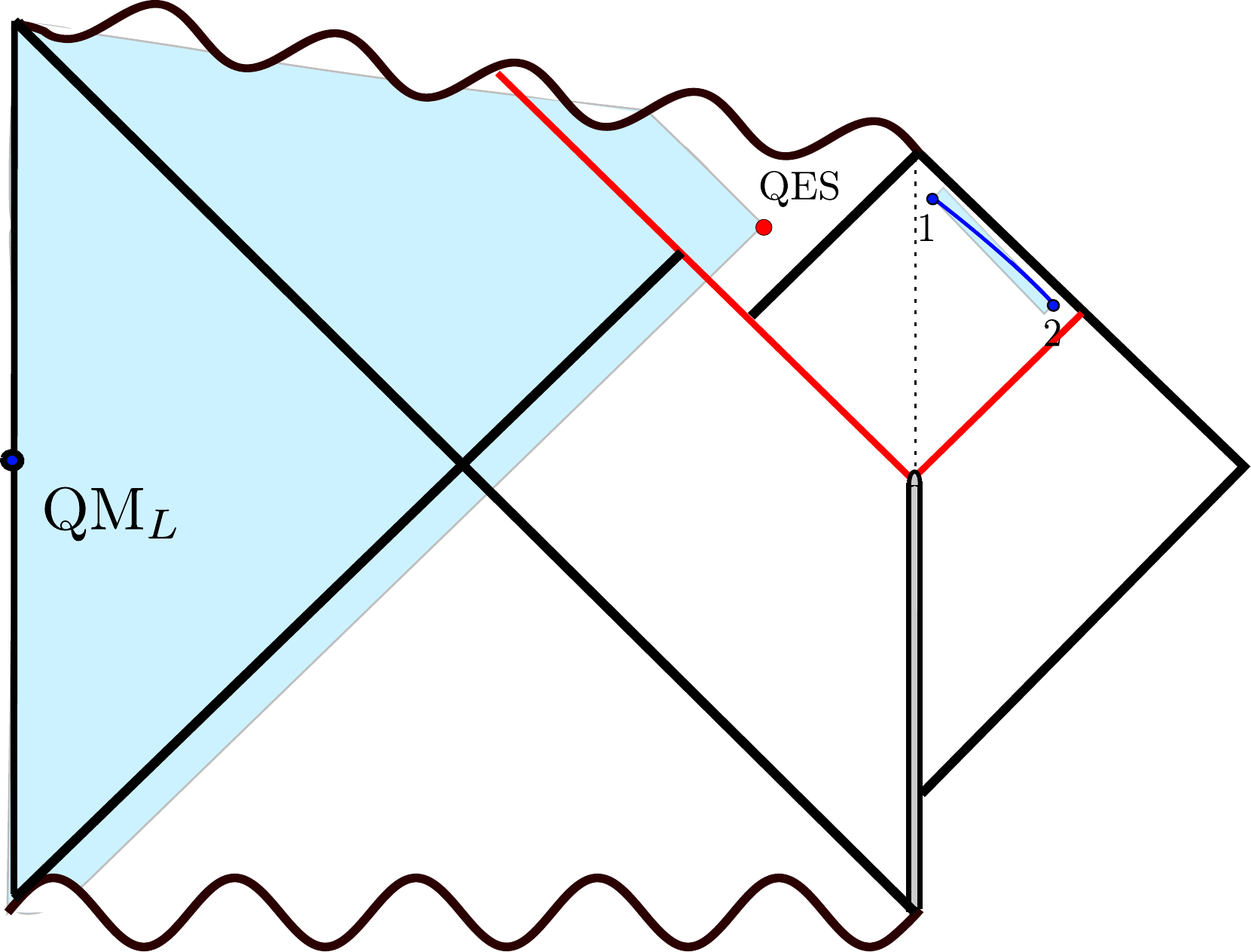}
          \caption{LateQuench phase}
          \label{figure: EW latequench}
      \end{subfigure}
      \caption{ A sketch of the entanglement wedges for the different phases the entanglement entropy  for QM$_L$+finite segment during late times. }
    \label{figure: entanglement wedges of QML+finite segment at late times}
\end{figure}

There are numerous QESs to consider for the computation of the entanglement entropy of finite intervals, due to the fact that there are two endpoints, which can each anchor an HRT surface in one of the four configurations discussed in section~\ref{sec:semi-infinite}. Additionally, there is a possibility of an HRT surface connecting the two interval endpoints. To facilitate the discussion of the different QESs that can dominate the Page curve of $\mathbf{R}\cup QM_L$, we will group them according to the relative position of the two interval endpoints with respect to the energy shock which is falling into the bath. That is, for an interval with endpoints $y_1^\pm = u \mp \sigma_1$ and $y_2^\pm = u \mp \sigma_2$ with $\sigma_1<\sigma_2$, the \textit{early time} QESs are those that can dominate when both interval endpoints are below the shock, as in figure~\ref{fig: entanglementwedge of Early phase}, and correspond to early times $u < \sigma_1 < \sigma_2$. The \textit{intermediate time} QESs have the first endpoint above the shock and the second below as in figure~\ref{figure: entanglement wedges of QML+finite segment at intermediate times} and occur at intermediate times $\sigma_1<u<\sigma_2$. Lastly, the \textit{late time} QESs have both endpoints above the shock as in figure~\ref{figure: entanglement wedges of QML+finite segment at late times} and occur at later times $u>\sigma_2>\sigma_1$.
\\
\\
\textbf{Early time} QESs correspond to intervals $\mathbf{R}$ with both endpoints below the shock. The bath segment is therefore below the shock and is causally disconnected from the gravity region, so no nontrivial QES can form yet. As explained below eq.~\eqref{eq: vonNeumannn entropy via onepoint function}, the matter entanglement entropy matches the vacuum value of entanglement entropy of a single interval on the half-plane \cite{Cardy06}. In the holographic BCFT this value is determined by the competition between two possible configurations of RT geodesics, plus the contribution from the trivial bifurcation surface QES corresponding to QM$_L$. We can schematically write down the corresponding saddles as
\begin{equation}
    \begin{aligned}
       &\text{EarlyDisconnected saddle} & S^{\rm ED}_{\rm gen}&=S_{\rm vN}\left(x_1\in \text{IV}\right)+S _{\rm  vN}\left(x_2\in \text{IV}\right)+S_{\rm BH},\\
        &\text{EarlyConnected saddle} & S^{\rm EC}_{\rm gen}&=S _{\rm vN }\left(x_1\in \text{IV},x_2\in \text{IV}\right)+S_{\rm BH}.
    \end{aligned}
\end{equation}
Explicitly, these correspond to
\begin{equation}
    \begin{split}
         S^{\rm ED}_{\rm gen}&=\frac{c}{6} \log \left(y^-_1-y^+_1\right)+\frac{c}{6} \log \left(y^-_2-y^+_2\right) +\frac{2 \pi  T_0 \phi_r+\Tilde{\phi}_0}{4 G_N}+\frac{c}{6}\log 2,\\
         S^{\rm EC}_{\rm gen}&=\frac{c}{6} \log \left(-(y^-_1-y^-_2)(y^+_1-y^+_2)\right)+\frac{2 \pi  T_0 \phi_r+\Tilde{\phi}_0}{4 G_N}+\frac{c}{6}\log 2.
    \end{split}
    \label{eq: competing phases at early times for finite segments}
\end{equation}
The entanglement entropy at early times is thus time-independent and is given by the sum of the Bekenstein-Hawking entropy of the initial black hole with temperature $T_0$ plus the vacuum entanglement entropy of $\mathbf{R}$. Which of the two saddles dominates the early times of the entanglement entropy depends on the length and location of the considered interval. The entanglement wedge in this early time regime is shown in figure \ref{fig: entanglementwedge of Early phase}, and unsurprisingly, does not reach inside the black hole. Analogously to the Trivial phase for semi-infinite segments, the mutual information between QM$_L$ and $\mathbf{R}$ vanishes because there has been no entanglement transferred by Hawking radiation yet. 
\\
\\
\textbf{Intermediate time} QESs occur when the first endpoint of the interval crossed the shock, and are therefore sensitive to the nontrivial dynamics associated to the coupling of the two systems, and the black hole evaporation. The first endpoint, which is above the shock ($x^\pm_1\in\text{II}$), can be accompanied by a nontrivial QES at an AdS$_2$ location $x_Q^\pm$ away from the bifurcation point. In the bulk perspective, this nontrivial QES corresponds to a candidate HRT surface connecting $x^\pm_1$ to $x_Q^\pm$, as shown on figure~\ref{figure: HRT at intermediate times}. There are four possible saddles that we write down schematically as
\begin{equation}
    \begin{aligned}
        &\text{CrossDisconnected saddle} & S^{\rm CD}_{\rm gen}&=S_{\rm vN} \left(x _1\in \text{II}\right)+S _{\rm vN}\left(x _2\in \text{IV}\right)+S_{\rm BH},\\
        &\text{CrossConnected saddle} & S^{\rm CC}_{\rm gen}&=S _{\rm vN}\left(x _1\in \text{IV},x _2\in \text{IV}\right)+S_{\rm BH},\\
        &\text{ScramblingTrivial saddle} & S^{\rm ST}_{\rm gen}&=S _{\rm  vN}\left(x _1\in \text{IV},x _{QS}\in\text{III}\right)\\
       && &\qquad+S _{\rm  vN}\left(x _2\in \text{IV}\right)+S_{\phi _{QS}},\\
        &\text{LateTrivial saddle}& S^{\rm LT}_{\rm gen}&=S _{\rm  vN}\left(x _1\in \text{IV},x _{QL}\in\text{I}\right)\\
      &&  &\qquad+S _{\rm vN}\left(x _2\in \text{IV}\right)+S_{\phi _{QL}}.
    \end{aligned}
\end{equation}
Of the four QESs, two of them correspond to trivial QES saddles
\begin{subequations}
    \label{eq: competing phases in reconstruction window at intermediate times for finite segments}
\begin{equation}\label{eq:inter trivial}
    \begin{split}
         S^{\rm CD}_{\rm gen}&=\frac{c}{6} \log \left(\frac{12 \pi E_S}{c } y^-_1 \frac{f\left(y^+_1\right)}{\sqrt{f'\left(y^+_1\right)}}\right)+\frac{c}{6} \log \left(y^-_2-y^+_2\right)\\
        &+\frac{2 \pi  T_0 \phi_r+\Tilde{\phi}_0}{4 G_N}+\frac{c}{6}\log 2,\\
         S^{\rm CC}_{\rm gen}&=\frac{c}{6} \log \left(\frac{12 \pi E_S y^+_2 f(y^+_1)(y^-_1-y^-_2) }{c\sqrt{f'(y^+_1)}}\right)\\
        &+\frac{2 \pi  T_0 \phi_r+\Tilde{\phi}_0}{4 G_N}+\frac{c}{6}\log 2\,,
    \end{split}
\end{equation}
in which the QES is located at the bifurcation surface. The other two nontrivial QES saddles are given by 

\begin{equation}\label{eq:inter non trivial}
    \begin{split}
         S^{\rm ST}_{\rm gen}&=\frac{c}{6} \log \left(\frac{24 \pi E_S x^-_{QS}y^-_1(x^+_{QS}-f(y^+_1))}{c (x^-_{QS}-x^+_{QS})\sqrt{f'(y^+_1)}}\right)\\
        &+\frac{c}{6} \log \left(y^-_2-y^+_2\right)+\frac{\Tilde{\phi}_0+\phi(x^+_{QS},x^-_{QS})}{4 G_N},\\
         S^{\rm LT}_{\rm gen}&=\frac{c}{6} \log \left(\frac{2(y^-_1-y^-_{QL})(x^+_{QL}-f(y^+_1))}{(x^-_{QL}-x^+_{QL})}\sqrt{\frac{f'(y^-_{QL})}{f'(y^+_1)}}\right)\\
        &+\frac{c}{6} \log \left(y^-_2-y^+_2\right)+\frac{\phi (x^+_{QL},y^-_{QL})+\Tilde{\phi}_0}{4 G_N}\,.
    \end{split}
\end{equation}
\end{subequations}
The QES is located away from the bifurcation surface because of quantum corrections associated to the von Neumman entropy of the bulk matter fields.

The entanglement wedge of each phase is illustrated in figure \ref{figure: entanglement wedges of QML+finite segment at intermediate times}. 
For the first two saddles in eq.~\eqref{eq:inter trivial}, the first endpoint is not associated to a nontrivial QES. During the corresponding phases the entanglement wedge does not contain the black hole interior, so the interior is not accessible for the reconstruction. The CrossDisconnected saddle has a three-dimensional description that involves two disconnected HRT surfaces, each attaching one endpoint of the segment to the end of the world brane, see figure \ref{figure: HRT-disc}. The HRT surfaces stretch with time due to the ETW brane falling into the bulk, which leads to a characteristic rapid increase in the entanglement entropy when the segment starts to cross the shock wave. This behaviour is very similar to the Quench saddle for semi-infinite intervals. The three-dimensional description of CrossConnected saddle is characterized by an HRT surface connecting the two endpoints, see figure \ref{figure: HRT-conn}. During the corresponding phase, the generalized entropy also increases, but at a lower rate than that of the disconnected phase. The generalized entropy in these trivial QES phases is factorized between a $QM_L$ contribution and a $\mathbf{R}$ contribution at leading order in large central charge, implying that the initial Hawking radiation captured by $\mathbf{R}$ is not very entangled with the purification $QM_L$. The entanglement dynamics is dominated by the joining of the two systems, rather than the evaporation process. Hence, the behaviour of entanglement entropy is in direct correspondence with previous results in local (joining) quenches \cite{Calabrese07,Calabrese09}, up to the addition of the constant Bekenstein-Hawking entropy of the black hole.

For the other two saddles in eq.~\eqref{eq:inter non trivial}, the QES is located away from the bifurcation point in a location entirely determined by the position of the first endpoint. The bulk description includes an HRT surface connecting the first endpoint of the segment, $x_1^\pm$, to the QES, see figure~\ref{figure: HRT-qes}. The second endpoint, which remains below the shock, anchors a bulk RT surface which connects to the ETW brane, in a similar configuration to the bulk RT surface in the Trivial saddle of section~\ref{sec:semi-infinite}. Since the QES moves away from the bifurcation surface, the entanglement wedge in both nontrivial QES phases includes a portion of the black hole interior, as can be seen in figure~\ref{figure: entanglement wedges of QML+finite segment at intermediate times}. When the ScramblingTrivial saddle is dominant, the QES is very close to the bifurcation surface. Similarly to the Scrambling phase for semi-infinite intervals, this is the only phase in which the evolution of entanglement entropy can be dominated by scrambling physics, characterized by an initial dip in the entanglement entropy, followed by a period of linear growth caused by the evaporation process. During this phase, one is allowed a peak into the black hole since there is a small portion of the black hole interior which is reconstructable by $\mathbf{R} \cup QM_L$. The LateTrivial phase on the other hand, allows for a drastically larger portion of the black hole to be reconstructed. In this phase, the QES is located above the shock and at a much larger distance from the bifurcation point compared to the Scrambling QES, instead being located exponentially close to the final event horizon. Similarly to the Late phase for semi-infinite intervals, the entanglement entropy decreases in this phase.
\\
\\
\textbf{Late time} QESs occur once the interval has crossed the shock, so the first and second endpoint are both above the shock in the bath region ($x^\pm_1,x^\pm_2\in\text{II}$). In principle, each endpoint can be connected to a nontrivial QES by a HRT surface, in what would constitute an island configuration. In addition, the homology condition further requires a third RT surface anchored at the bifurcation point. However, these island configurations have much greater generalized entropy than the non-island configurations since the bulk HRT surface has three anchor points on the JT brane, each contributing a large amount of entropy through the length of the geodesics approaching the brane. The island phases can only occur at very late times, and for very large intervals as we will explain later in footnote~\ref{foot:islands}.

For finite intervals, there are two competing saddles that we write down schematically as
\begin{equation}
    \begin{aligned}
        &\text{\rm FutureConnected saddle} & S^{\rm FC}_{\rm gen}&=S_{\rm vN}\left(x _1\in \text{II},x _2\in \text{II}\right)+S_{BH}\,,\\
        &\text{\rm LateQuench saddle} & S^{LQ}_{\rm gen}&=S_{\rm vN}\left(x _1\in \text{II},x _{QL}\in \text{I}\right)+S_{\rm vN}\left(x _2\in \text{II}\right)+S_{\phi_{QL}}\,,
    \end{aligned}
\end{equation}
whose generalized entropy is more explicitly given by
\begin{equation}
    \begin{split}
         S^{\rm FC}_{\rm gen}&=\frac{c}{6}\log \left(\frac{(y^-_1-y^-_2)(f(y^+_2)-f(y^+_1))}{\sqrt{f'(y^+_1)}\sqrt{f'(y^+_2)}}\right)\\
        &+\frac{2 \pi  T_0 \phi_r+\Tilde{\phi}_0}{4 G_N}+\frac{c}{6}\log 2\,,\\
         S^{\rm LQ}_{\rm gen}&=\frac{c}{6} \log \left(\frac{2(y^-_1-y^-_{QL})(x^+_{QL}-f(y^+_1))}{x^-_{QL}-x^+_{QL})}\sqrt{\frac{f'(y^-_{QL})}{f'(y^+_1)}}\right)\\
        &+\frac{c}{6} \log \left(\frac{12 \pi Es y^-_2 f(y^+_2)}{c\sqrt{f'(y^+_2)}}\right)+\frac{\phi (x^+_{QL},y^-_{QL})+\Tilde{\phi}_0}{4 G_N}\,.
    \end{split}
    \label{eq: competing phases in at late times for finite segments}
\end{equation}
The entanglement wedges of the corresponding saddles are depicted in figure~\ref{figure: entanglement wedges of QML+finite segment at late times}.
As reflected in the names, these phases naturally follow up on those discussed for intermediate times. There is one saddle where the first endpoint is not accompanied by a nontrivial QES, the FutureConnected saddle, which resembles the CrossConnected saddle. The three-dimensional description is similar to the one shown in figure~\ref{figure: HRT-conn}, with the difference being that the ETW brane has fallen more into the bulk compared to intermediate times. For LateQuench, the first endpoint is connected to a nontrivial QES on the JT brane by a bulk HRT surface in the three-dimensional description similarly to their counterpart at intermediate times, see figure~\ref{figure: HRT-qes}. The second endpoint is connected to the ETW brane, but is now above the shock, in a similar configuration as the Quench saddle in section~\ref{sec:semi-infinite}. Finally, ScramblingQuench and CrossDisconnected  (the counterparts of ScramblingTrivial and FutureDisconnected, respectively) will never dominate and thus will be left out of the discussion from now on. Similarly, any potential island configuration (such as ScramblingScrambling, LateScrambling and LateLate) will be greatly suppressed.

\subsection{The Page curve}
\label{ssection: page curve}

\begin{figure}[h!]
\begin{center}
    \includegraphics[scale=0.52]{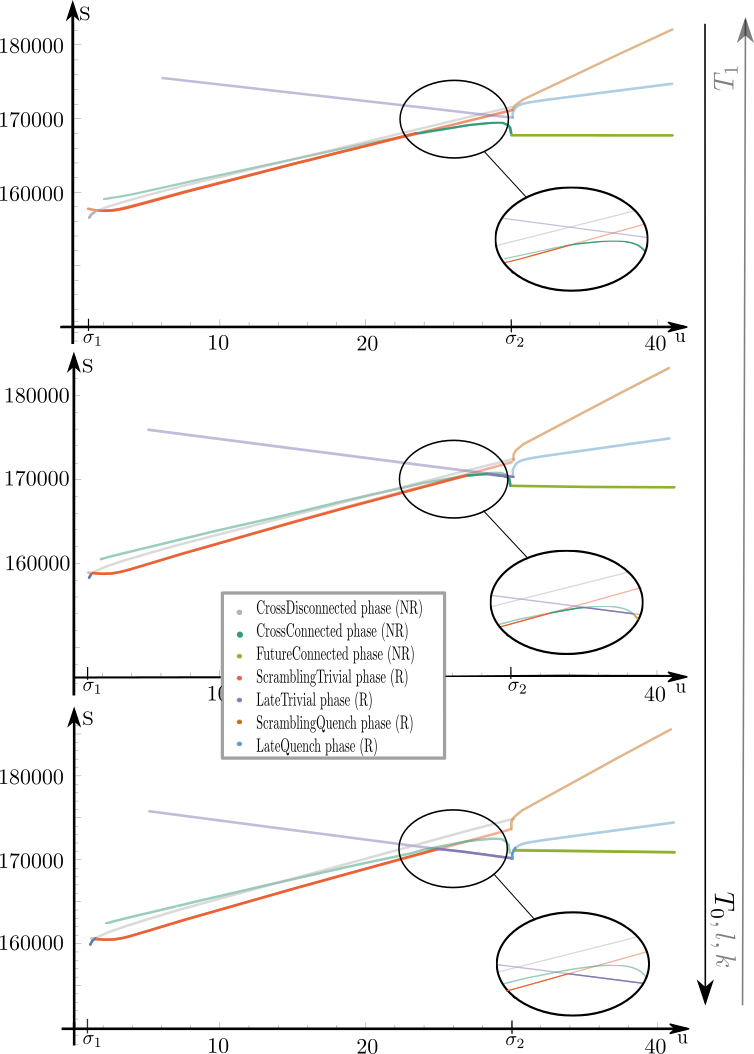}
    \caption{ This figure shows the general shape of the competing saddles at intermediate and late times for the entanglement entropy of QM$_L$+finite segment of radiation in the bath depending on the values of the free parameters in the theory. Specifically, we used different values of $T_0$. Namely, we used $T_0=\frac{449}{512\pi}$ for the upper plot, $T_0=\frac{453,5}{512\pi}$ for the middle plot and $T_0=\frac{459}{512\pi}$ for the lower plot, in order to illustrate the different scenarios. The entanglement entropy is obtained by minimizing between the different saddles. Transparent colors are used to illustrate saddles outside of their regimes of dominance. The values of the other parameters used in this plot can be found in table~\ref{table section4}.}
	\label{figure: finite segments intermediate times}
\end{center}
\end{figure}

\begin{figure}[t]
\centering
\includegraphics[scale=0.9]{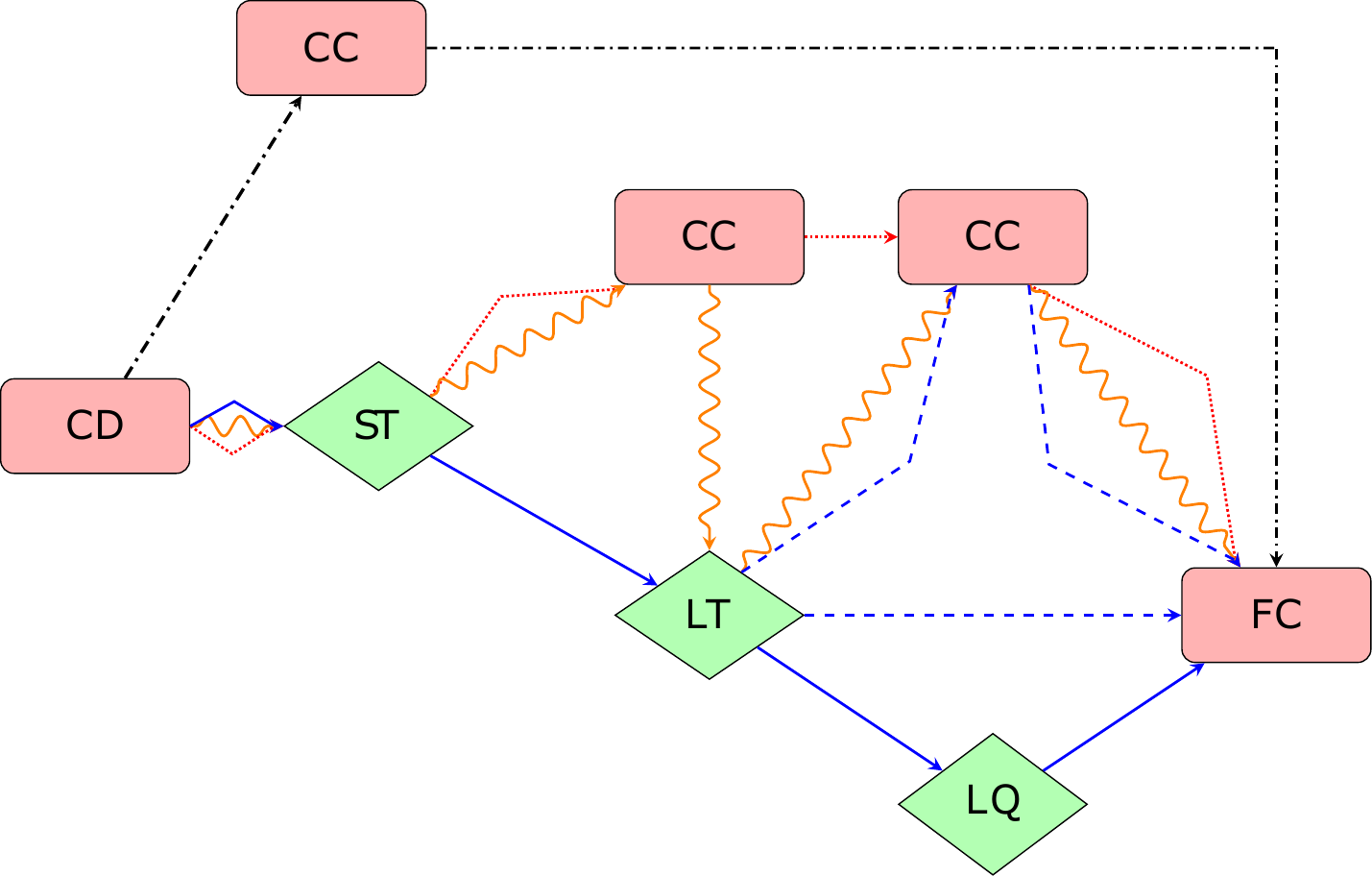}
\caption{Flowchart of the possible sequences of phases  in the entanglement evolution of $\mathbf{R} \cup QM_L$. Phases in which interior reconstruction is (not) possible are in green diamonds (red rectangles). The densely dashed red (solid blue) path corresponds to the small (large) reconstruction path as illustrated in the top (bottom) of figure~\ref{figure: finite segments intermediate times} which we label as scenario 1 (2). The dashed blue paths correspond to limiting behaviours which require a lot of fine tuning and provide a similar large reconstruction as the blue path. The wavy orange path corresponds to one in which interior reconstruction is temporarily interrupted as shown in the middle of figure~\ref{figure: finite segments intermediate times} corresponding to scenario 3. The dash-dotted black path does not allow for interior reconstruction and is referred to as scenario 0.}
\label{fig:flowchart} 
\end{figure}

\begin{figure}[t]
\begin{center}
    \includegraphics[scale=0.45]{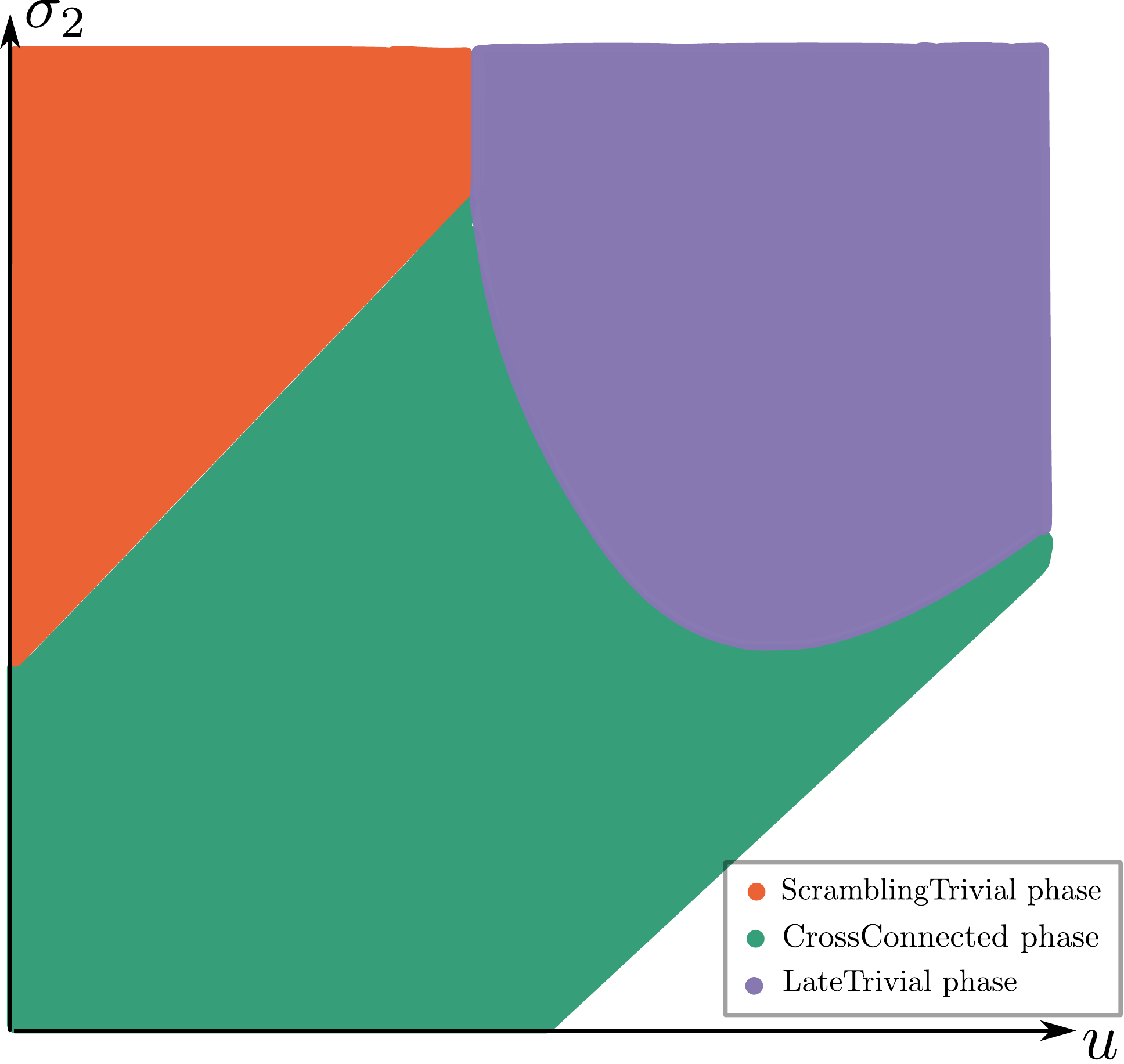}
	\caption{This figure is a phase diagram that shows the saddles that dominate the generalized entropy for each time and $\sigma_2$ in the region around the temporary blindness in the interior reconstruction window. We used $T_0=\frac{453,5}{512\pi}$ for this plot. The values of the other parameters used, can be found in table~\ref{table section4}.}
	\label{figure: phase diagram for blind spot}
\end{center}
\end{figure}

The Page curve is found by minimizing over the generalized entropy of the QESs catalogued in section~\ref{ssection: time windows}, since these provide the saddle points in the path integral computing the entropy of $\mathbf{R} \cup QM_L$ \cite{Almheiri194,Penington192,Goto20}. The overall behaviour of the entanglement entropy is as follows: while the interval is below the shock, the entropy is constant and given by the vacuum entropy of $\mathbf{R}$ plus the Bekenstein-Hawking entropy of the black hole $S_{\rm BH}$. As soon as the first endpoint crosses the shock, the entropy rapidly increases due to the large amount of entropy introduced by the local quench \cite{Calabrese07,Calabrese09}. The middle stages of the evaporation show some transient behaviour which depends on the parameters of the theory and the length of the interval. Overall, the entropy will increase due to the increasing amount of Hawking radiation collected in the interval until it transitions to a decreasing phase. This can be a sharp transition to a linearly decreasing behaviour if the interval is larger than the Page time, mimicking the behaviour of semi-infinite intervals. The transition can also be to a logarithmically decreasing behaviour as the second endpoint approaches the shock, mimicking the standard results from quenched systems \cite{Calabrese09}. After the second endpoint crosses the shock, the entropy slowly asymptotes to a value very close the vacuum entropy of the interval plus the Bekenstein Hawking entropy of the black hole. However, despite the overall behaviour of the entropy being relatively robust to changes of parameters and of the length of the interval, we will see shortly that the sequence of phases is not. This has a significant influence on the possibility of reconstructing the black hole interior from $\mathbf{R}  \cup QM_L$, since the size of reconstructable portions of black hole interior can be drastically different between phases. The main results presented in this subsection can be summarized in figures~\ref{figure: finite segments intermediate times}, \ref{fig:flowchart} and \ref{figure: phase diagram for blind spot}, which will be frequently referred to throughout the text.

Figure~\ref{figure: finite segments intermediate times} provides three numerical plots of the Page curve, each for a different value of $T_0$.\footnote{As in section \ref{sec:semi-infinite}, we use the analytic approximations of the QES locations for the ScramblingTrivial/Quench saddle given in eqs.~\eqref{eq: scrambling QES} as well as for the LateTrivial/Quench saddle given in eqs.~ \eqref{eq: late QES} as numerical input.} As is clear from this figure, the overall behaviour of the entropy is robust to small variations in $T_0$ since the shape is very similar for the three examples and follows the behaviour described above. However, the sequence of phases is different for each plot. These sequences are represented by paths through the flowchart on figure~\ref{fig:flowchart}. There are 4 main paths, each in different color, which represent 4 physically distinct scenarios of the entanglement evolution. Besides those, there are 2 additional paths, each of them indicated by blue dashed arrows, which have similar entanglement wedge reconstruction properties to the blue path. 

In what follows we provide a brief overview of the possible sequence of dominant QESs in the Page curve. We will discuss reconstruction properties of the various possibilities in more detail further below. The first possibility corresponds to the  dash-dotted black path in figure~\ref{fig:flowchart}. This scenario occurs when the finite segment is much shorter compared to $u_P$ defined in eq.~\eqref{eqn: page time with uhp}. After the first endpoint crosses the shock, $u=\sigma_1$, the generalized entropy evolves through the CrossDisconnected phase. During this phase, the HRT surface connecting the endpoint above the shock to the ETW brane keeps stretching as the ETW brane falls into the bulk. Hence, the corresponding generalized entropy of this QES keeps increasing, until eventually it surpasses the generalized entropy of the CrossConnected saddle. At this point, the CrossConnected phase takes over. The generalized entropy still increases but at a lower rate than before. The corresponding saddle dominates the entanglement entropy evolution until the second endpoint crosses the shock $u=\sigma_2$. Once the interval is completely above the shock, the FutureConnected phase takes over and the entropy eventually equilibrates to the vacuum value, as we will see below in eq.~\eqref{eq:FCverylate}. This scenario does not allow for black hole reconstruction. Intuitively, this can be understood from the size of the interval. Since it is so small, it can never capture enough Hawking radiation in order to reconstruct any portion of the black hole interior. The other possibilities occur when the segment length is comparable to $u_P$. The generalized entropy for segments of such lengths is shown in figure \ref{figure: finite segments intermediate times}. For each scenario in figure \ref{figure: finite segments intermediate times}, there is a reconstruction window for $\mathbf{R}\cup QM_L$ which always starts before the second endpoint crosses the shock. However, the details of this interior reconstruction window depend on the choice of parameters. In particular, the upper figure shows a scenario where the interior reconstruction window ends well before the segment crosses the shock. The lower figure shows a scenario where the interior reconstruction window may continue even after the segment has crossed the shock without interruption. The middle figure shows a scenario where the interior reconstruction window is interrupted by a period of temporary blindness. In this case, the reconstruction window will end shortly before the segment has crossed the shock. The free parameters that control the interruption of the interior reconstruction window are $\left(l,T_0,T_1,k\right)$, where $l$ is length of the segment, $T_0$ and $T_1$ are the initial and final temperature of the black hole and $k$ is the back-reaction parameter. The other parameters in this model are either not free, such as $E_S$ and $\phi_r/G_N$ which are fixed by the above mentioned parameters via eq.~\eqref{eq: eqn for k related to other parameters} and eq.~\eqref{eq: formula for energy of the shock}, or will have no implication on the interior reconstruction window, such as $\phi_0$ and $\sigma_1$. Theses last two parameters will move the whole Page curve up and offset the curve to the right, respectively if their values are increased.

Let us now discuss each reconstruction scenario with more care. All scenarios share one common feature, the interior reconstruction window always begins with a period where the ScramblingTrivial saddle dominates the entropy evolution. For the  solid blue path in figure~\ref{fig:flowchart}, which corresponds to the bottom scenario in figure~\ref{figure: finite segments intermediate times}, this phase is followed by the LateTrivial phase during which a large portion of the black hole interior is accessible. For a substantial region in the parameter space, this reconstruction window continues even after the second endpoint crosses the shock by the evolution through the LateQuench phase before the FutureConnected phase takes over. Avoiding the LateQuench phase requires a high degree of fine-tuning of the parameters in the theory but is nevertheless possible. These very fine-tuned paths are shown in blue dotted lines in figure~\ref{fig:flowchart}. All of these cases lead to an uninterrupted interior reconstruction window with the large amount of the black hole being reconstructable. This only happens for segments larger than $u_P$, or equivalently when $T_0$ or $k$ are large or $T_1$ is small. For very large segments, the sequence of phases remains the same, but the LateQuench to FutureConnected transition gets pushed to very late times eventually mimicking the behaviour of semi-infinte intervals described in section~\ref{sec:semi-infinite}.

The ScramblingTrivial phase can also be followed by a period of no insight into the black hole as shown in the upper plot in figure~\ref{figure: finite segments intermediate times}. This is represented by the densely dashed red path in figure~\ref{fig:flowchart}. The nontrivial QES surface that appeared in the ScramblingTrivial phase bounces back to the bifurcation point as the generalized entropy evolves through the CrossConnected phase. The corresponding saddle can dominate all the way until the second endpoint crossed the shock, after which FutureConnected saddle dominates the evolution. In this scenario, only a small portion of the black hole interior is accessible to $\mathbf{R}\cup QM_L$. This scenario requires the segment to be shorter than the Page time, or equivalently $T_0$ or $k$ should be small or $T_1$ is large. Of course, the reduced length of the interval relative to the blue path explains why only a small portion of the black hole interior is accessible to $\mathbf{R}\cup QM_L$.

The last possibility, the middle plot in figure~\ref{figure: finite segments intermediate times}, which corresponds to the wavy orange path in figure~\ref{fig:flowchart}, is for the ScramblingTrivial phase to be followed by the CrossConnected phase, during which the black hole interior is no longer accessible. However, after a short while, the LateTrivial phase takes over, in which a large portion of the black hole becomes reconstructable. This scenario describes a period of temporary blindness during the interior reconstruction window when transitioning from a phase with access to a small portion of the black hole interior to a phase with access to a large portion of the black hole interior. The LateTrivial phase is then followed by the CrossConnected phase until the second endpoint crosses the shock. Finally, the entropy asymptotes to the vacuum value during FutureConnected phase, see eq.~\eqref{eq:FCverylate}. 

This last scenario is most peculiar since the interior reconstruction window gets interrupted for a finite amount of time. The emergence of the temporary blindness in the interior reconstruction window strongly depends on the choice of parameters $k, T_0,T_1$ and $l$. The phase diagram in figure~\ref{figure: phase diagram for blind spot} shows this sensitive dependence specifically for the parameter $\sigma_2$ which determines $l$. An important feature of this phase diagram is the fact that the separation between the two reconstructing phases (red and purple regions in figure \ref{figure: phase diagram for blind spot}) remains vertical at larger $\sigma_2$. This is consistent with the entanglement wedge nesting: when increasing $\sigma_2$, the entanglement wedge should never become smaller.

Similar phase diagrams can be obtained for the other parameters. To build some intuition about the occurrence of this phenomenon, we derive the bounds on a dimensionless parameter $\frac{E_S}{c(T_1+T_0)}=\frac{\pi}{12}\frac{T_1-T_0}{ k}$, allowing us to determine the onset and the end of the temporary blindness in parameter space. This parameter intuitively measures the effects of the radiation carried by the shock $\propto E_S$ compared to the effect of Hawking radiation $\propto c T$. The lower bound corresponds to when the CrossConnected saddle matches the entropy of both reconstructing saddles at the Page transition
\be\label{eq:triple}
S^{\rm CC}_{\rm gen}=S^{\rm ST}_{\rm gen}=S^{\rm LT}_{\rm gen}
\ee
and it is the limiting case between the middle and lower scenario in figure~\ref{figure: finite segments intermediate times}. The upper bound corresponds to the boundary of when the LateTrivial QES never dominates and can be found by equating both the value and the derivative of the generalized entropy of the CrossConnected saddle and the LateTrivial saddle
\be\label{eq:upper bound}
S^{\rm CC}_{\rm gen}=S^{\rm LT}_{\rm gen},\qquad \partial_u S^{\rm CC}_{\rm gen}=\partial_u S^{\rm LT}_{\rm gen}\,.
\ee
This is the limiting case between the middle and upper scenario in figure~\ref{figure: finite segments intermediate times}. The derivation of these bounds can be found in appendix~\ref{section: Bounds on blind spot} and in particular, the first approximations for the bounds can be found in eq.~\eqref{eqn:first bound on param} and eq.~\eqref{eqn: second bound on param}. If the dimensionless parameter $\frac{E_S}{c(T_1+T_0)}$ is less than the lower bound given in eq.~\eqref{eqn:first bound on param}, the interior reconstruction window is never interrupted and a large portion of the black hole interior is accessible. Intuitively, this means that the physics due to Hawking radiation overrules the information carried by the shock radiation. On the other hand, if the dimensionless parameter $\frac{E_S}{c(T_1+T_0)}$ is greater than the upper bound given in eq.~\eqref{eqn: second bound on param}, the saddle in which a large portion of the black hole interior is accessible never dominates. In this case, the entanglement carried by the shock overshadows the entanglement carried by the Hawking radiation. This understanding of $\frac{E_S}{c(T_1+T_0)} = \frac{\pi}{12}\frac{T_1-T_0}{k}$ directly implies the behaviour of figure~\ref{figure: finite segments intermediate times} under changes of $T_0$, $T_1$ and $k$ as indicated by the arrows on the right. The remaining parameter in the figure is the length $l$ of the segment. This parameter reflects the dependence of figure~\ref{figure: finite segments intermediate times} on how much of the Hawking radiation can be captured by the interval $\mathbf{R}$ as was explained for each scenario above.

An interesting feature of the scenarios 2 and 3, the bottom and middle plots of figure~\ref{figure: finite segments intermediate times}, is the appearance of a python's lunch region \cite{Python1,Python2,Python3}. A python's lunch is a spatial region contained between the minimal QES and a subleading QES. Moreover, this region should also contains a third QES which has the largest generalized entropy of the three. A python's lunch region is interesting because it is stated to be exponentially complex to reconstruct by the restricted complexity conjecture of \cite{Python1}. In application to the present model, the minimal QES has to be realized by the LateTrivial phase, the subleading one is the trivial QES of the CrossConnected saddle, while the largest QES belongs to the ScramblingTrivial saddle. The python's lunch appears only when we are in the LateTrivial phase, and the three competing saddles are in the following hierarchy:
\be
S^{\rm ST}_{\rm gen}> S^{\rm CC}_{\rm gen} > S^{\rm LT}_{\rm gen}\,. \label{python}
\ee
Note that in both scenarios the ability to reconstruct is lost after a transition from the LateTrivial or LateQuench phase to a trivial QES phase, such as the CrossConnected or the FutureConnected phase. This transition is always preceded by the appearance of a python's lunch. The appearance of the python's lunch before the loss of reconstruction signals that interior reconstruction becomes exponentially complex some time before access to the interior is lost. Notice that the python's lunch for scenario 3, the middle plot of figure~\ref{figure: finite segments intermediate times}, occurs right after the reconstruction is interrupted by the CrossConnected phase. In this case, the hierarchy given in eq.~\eqref{python} is immediately established at the change of dominance to the LateTrivial phase. However, for scenario 2, the bottom  plot of figure~\ref{figure: finite segments intermediate times}, this hierarchy is not immediately established in the LateTrivial phase. There is some time where the ScramblingConnected curve is still below the CrossConnected curve when the LateTrivial curve dominates. This implies that the python's lunch does not emerge immediately in the reconstructable portion of the black hole interior in this scenario.

\begin{figure}[t]
     \centering
     \begin{subfigure}[b]{0.4\textwidth}
         \centering
         \includegraphics[width=\textwidth]{figures/EW-Future.pdf}
         \caption{FutureConnected phase}
     \end{subfigure}
     \hfill
     \begin{subfigure}[b]{0.41\textwidth}
         \centering
         \includegraphics[width=\textwidth]{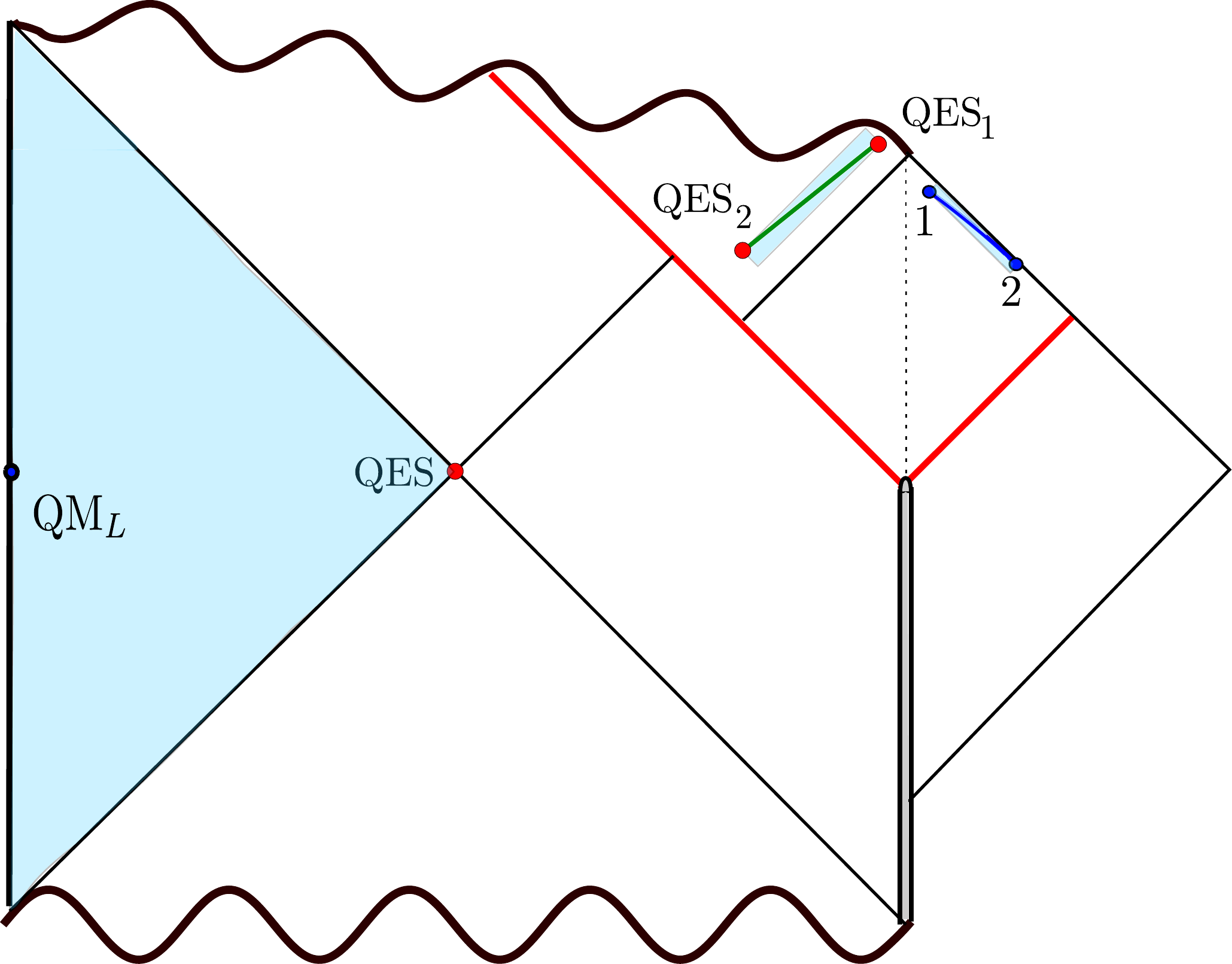}
         \caption{LateLate phase}
         \label{figure: EW latelate}
     \end{subfigure}
     \caption{ A sketch of the entanglement wedges for the different phases of the entanglement entropy for QM$_L$+finite segment during very late times.}
    \label{figure: entanglement wedges of QML+finite segment at very late times}
\end{figure}

Let us now return to the plots in figure~\ref{figure: finite segments intermediate times} to discuss the later times. First of all, we notice that for late times, the analogue of ScramblingTrivial, which could be dubbed ScramblingQuench, is never dominant in any scenario. This is because there is an upper limit on the length of the interval in order for the Scrambling QES to dominate over the Late QES after the second endpoint crosses the shock. However, the CrossConnected QES becomes favorable over the reconstructing QESs for intervals of such lengths. Moreover, for intervals that are not parametrically large in $k$, islands never form within the range of applicability of our theory.\footnote{ \label{foot:islands} More precisely, the leading contribution to the generalized entropy of the LateLate QES is given by the dilaton terms
\begin{equation}
S^{\rm LL}_{\rm gen}-S_{\rm BH} \approx \frac{\pi \phi_r T_1 e^{- \frac{ku}{2}}}{G_N}\,,
\end{equation}
while the generalized entropy for the FutureConnected QES asymptotes to the vacuum entropy of the interval plus the Bekenstein-Hawking entropy of the black hole as in eq.~\eqref{eq:FCverylate}. Hence, for generically sized intervals, the LateLate saddle cannot compete with FutureConnected. On the other hand, when the size of the segment scales as $l \sim e^{\frac{T_1}{k}}$, the LateLate QES can dominate over FutureConnected QES, which would lead to an island phase at very late times. Notice that this statement holds true for $\mathbf{R}$ and $\mathbf{R}\cup QM_L$.} Shortly after the interval is completely above the shock, island saddles such as ScramblingScrambling and LateScrambling both have a generalized entropy which is approximately $2 S_{\rm BH}$ higher than that of FutureConnected and therefore never prevail. However, as time evolves, the LateLate QES could potentially accommodate for this gap since it has a decreasing generalized entropy. The LateLate saddle is a true island saddle as is shown in figure \ref{figure: EW latelate}. One can check that even for very late times, near the end of the regime of applicability of the model, this decrease does not overcome this gap, confirming that there are no islands, at least for intervals that are not parametrically large in $k$. 

Finally, let us look the very late time behaviour of the entropy in detail. We will need the following expansions for $u\to\infty$
\bea
f(u) &=& t_\infty - \frac{A}{u}  + O\left(\frac{1}{u^2}\right)\,,\label{fasy}\\ 
f'(u)&=& \frac{A}{u^2} + O\left(\frac{1}{u^3}\right)\,, \label{f'asy}
\eea
where the coefficients of the power expansion are given by 
\bea
t_\infty &=& \frac{I_0 \left(\frac{2\pi T_1}{k}\right)}{\pi T_1 I_1 \left(\frac{2\pi T_1}{k}\right)}\approx \frac{1}{\pi T_1}\,, \label{A0}\\
A &=& \frac{1}{\left(\pi T_1 I_1 \left(\frac{2\pi T_1}{k}\right)\right)^2} \approx \frac{4 e^{-\frac{4\pi T_1}{k}}}{k T_1}\,. \label{A1}
\eea
Using these expansions and dropping fast decaying terms, we get 
\bea\label{eq:FCverylate}
 S^{\rm FC}_{\rm gen}-S_{\rm BH} &=& \frac{c}{3} \log l + \frac{c}{36} l^2 \pi^2 T_1^2 e^{-ku} = S_{\rm vac}+ \frac{c}{36} l^2 \pi^2 T_1^2 e^{-ku} . 
\eea
We see that at very late times the generalized entropy of the FutureConnected QES asymptotes to the entanglement entropy of an interval of length $l$ plus the Bekenstein-Hawking entropy. Let us stress the fact that the approximations in eq.~\eqref{fasy} and eq.~\eqref{f'asy} are derived beyond the regime of applicability of our theory but still reproduce the expected results since the relevant HRT surfaces remain far from the Plank brane where large quantum fluctuations break down the theory. However, this does not guarantee that eq.~\eqref{eq:FCverylate} will correspond to the entanglement entropy of $\mathbf{R} \cup QM_L$ at very late times.

\subsection{Size of the entanglement wedge and subregion complexity}\label{sec: complex}

In section \ref{ssection: page curve} we have shown that the entanglement wedge of a finite segment and QM$_L$ can behave in a highly nontrivial way as the segment evolves forward in time. Moreover, as with any Hartman-Maldacena-like transition, the entanglement wedge exhibits a discontinuity across phase transitions between trivial and nontrivial QES saddles. We want to quantify this nontrivial behavior of the entanglement wedge using different tools than those presented so far. To this end, we look at a quantity in the CFT state that captures geometric properties of the entanglement wedge, namely the holographic complexity \cite{Stanford:2014jda,Susskind16}. The doubly holographic nature of the setup makes the calculation of this quantity a feasible task. From the technical point of view, we are going to compute the holographic complexity of subregions on a 2 dimensional boundary. 

\begin{figure}[t!]
	\centering
	\includegraphics[scale=0.8]{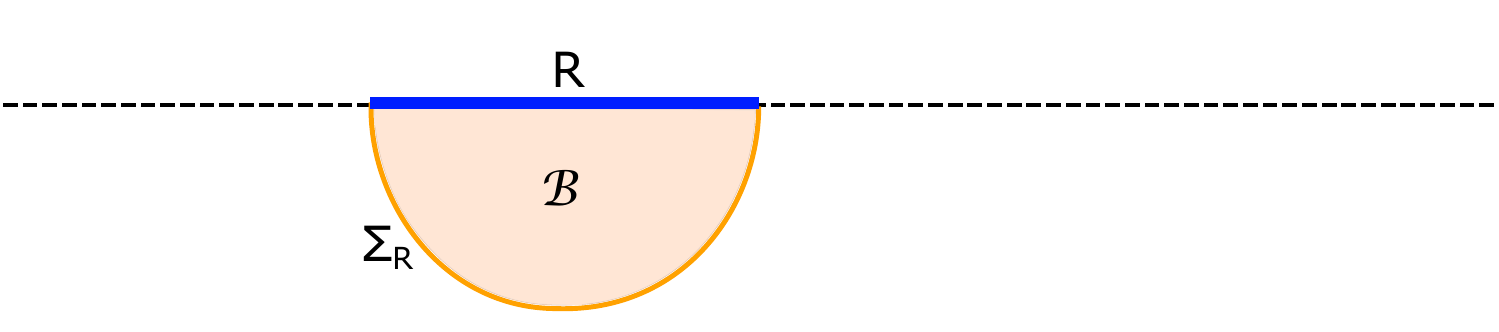}
	\caption{The subregion complexity of the boundary subregion $\mathbf{R}$ is given by the volume of the maximal bulk codimension one surface $\cal B$ anchored at $\mathbf{R}$ and its HRT surface $\Sigma_\mathbf{R}$. The maximization is done over the time direction.}
	\label{fig:CV}
\end{figure}
We employ the holographic complexity = volume (CV) prescription introduced by \cite{Susskind16,Stanford:2014jda}, which proposes that the holographic dual to the complexity of the state in the boundary CFT defined on a boundary time slice $\mathbf{S}$ is given by
\begin{equation}
    C_{V}\left(\mathbf{S} \right) = \max_{\partial {\cal B} = \mathbf{S}} \frac{V\left({\cal B} \right)}{G_N \ell}\,,
\end{equation}
where $G_N$ is the Newton's constant of the bulk gravity theory, and $\ell$ is a length scale which is introduced to make the right-hand side dimensionless. More generally, motivated by entanglement wedge reconstruction, the CV conjecture is generalized to the reduced density matrices corresponding to finite subregions in the boundary CFT by \cite{Alishahiha15,Ben-Ami16,Carmi2017}. The subregion CV conjecture proposes that the complexity of the mixed state defined on a boundary subregion $\mathbf{R}$ is given by
\begin{equation}\label{eq:CVsub}
    C_{V}^{\rm sub}\left(\mathbf{R} \right) = \max_{\partial {\cal B} = \mathbf{R}\cup \Sigma_{\mathbf{R}}} \frac{V\left({\cal B} \right)}{G_N \ell}\,,
\end{equation}
where $\Sigma_{\mathbf{R}}$ is the HRT surface associated to the boundary subregion $\mathbf{R}$. That is, the subregion CV prescription instructs to find the maximum volume codimension one surface in the bulk anchored between the boundary subregion and the HRT surface, and identifies its volume with the complexity of the mixed state associated to the boundary subregion -- see figure~\ref{fig:CV}.
\begin{figure}[t!]
	\centering
	\includegraphics[scale=0.6]{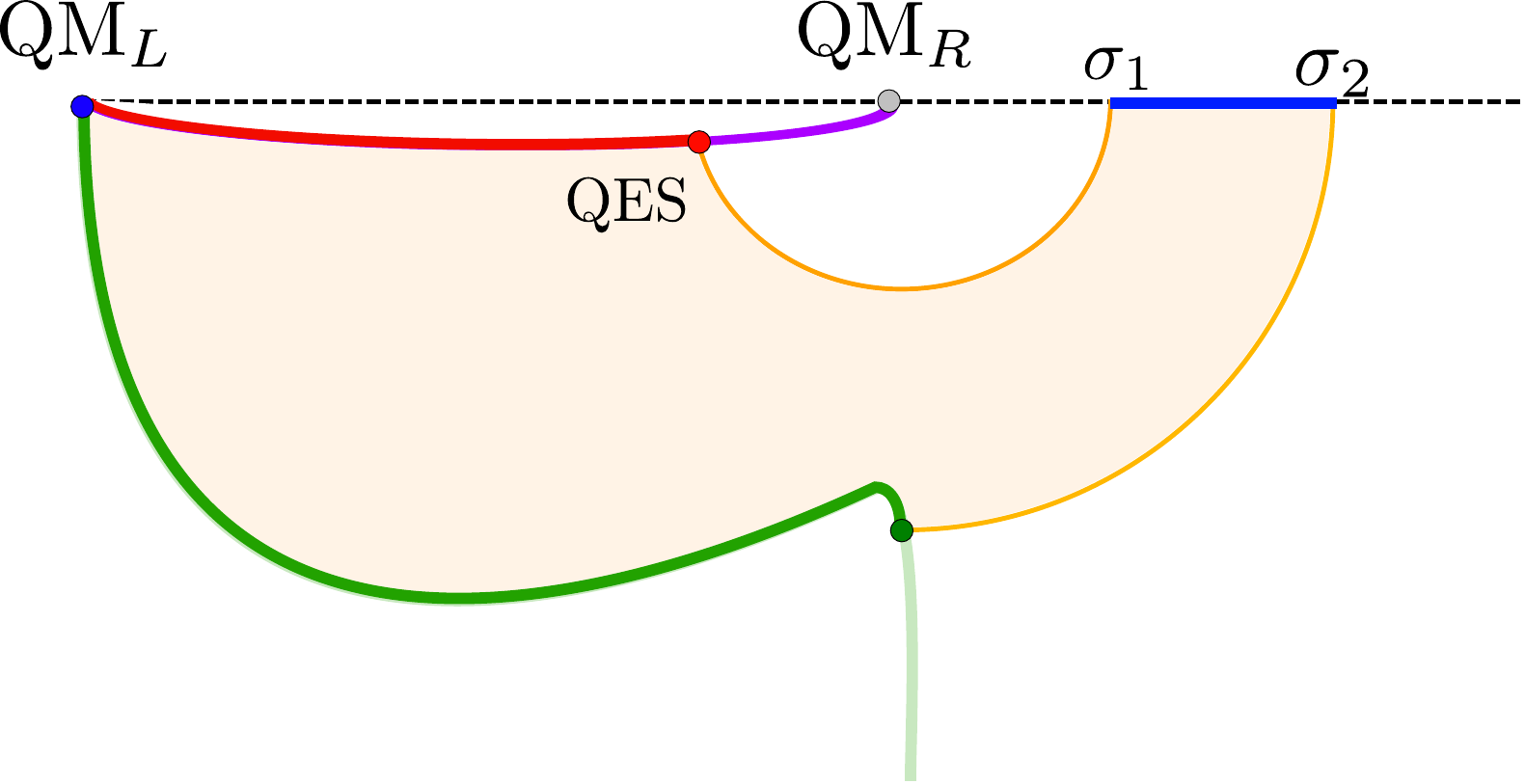}
	\caption{The shaded region is the volume which computes complexity of the bath subregion $[\sigma_1, \sigma_2] \ \cup\ $QM$_L$ in a reconstructing phase.}
	\label{fig:CV2}
\end{figure}

There are some subtleties involved in the computation of holographic CV in brane gravity + bath models due to the presence of the different branes in the bulk. In particular, the HRT surfaces are allowed to end at the ETW brane or the Planck brane, and so the boundary of the entanglement wedge of a boundary subregion can include parts of the ETW brane and of the Planck brane. In these cases, the boundary of the bulk codimension one surface $\cal B$ will include those sections of the branes as well -- see figure~\ref{fig:CV2}. However, while the entanglement entropy computation fixes the HRT surface (orange lines) and therefore its intersection with the Planck brane and ETW brane (red and green dots respectively), it does not fix the particular time slice on the Planck brane and ETW brane (red and green lines) which correspond to the remaining boundary of the bulk codimension one surface $\cal B$ (shaded orange region) in the complexity formula in eq.~\eqref{eq:CVsub}. These particular time slices, which are the intersection of $\cal B$ with the Planck brane and the ETW brane, maximize the total volume of the corresponding bulk surface $\cal B$.

We want to use subregion complexity to provide a different perspective on the subregion state. To this end, we study the holographic subregion complexity of the state in $\mathbf{R} \cup QM_L$ throughout the evaporation. However, there are two important obstacles to this end. First, due to the UV entanglement structure of the state in a local QFT, the holographic complexity=volume is UV divergent~\cite{Carmi2017}
\begin{equation}\label{eq:CV-UV}
    C_V^{\rm sub} (\mathbf{R}) = \frac{L}{\delta} \frac{\tilde{V}(\mathbf{R})}{G_N \ell} +\cdots\,,
\end{equation}
where $\Tilde{V}(\mathbf{R})$ is the volume of the subregion $\mathbf{R}$ on the boundary theory and $\delta$ is the holographic UV cutoff. In eq.~\eqref{eq:CV-UV}, we have omitted all terms that do not diverge with the cutoff $\delta/L\to 0$. The second obstacle is that computing the maximum volume bulk slice in a nontrivial, time-dependent geometry is in general a very complicated task, especially taking into account that the boundary subregions we are considering are not symmetric.

The first issue can be resolved by instead computing the increase in complexity
\begin{equation}\label{eq:DCsub}
    \Delta C^{\rm sub}_V\left(\mathbf{R}\right) = C^{\rm sub}_V\left(\mathbf{R}\right) - C^{\rm sub}_V\left(\mathbf{R}\right)\Big|_{u=0}\,,
\end{equation}
since the UV divergences will cancel out. The second obstacle can be overcome by considering the leading contributions to the subregion complexity following the strategy outlined in \cite{EasyIslands3}. Since the induced volume form diverges near the asymptotic boundaries, the leading contribution to the volume of $\cal B$ is a universal UV divergence proportional to the volume of the boundary region (blue line)~\cite{Carmi2017}. This universal UV divergence is cancelled in the increase in complexity in eq.~\eqref{eq:DCsub}. When the Planck brane is located very close to the asymptotic boundary, there is another large contribution associated to the volume of the time slice on the Planck brane connecting $QM_L$ to the QES (red line). This can be formalized with the use of a Fefferman-Graham (FG) expansion around the asymptotic boundary and computing the leading contribution to holographic CV in powers of the distance between the Planck brane and the QES~\cite{EasyIslands3}. In the limit where the Planck brane is located very close to the asymptotic boundary, the increase in complexity can be approximated by the first terms in the FG expansion, and is given by
\begin{equation}\label{eq:CV-FG}
    \Delta C_V^{\rm sub} (\mathbf{R}\cup QM_L) = \frac{L}{z_B} \frac{W({\cal I})}{G_N \ell} +\cdots\,,
\end{equation}
where $z_B \ll L$ is the position of the Planck brane in FG coordinates, $W({\cal I})$ is the generalized volume of the region $\cal I$ on the brane, and we omit higher order powers of $z_B/L$. This $z_B/L$ parameter serves a similar role as the UV cutoff $\delta/L$ in eq.~\eqref{eq:CV-UV}, and in higher dimensional doubly holographic models sets the scale of how local the brane effective action is~\cite{EasyIslands1,EasyIslands2,EasyIslands3}. This reduces the task to finding a codimension one surface on the Planck brane which maximizes the generalized volume, which is a much easier task. The generalized volume in JT gravity is given by an integral of the dilaton, $\phi_0 + \phi$, along the considered interval. We will further assume that $\phi_0 \gg \phi$, in which case the generalized volume is simply proportional to the length of the interval connecting $QM_L$ and the QES. The latter is precisely the distance between the left asymptotic boundary and the QES. In this limit, the leading contribution to the increase in complexity $\Delta C^{sub}_V$ in the FG expansion corresponds to the increase in size of the entanglement wedge in the brane perspective. 

This method greatly simplifies the computation of holographic CV, at the cost of losing track of the higher order terms in the FG expansion. Still, we can use this approximation as a complementary tool to the Page curve to enrich our understanding of the information properties of the system during the evaporation. In the brane perspective, the entanglement wedge corresponds to the domain of dependence of intervals anchored at the QES (and at the left asymptotic boundary when the purifier $QM_L$ is included) -- for example, see figure~\ref{figure: entanglement wedges of QML+finite segment at intermediate times}. We define the size of the entanglement wedge as the distance between the two endpoints of such an interval. The distance between two points in AdS$_2$ is
\begin{equation}
    d(x_1,x_2) = 2\, {\rm ArcTanh} \sqrt{\frac{(x_1^+-x_2^+)(x_1^--x_2^-)}{(x_1^+-x_2^-)(x_1^--x_2^+)}}.
\end{equation}

When considering one of the two endpoint at the left asymptotic boundary, $\frac{x^+_2-x^-_2}{2}=s_2 = s_\infty$ where $s_\infty$ is the UV regulator that goes to infinity, we find the size of the entanglement wedge of $\mathbf{R}\cup QM_L$:
\begin{equation}
    d(QM_L,x_1) = \log \frac{s_\infty}{s_1}\,.
\end{equation}
We define $d(QM_L,x_1)$ as the distance between the left asymptotic boundary and the point $x_1^\pm$. This distance is UV divergent, and so in what follows we will be interested in the increase of size of entanglement wedges
\begin{equation}\label{eq: Delta d}
    \Delta d(x_1,x_2) \equiv  \left| d(QM_L,x_1) - d(QM_L,x_2) \right| = \left| \log \frac{s_1}{s_2}\right|\,.
\end{equation}

The increase in size of the entanglement wedge between the trivial QES phases and the Scrambling phases is found by using the position of the QES in eqs.~\eqref{eq: scrambling QES} to find
\begin{equation}
\begin{aligned}
    \Delta d(x_{\rm bif},x_{QS}) &= \log \frac{2 \pi^2 T_0^2 (\pi T_0 t-1)}{2 \pi^2 T_0^2 (\pi T_0 t-1) - 2k \pi T_0 + k^2 (\pi T_0 t-1)}\\ 
    & = \frac{k}{\pi T_0- \pi^2 T_0^2 t}+ {\cal O}(k^2)\\
    & = \frac{k T_1}{\pi T_0 (T_1-T_0) + 2 \pi T_0^2 e^{-2\pi T_1 u}} + {\cal O}(k^2)\,,
\end{aligned}
\label{eq: deltaD-scrambling}
\end{equation}
where we have used eq.~\eqref{eq:fapprox} in the last equality. This increase in size is very small, and asymptotes to $\frac{k T_1}{\pi T_0(T_1-T_0)}$.

The increase in size of the entanglement wedge between the trivial QES phase and the Late phases can be found using eqs.~\eqref{eq: late QES}
\begin{equation}
\begin{aligned}
    \Delta d(x_{\rm bif},x_{QL}) &= \log \frac{6 k}{\pi T_0 (t_\infty-t) (8\pi T_1+ k)}\\ 
    & = 2\pi T_1 u - \log \frac{8\pi T_0}{3k} + {\cal O}(k)\,.
\end{aligned}
\end{equation}
where we have used eq.~\eqref{eq:fapprox} in the last equality. The linear increase in the size of the entanglement wedge is due to the linear growth of the Einstein-Rosen bridge. The offset $\log \frac{8\pi T_0}{3k}$ can be thought of as $2\pi T_0 u_{\rm HP} (T_0)$ where $u_{\rm HP} (T_0)$ is the Hayden-Preskill time of the unperturbed black hole. Alternatively, comparing the increase in size of the entanglement wedge in the Late phases to the entanglement wedge of a non-evaporating black hole with temperature $T_1$
\begin{equation}
\begin{aligned}
    \Delta d(x^\pm = \frac{\pm 1}{\pi T_1},x_{QL}) &= \log \frac{6 k}{\pi T_1 (t_\infty-t) (8\pi T_1+ k)}\\ 
    & = 2\pi T_1 (u - u_{\rm HP})\,,
\end{aligned}
\end{equation}
where $u_{\rm HP}$ is the Hayden-Preskill time of the black hole with temperature $T_1$.
\begin{figure}[t!]
\begin{center}
	\includegraphics[scale=0.55]{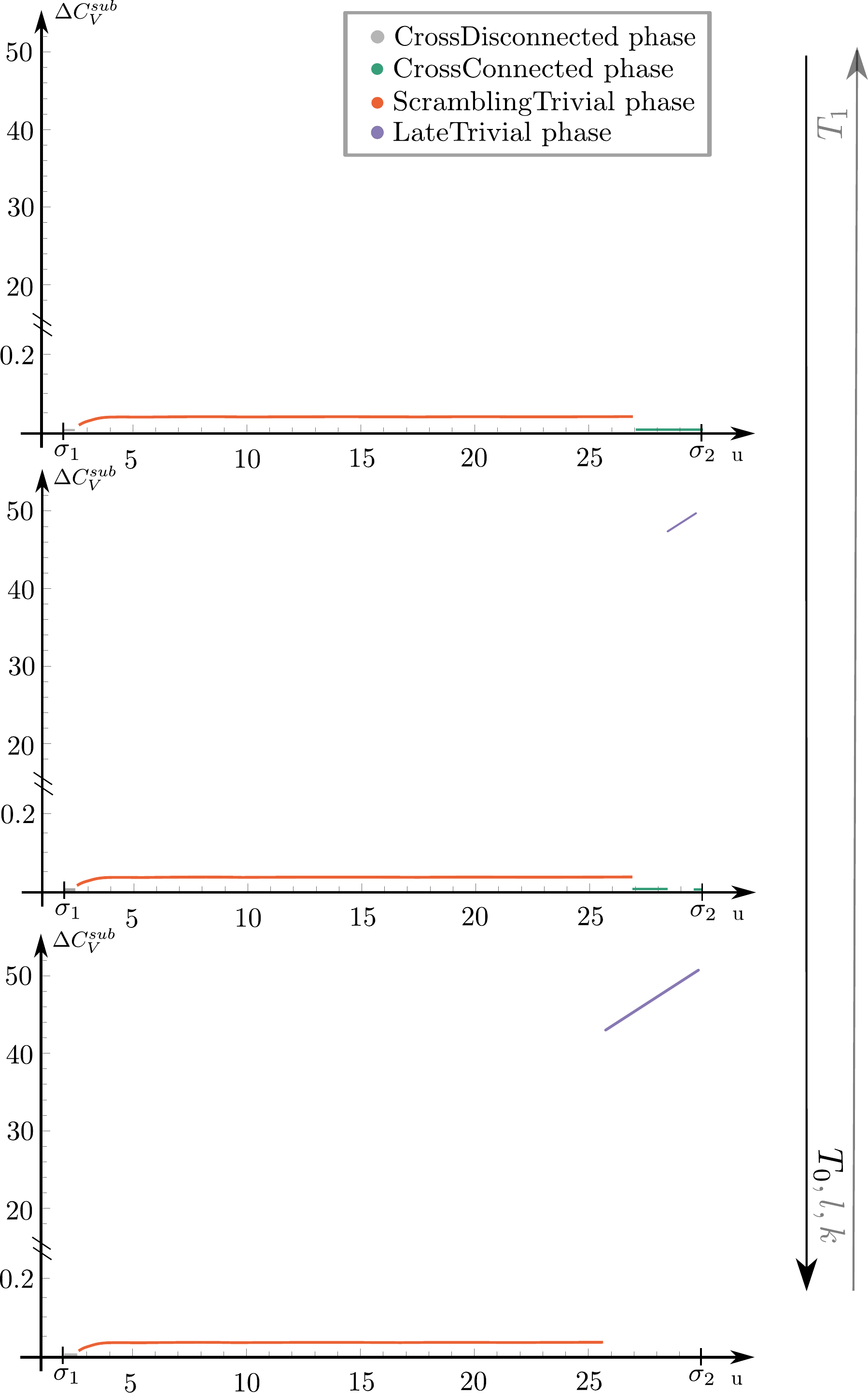} 
	\caption{This figure shows the evolution of sub-region complexity at intermediate for QM$_L$+finite segment of radiation in the bath depending on the values of the free parameters in the theory. Specifically, we used different values of $T_0$. Namely, we used $T_0=\frac{449}{512\pi}$ for the upper plot, $T_0=\frac{453,5}{512\pi}$ for the middle plot and $T_0=\frac{459}{512\pi}$ for the lower plot, in order to illustrate the different scenarios. The values of the other parameters used in this plot can be found in table~\ref{table section4}.  We have set $\ell = L_{AdS}/G_N$.  }
	\label{figure: subregion complexity finte inteval with QML intermediate times}
\end{center}
\end{figure}

The increase in subregion complexity for the three reconstruction scenarios described above are plotted in figure~\ref{figure: subregion complexity finte inteval with QML intermediate times}. This plot confirms that the leading behaviour of the subregion complexity captures the behaviour of the entanglement wedge, from the characteristic increase during the LateTrivial phase due to the growth of the Einstein-Rosen bridge, to the non-monotonicity for scenario 3 due to temporary blindness. The CrossConnected phase has no access to the black hole interior, and the entanglement wedge remains the same up to the time dependence of $\mathbf{R}$ in the bath. For the reconstructing phases, the intersection of the entanglement wedge with the Plank brane evolves nontrivially. The small increase in subregion complexity in the ScramblingTrivial phase reflects the small amount of extra information contained in $\mathbf{R}\cup QM_L$ which enables it to reconstruct a small portion of the black hole interior. Similarly, the large and linearly increasing complexity in the LateTrivial phase reflects the fact that $\mathbf{R}\cup QM_L$ encodes a large and increasing portion of the black hole interior. Indeed, the throat of the Einstein-Rosen bridge is in the entanglement wedge of $\mathbf{R}\cup QM_L$, and it is known to grow linearly with time. The increase of subregion complexity in both reconstructing phases is related to the short distance correlations in the portion of the black hole interior which is accessible. This can be understood in the brane perspective. For $\mathbf{R}\cup QM_L$ to accurately reconstruct a portion of the black hole interior $\cal I$, it must contain information of the entanglement structure of the degrees of freedom in the aforementioned region. In a physical state of a local QFT, there is a large amount of short distance entanglement, so there is a large amount of short range entanglement in the degrees of freedom in the region $\cal I$. Indeed, the leading contribution to the increase in complexity in the FG expansion in eq.~\eqref{eq:CV-FG} suggests that the leading order contribution in the increase in complexity of $\mathbf{R}\cup QM_L$ is due to the complexity of establishing correlations at distances of order $z_B \ll L$ in the now reconstructable region $\cal I$ inside the black hole. Finally, let us note that the leading contribution to the holographic subregion complexity experiences discontinuities across phase transitions of the entanglement wedge. Such discontinuities in holographic settings are generic to Hartman-Maldacena-type transitions~\cite{Alishahiha15}, and were observed in the context of Page transitions before in \cite{Bhattacharya21,Bhattacharya212,Bhattacharya213}.

Figure~\ref{figure: subregion complexity finte inteval with QML intermediate times} provides a different perspective on the three reconstructing scenarios described in section~\ref{ssection: page curve}. The upper plot corresponds to scenario 1 for which the complexity increases by a very small amount at the ScramblingTrivial phase. Eventually the interior becomes inaccessible and the leading contribution to the subregion complexity returns to its initial value at the CrossConnected phase. The lower plot is for scenario 2. Here, the interior reconstruction window is longer and after the Page transition the subregion complexity is much larger and keeps increasing linearly in time. Eventually, at the end of the interior reconstruction window, the leading contribution to the subregion complexity drops down to its initial value. As mentioned above, this linear increase of the complexity in the LateTrivial phase is due to the growth of the Einstein-Rosen bridge, which is now inside the reconstructable region of the black hole interior. The middle plot corresponds to scenario 3, in which the interior reconstruction window is temporarily interrupted. In this case, there is a temporary drop in the subregion complexity in the CrossConnected phase as access to the interior is lost between the ScramblingTrivial phase and the LateTrivial phase. Similarly to scenario 2, the complexity in the LateTrivial phase increases linearly in time until the end of the interior reconstruction window, at which point the leading contribution drops to its initial value.

\section{Discussion and outlook}
\label{sec:Discussion}

We have studied in detail the entanglement dynamics and the ability to reconstruct the black hole interior of different portions of Hawking radiation of an AdS$_2$ black hole evaporating into the external flat bath. For this purpose we have constructed the full numerical Page curves for portions of radiation gathered by different segments in the external bath, and examined its properties analytically. The main results are the following. 

\paragraph{1.} For semi-infinite segments, one can recover information about the black hole interior during an island phase at very late times in a certain range of parameters of the model. Specifically, the dilaton constant $\Tilde{\phi}_0$ cannot be too large and must obey the inequality in eq.~\eqref{eq:phi0-condition}. Violation of this inequality moves the potential island phase to times parametrically larger than the order of $1/k$, beyond the regime in which the semiclassical QES computation is applicable. If this happens, the island phase does not occur within the regime of applicability of the model.

\paragraph{2.} For finite segments in the bath we establish that the black hole purifier is essential for the interior reconstruction, unless the interval scales exponentially in $T_1/k$. Furthermore, we find significant dependence of the saddle point structure and physics of the Page curve on the parameters of the model. We identify four main scenarios for the entropy evolution, and we analyse the possibilities of interior reconstruction within them. For each of these scenarios, the entanglement wedge has drastically different temporal behaviours. Scenario 0 applies to small segments -- in this case the evolution of the entanglement entropy is completely analogous to the local quench scenario in flat space CFT \cite{Calabrese07,Calabrese09}, and there is no interior access. The other 3 scenarios apply to segments of size comparable to the Page time: these segments have the most nontrivial pattern of interior reconstruction. The corresponding Page curves are presented on figure~\ref{figure: finite segments intermediate times}. In scenario 1 the Page curve evolves through the ScramblingTrivial phase, during which the entanglement wedge contains a small portion of the black hole interior available for reconstruction. This entanglement wedge is bounded by a QES, whose distance from the bifurcation surface is of order $k$. This reconstruction window gets interrupted before the second endpoint of the interval crosses the shock as we transition into the non-reconstructing phases similar to scenario 0. In scenario 2, the size of the interior portion of the entanglement wedge grows in a discontinuous manner with a jump, as the shock moves through the radiation segment $\mathbf{R}$. This jump occurs on the phase transition from ScramblingTrivial phase to LateTrivial phase. After the sudden growth of the entanglement wedge and until the end of the interior reconstruction window, there is a time region where this segment reconstructs a large and linearly increasing portion of the black hole interior. Finally, scenario 3 is the most nontrivial one: the size of the entanglement wedge of $\mathbf{R}$ behaves non-monotonically: the reconstructable interior region is small, then becomes inaccessible for a time, after which a much larger interior region can be accessed until the end of the interior reconstruction window. 

\paragraph{3.} We identify which parameters of the model control the emergence of the non-monotonic behaviour in the size of the entanglement wedge which manifest as a temporary interruption of the interior reconstruction window that occurs in scenario 3. When varying the size of the segment $l = \sigma_2 - \sigma_1$, this non-monotonicity is caused by the competition between the generalized entropy of the nontrivial QES saddles, which depend on the second endpoint location in a linear way, with the CrossConnected saddle, which is determined by locations of both endpoints of $\mathbf{R}$ and has a logarithmmic divergence when the shock goes through the second endpoint at $u=\sigma_2$. We checked that this non-monotonicity in the size of the entanglement wedge is consistent with entanglement wedge nesting when the size of the segment is varied. The other parameters, namely the temperatures $T_0$, $T_1$ and the coupling parameter $k$, quantify the ratio between the energy of the shock and the temperature of the evaporating black hole. This combination of parameters controls the impact of the entanglement created by the shock relative to the effect of the Hawking radiation during the evaporation. When the effect of the Hawking radiation is more important, we find the large reconstruction scenario 2. On the other hand, when the entanglement created by the shock dominates, the Page curve evolves through scenarios 0 or 1, which allows for much less, if any, interior reconstruction. When transitioning between large reconstruction to small reconstruction in phase space, the curious non-monotonic scenario 3 occurs. Comparing the scenarios 2 and 3, the emergence of the python's lunch is delayed in scenario 2 but is instant in scenario 3. For late time $u > \sigma_2$, we find that the island saddles (which would have 2 nontrivial quantum extremal surfaces) can never dominate over the FutureConnected saddle within the regime of applicability of the model, unless the segment size scales exponentially in $T_1/k$. We have also checked that this late-time behaviour holds true in the version of the model where the bath has finite temperature~\cite{Chen20}. 

\paragraph{4.} We compute the leading contribution to holographic subregion complexity=volume for the radiation regions to leading order in the Fefferman-Graham expansion. We find that the discontinuities and non-monotonicity in the subregion complexity reflect the behaviour of the entanglement wedge throughout the phase evolution of the Page curve.\footnote{While we only presented the computation for finite bath regions, the relation between subregion complexity and geometric properties of the entanglement wedge is more general and applies not only to semi-infinite intervals in this model, but also to any holographic model.} The results of the computation are presented in figure~\ref{figure: subregion complexity finte inteval with QML intermediate times}, which illustrates that the holographic subregion complexity growth can be discontinuous as opposed to the expectation for the complexity of a pure state from \textit{e.g.}\ circuit complexity \cite{Nielsen05,Nielsen06,Dowling08}. Specifically, we have shown that the subregion complexity of radiation segments in this model is not continuous due to phase transitions of the entanglement wedge. This discontinuity is most easily understood from the brane perspective. In this perspective, there is an additional region $\cal I$ on the brane which becomes accessible by $\mathbf{R}\cup QM_L$ after the transition to a reconstructing phase. It is therefore natural to associate the discontinuous increase in complexity $\mathbf{R}\cup QM_L$ to the complexity of establishing the short distance correlation of the state in $\cal I$. The boundary perspective provides another angle on this discontinuity in the subregion complexity. From this point of view it is clear that this discontinuity is not exclusive to transitions between reconstructing and non-reconstructing phases in black hole backgrounds. On the contrary, it is a generic feature in holographic models when considering disconnected subregions $A \cup B$ \cite{Ben-Ami16}, with separate components $A$ and $B$, that can exhibit a Hartman-Maldacena type phase transition \cite{HM}. The discontinuity in $C^{\rm sub}_V(A\cup B)$ can be understood from the point of view of the entanglement/information structure of the state in the boundary CFT. Specifically,  the jump in subregion CV quantifies the discontinuity in the entanglement wedge between a connected and a disconnected configuration of $A\cup B$ in which different HRT surfaces dominate the entanglement entropy $S(A\cup B)$. For the boundary state, a manifestation of this transition is the appearance of correlation between $A$ and $B$ at large $N$. In the specific evaporating black hole model we have studied, the two disconnected subregions in question are the black hole purifier $QM_L$ and the bath interval $\mathbf{R}$. This suggests that the discontinuous increase of subregion complexity at the transition to a reconstructing phase may be related to the sudden emergence of nontrivial correlation at large $N$ between $QM_L$ and $\mathbf{R}$. We leave it to the future work to establish the precise connection between the long range entanglement in the boundary perspective and the short range entanglement withing the region $\cal I$ in the brane perspective. 
\\
\\
The main point which stands out in the present study is the strong dependence of the phase structure of the subregion Page curves and of the evolution of entanglement wedges on parameters of the model. These parameters include $T_0$ and $T_1$, which are related to the energy of the local quench $E_S$ through eq.~\eqref{eq: formula for energy of the shock}. In other words, these parameters can be shifted by perturbations in the bath. It is commonly accepted \cite{Penington191,Penington192,Almheiri192} that the Page curve of the entire radiation of an evaporating black hole, as well as the interior reconstruction are qualitatively robust to any unitary dynamics applied to the bath. The parameter dependence of the phase structure of the Page curves of finite portions of radiation indicates that these Page curves are devoid of this robustness property. Indeed, generic unitary dynamics in the bath changes entanglement structure within the bath, and finite segments will be sensitive to these effects, unlike the semi-infinite ones. Therefore we see that this quantum information about the black hole interior in the finite radiation portions is much more fragile, which comes as a trade-off for having a more fine grained description of where the information about the black hole interior is located, compared to probing only larger radiation portions. 

A related point was brought up in a recent work \cite{Balasubramanian22}, where it was shown in the setting of the PSSY (west coast) model \cite{Penington192} that the Page curve phase structure of a portion of radiation is robust to quantum errors. One can think that if slight parameter changes of the semiclassical state could be interpreted as quantum errors, our observations of parameter dependence would be at odds with those results. However, it is important to note that the quantum error correction results of \cite{Balasubramanian22} only involve very large radiation regions at very late times. This regime corresponds to the semi-infinite segments at very late times in our study, which are consistent with robustness of the Page curve. Nevertheless, it would be interesting to investigate the quantum error correcting properties of the fine-grained Page curves of finite portions of radiation throughout the entire time of evaporation, and to make contact with random unitary toy models of QEC in Hawking radiation, such as in \cite{Akers22}. We leave this question for future work.

In this paper, we have modeled Hawking radiation by a holographic CFT. We speculate that, moving away from this holographic limit, the details of the Page curves for finite segments would be sensitive to the specifics of the model. This would introduce an additional degree of sensitivity, beyond the dependence on parameters observed in the current paper. We can motivate this expectation by comparing to the case of eternal black holes. In the work \cite{Hollowood21} it was shown that Page curves of finite radiation segments represented by the free fermion CFT are sensitive to memory effects of the entanglement entropy \cite{Asplund15}, whereas the work \cite{Balasubramanian212} had shown that the holographic radiation does not exhibit such effects. In principle, the computations of the present work could be repeated e.g.\ for the non-holographic free fermion CFT, and we expect that memory effects would cause the resulting Page curves to have even more intricate structure than the results we have presented.

Other interesting future directions involve going away from the AdS$_2$ evaporating black hole model of \cite{Almheiri191} to more realistic settings. One interesting generalization is replacing the flat external bath with a dynamical observer in the bulk, which can only absorb the Hawking radiation for a finite period of time, along the lines of the recent work \cite{DeVuyst22}. Another example is higher-dimensional generalizations with emphasis on dependence of the Page curve on the shape of the radiation segments.


\section*{Acknowledgements}
We would like to thank Sergio E.~Aguilar-Gutierrez for collaboration during the initial stage of this work, and Irina Aref'eva, Vijay Balasubramanian, Guglielmo Grimaldi, and Thomas Mertens for useful discussions. This research has been supported by FWO-Vlaanderen projects G006918N and G012222N, and by Vrije Universiteit Brussel through the Strategic Research Program High-Energy Physics.

\appendix

\section{Dilaton solution after the quench}
\label{section: Dilaton Solution}
In this appendix we review the derivation of the dilaton solution after the quench as is given by eq.~\eqref{eq: dilaton-closed}. The general inhomogeneous solution with a nontrivial stress tensor source was found in \cite{Almheiri14,Engelsoy16} as is given in eq.~\eqref{eq: dilaton-integral}. We copy this equation for convenience 
\be
\phi = 2 \phi_r \frac{1-(2\pi T_0)^2x^+x^-+kI}{x^+-x^-}\,,
\ee
where 
\be
I = - \frac{24\pi}{c} \int_{0^-}^{x^-} dt (x^+-t)(x^--t) \langle T_{x^-x^-}(t)\rangle\,. \label{eq: dilaton-I}
\ee
This expression can be rewritten in a more convenient form. For this we use the form of the stress tensor after the quench given in eq.~\eqref{eq:TAdS}. We also use the reparametrization function $f$ which is determined by the energy balance equation in eq.~\eqref{eq:SchwEq}. This information allows to rewrite the stress tensor as a total derivative
\begin{equation}
	\begin{aligned}
		T_{x^+x^+}(x^+)&=0,\\
		T_{x^-x^-}(x^-)&=\frac{\partial_{y^-}E(y^-)}{(f'(y^-))^2}=-\frac{\phi_r}{8\pi G_N}\frac{\partial_{y^-}\qty{f(y^-),\: y^-}}{(f'(y^-))^2}=-\frac{\phi_r}{8\pi G_N}\partial_{x^-}^3 f'(y^-),
	\end{aligned}\label{eq:Tmunu_3ple}
\end{equation}
where $x^-=f(y^-)$, and the last equality follows from the chain rule. Substituting this into eq.~\eqref{eq: dilaton-I} and computing the integral, we get
\begin{equation}
	\begin{aligned}
		\int_{0^-}^{x^-} dt\: (t-x^+)(t-x^-)\partial_{t}^3 h(u)=&-x^+x^- \partial^2_{t}h(u)\Big|_{u=0^-}+(x^+-x^-)\partial_{x^-}h(y^-)\\
		&-(x^++x^-)\partial_{t}h(u)\Big|_{u=0}+2[h(y^-)-h(0)],
	\end{aligned} \label{eq: int-I}
\end{equation}
where $h(y^-)=f'(y^-)$ and $t=f(u)$. Using the explicit form of the solution for $f$ given in eq.~\eqref{f(u)} (and its derivatives) we can establish the following identities:
\begin{align}
	h(0)&=\eval{f'(u)}_{u=0}=1,\\
	\partial_{t}h(u)\Big|_{u=0}&=\eval{\frac{f''(u)}{f'(u)}}_{u=0}=0,\\
	\partial^2_{t}h(u)\Big|_{u=0^-}&=\eval{\qty[\frac{f'''(u)}{(f'(u))^2}-\frac{(f''(u))^2}{(f'(u))^3}]}_{u=0^-}=-2(\pi T_0)^2.
\end{align}
Using these in eq.~\eqref{eq: int-I} and putting it all back into the general solution given in eq.~\eqref{eq: dilaton-integral}, the dilaton profile becomes:
\begin{equation}
	\phi(x)=\phi_r\qty[\partial_{x^-}h(y^-)+2\frac{h(y^-)}{x^+-x^-}]=\phi_r\qty[\frac{f''(y^-)}{f'(y^-)}+2\frac{f'(y^-)}{x^+-f(y^-)}],\label{eq:dil_back}
\end{equation}
where $f(y^-)$ is in eq.~\eqref{f(u)}. This form of the solution is more convenient for practical computations of the locations of the QESs, as established in \cite{Hollowood20}.

\section{Approximate analytic solutions for QES at intermediate and late times}
\label{section: QES}

In this appendix, we provide the necessary details needed for the analytic derivation of the QES solutions for the nontrivial QES phases.

\subsection{Scrambling saddles}
\label{section: QES solution for Scrambling Phases}
We begin with the Scrambling saddles with a nontrivial QES located below the shock. The location of the QES is denoted by $x_{QS}$ and is a solution to the following two equations 
\begin{equation}
\begin{split}
    0=\frac{\partial S^{S.}_{gen}\left(x^+_{QS},x^-_{QS}\right)}{\partial x^+_{QS}}&\propto k x^-_{QS} \left(x^-_{QS} - 
     x^+_{QS}\right) + \left(-1 + \pi^2 T_0^2 x^{2-}_{QS}\right) x^+_{QS} \\
     &\quad + \left(1 - \pi^2 T_0^2 x^{2-}_{QS} + 
     k (-x^-_{QS} + x^+_{QS})\right) f(y^+),
\end{split}
\end{equation}
\begin{equation}
\begin{split}
    0=&\frac{\partial S^{S.}_{gen}\left(x^+_{QS},x^-_{QS}\right)}{\partial x^-_{QS}}\propto\left(k x^{2+}_{QS} - x^-_{QS} \left(-1 + k x^+_{QS} + \pi^2 T_0^2 x^{2+}_{QS}\right)\right).
    \end{split}
\end{equation}
There is an exact and unique solution to these equations. However, for our purposes it is sufficient to find the approximate solutions for $x^+_{QS}$ and $x^-_{QS}$. To this end, we expand these equations in the small parameter $k$ and solve the equations order-by-order. There are several solutions but only one which is located in the AdS$_2$ region below the shock. The only solution that lies within the correct region of spacetime is 
\begin{subequations}\label{eq:QESscram}
\begin{align}
    x^-_{QS}&= \frac{-1}{\pi  T_0}+ \frac{k (\pi  T_0 f(u-\sigma)+1)}{\pi^2  T_0^2 (1-\pi  T_0 f(u-\sigma))}+\mathcal{O}\left(k^2\right)\,,
    \label{eq: solution of x^-_Q for scrambing phases}\\
    x^+_{QS}&= \frac{1}{\pi  T_0}-\frac{k}{\pi^2  T_0^2}+\mathcal{O}\left(k^2\right)\,.
    \label{eq: solution of x^+_Q for scrambing phases}
\end{align}
\end{subequations}

\subsection{Late saddles}
\label{section: QES solution for Late Phases}
Next, we focus on the Late saddles with a nontrivial QES located above the shock. The location of the QES is denoted by $x_{QL}$ and is a solution to the following two equations 
\begin{equation}
    0=\frac{\partial S^{L.}_{gen}\left(x^+_{QL},y^-_{QL}\right)}{\partial x^+_{QL}}\propto f'(y^-_{QL}) (f(y^+)-x^+_{QL})+k (x^+_{QL}-f(y^-_{QL})) (f(y^+)-f(y^-_{QL})),
    \label{eq: first QES eqn, derivative to xplus}
\end{equation}
\begin{equation}
\begin{split}
    0=&\frac{\partial S^{L.}_{gen}\left(x^+_{QL},y^-_{QL}\right)}{\partial y^-_{QL}}\propto \left(y^-+y^-_{QL}\right)\left(f'''(y^-_{QL}) f'(y^-_{QL}) \left(x^+_{QL}-f(y^-_{QL})\right)^2\right.\\
    &\quad\left.+2 f''(y^-_{QL}) f'(y^-_{QL})^2 \left(x^+_{QL}-f(y^-_{QL})\right) -f''(y^-_{QL})^2\left(x^+_{QL}-f(y^-_{QL})\right)^2\right.\\
    &\quad\left.+2 f'(y^-_{QL})^4 \right)+k\left(f'(y^-_{QL}) \left(x^+_{QL}-f(y^-_{QL})\right)\left(\left(y^-+y^-_{QL}\right) f''(y^-_{QL}) \left(x^+_{QL}-f(y^-_{QL})\right)\right.\right.\\
    &\quad\left.\left.-2 f'(y^-_{QL}) \left(\left(y^-+y^-_{QL}\right) f'(y^-_{QL})+f(y^-_{QL})-x^+_{QL}\right)\right)\right).
    \end{split}
    \label{eq: second QES eqn, derivative to ymin}
\end{equation}
We aim to approximate the equations in order to find the approximate solutions for $x^+_{QL}$ and $y^-_{QL}$ that determine the location of the QES. 

For late times, we can use a small $k$ expansion while keeping $k y^+$ fixed. In this regime, one can find the following approximation of $f(y^+)$
\begin{equation}
    f(y^+)=t_\infty\left(1-2\lambda\right), \quad \lambda=\exp\left[-\frac{4 \pi  \left(1-e^{-\frac{k y^+}{2}}\right) T_1}{k}+\frac{k}{4 \pi  T_1}+\mathcal{O}\left(k^2\right)\right],
    \label{eq: approximation of f(u-sigma)}
\end{equation}
where $\lambda$ is another small parameter. Specifically, when $y^+$ is around the Hayden-Preskill time $u_{\rm HP}= \frac{1}{2\pi T_1} \log\frac{8\pi T_1}{3k}$, $\lambda \sim \left(\frac{3k}{8\pi T_1}\right)^{2\pi T_1}$ is polynomial in $k$. For later times, with $ky^+$ still finite as assumed for the rest of the derivation, it is exponential in $\frac{-1}{k}$ and therefore much smaller than $k$. Furthermore, we can use eq.~\eqref{eq: approximation of f(u-sigma)} to find approximations for the derivatives of $f(y^+)$
\begin{equation}
    \begin{split}
        f'(y^+)&=4 \pi  T_1 e^{-\frac{k y^+}{2}}t_\infty \lambda\left(1-\frac{k^2 }{32 \pi ^2 T_1^2 e^{-\frac{k y^+}{2}}}+\mathcal{O}\left(k^3\right)\right)  ,\\
        f''(y^+)&=-8 \pi ^2 T_1^2 e^{-k y^+} t_\infty \lambda\left(1+\frac{k e^{\frac{k y^+}{2}}}{4 \pi  T_1}-\frac{k^2 e^{\frac{k y^+}{2}}}{16 \pi ^2 T_1^2}+\mathcal{O}\left(k^3\right)\right) ,\\
        f'''(y^+)&=16 \pi ^3 T_1^3 e^{-\frac{3 k y^+}{2} }t_\infty\lambda \left(1+\frac{3 k e^{\frac{k y^+}{2}}}{4 \pi  T_1}\frac{k^2 e^{\frac{k y^+}{2}} \left(2 e^{\frac{k y^+}{2}}-3\right)}{32 \pi ^2 T_1^2}+\mathcal{O}\left(k^3\right)\right) .
    \end{split}
    \label{eq: approximations of derivatives of f(u-sigma)}
\end{equation}
Finally, defining another small parameter $\tau$, as
\begin{equation}\label{eq: f-tau}
     f(y^-_{QL})=t_\infty\left(1-2\tau\right)\,, \quad \tau=\exp\left[-\frac{4 \pi  \left(1-e^{-\frac{k y^-_{QL}}{2}}\right) T_1}{k}+\frac{k}{4 \pi  T_1}+\mathcal{O}\left(k^2\right)\right]\,,
\end{equation}
we find identical approximations for the derivatives as in eq.~\eqref{eq: approximations of derivatives of f(u-sigma)} but with $\lambda \to \tau$ and $y^+ \to y^-_{QL}$.

Having identified the small parameters, we expand the position of the QES as follows
\begin{equation}
\begin{split}
    x^+_{QL}&=t_\infty\left(1+\delta x^+\right)\,,\\
    y^-_{QL}&=y^+ -\delta y^-.
    \end{split}
\end{equation}
Note that $\delta x^+$ is dimensionless, while $\delta y^-$ has dimension length. Now, let us use these approximations in order to find the location of the QES. We assume that $\lambda$ is of the same order as $k\tau$. We will confirm this assumption later, when analyzing eq.~\eqref{eq: second QES eqn, derivative to ymin}. 

We begin by approximating eq.~\eqref{eq: first QES eqn, derivative to xplus}. To this end, we introduce two tracing parameters $\mu$ and $\nu$ where $\mathcal{O}\left(\mu\right)=\mathcal{O}\left(\tau\right)$ and $\mathcal{O}\left(\nu\right)=\mathcal{O}\left(k\right)$ such that $\mathcal{O}\left(\mu\nu\right)=\mathcal{O}\left(\lambda\right)$. We do this by rescaling $\lambda\to \mu\nu\Tilde{\lambda}$, $\tau\to \mu\Tilde{\tau}$ and $k\to \nu \Tilde{k}$, where the tilded parameters are kept finite.\footnote{Note that since we keep $k y$ fixed, we don't rescale that combination.} The equation up to leading order in $\mu$ reads
\begin{equation}
    0=t_\infty \delta x^+ \left(2 \nu \Tilde{k}  t_\infty (\Tilde{\tau} -\nu \Tilde{\lambda}   )-4 \pi  T_1 e^{-\frac{k y^-_{QL}}{2}}t_\infty \Tilde{\tau}\left(1-\frac{\nu^2 \Tilde{k}^2 }{32 \pi ^2 T_1^2 e^{-\frac{k y^-_{QL}}{2}}}+\Tilde{k}\mathcal{O}\left(\nu^3\right)\right)\right)\mu+\mathcal{O}\left(\mu^2\right).
\end{equation}
This tells us that $\delta x^+ = 0 + {\cal O}(\mu)$. With this knowledge, we can look at the next order expansion in small $\mu$, which is given by
\begin{equation}
\begin{split}
    0=2t_\infty^2 & \left[\nu \Tilde{k}  (\Tilde{\tau} -\nu\Tilde{\lambda} )   (\delta x^++2\Tilde{\tau})-2\pi  T_1 e^{-\frac{k y^-_{QL}}{2}} \Tilde{\tau}\left(1-\frac{\nu^2 \Tilde{k}^2 }{32 \pi ^2 T_1^2 e^{-\frac{k y^-_{QL}}{2}}}+\Tilde{k}\mathcal{O}\left(\nu^3\right)\right)\right.\\
    &\qquad \left.(\delta x^+ + 2\nu\Tilde{\lambda}  )\right]\mu^2+\mathcal{O}\left(\mu^3\right).
\end{split}
\end{equation}
This expression can now be expanded in small $\nu$, which leads to
\begin{equation}
\begin{split}
   0&=-4 \left(\pi  \Tilde{\tau}^2 T_1 t_\infty^2 \delta x^+ e^{-\frac{k y^-_{QL}}{2}}\right)\mu^2+ 2 \nu  \Tilde{\tau}  t_\infty^2 \left(\Tilde{k}  (\delta x^++2\Tilde{\tau})-4\pi  T_1 e^{-\frac{k y^-_{QL}}{2}} \Tilde{\lambda}  \right)\mu^2+\mathcal{O}\left(\nu^2\mu^2\right).
   \end{split}
\end{equation}
From this equation we find that $\delta x^+=0+\mathcal{O}\left(\mu\nu\right)$. 

The next order then tells us that
\begin{equation}
    0=\Tilde{k} \Tilde{\tau} -\pi  T_1 e^{-\frac{k y^-_{QL}}{2}} \left(  \delta x^++2\Tilde{\lambda}\right) + {\cal O}(\mu\nu^2).
\end{equation}
This equation is exactly solvable and leads to
\begin{equation}
    \delta x^+ = \frac{\Tilde{k} \Tilde{\tau}  e^{k y^-_{QL}/2}}{\pi   T_1}-2\Tilde{\lambda} + {\cal O}(\mu \nu^2).
\end{equation}
For early times, we can safely assume that $k y^-_{QL}$ is small, such that this solution becomes
\begin{equation}
    \delta x^+= \frac{\Tilde{k} \Tilde{\tau} }{\pi    T_1}-2\Tilde{\lambda}+ {\cal O}(\mu \nu^2).
\end{equation}
At this point we can summarize our results as follows
\begin{equation}
    x^+_{QL}=t_\infty+t_\infty\left(e^{k y^-_{QL}/2} \frac{k\tau}{\pi T_1}-2\lambda\right)+\mathcal{O}\left(k\lambda\right)\simeq t_\infty+t_\infty\left(\frac{k\tau}{\pi T_1}-2\lambda\right)+\mathcal{O}\left(k\lambda\right).
    \label{eq: first approx to x^+_Q for late times}
\end{equation}

Let us now proceed and tackle the second equation in eq.~\eqref{eq: second QES eqn, derivative to ymin} in a similar way. Expanding the equation in small $\mu$ and $\nu$, we find that the leading order is given by
\begin{equation}
\begin{split}
    0&=64 \pi ^2 \Tilde{\tau} ^3 T_1^2 t_\infty^4 e^{-2 k y^-_{QL}} \left(3 \pi  \Tilde{k} \Tilde{\tau}  T_1 e^{k y^-_{QL}/2} (\delta y^-+y^--y^+)\right.\\
    &\left.+2 \Tilde{k} \Tilde{\tau}  e^{k y^-_{QL}}-8 \pi ^2 \Tilde{\lambda} T_1^2 (\delta y^-+y^--y^+)\right)\nu\mu^4 +\mathcal{O}\left(\nu^2\mu^4\right).
    \end{split}
\end{equation}
Due to the presence of $\tau$ in this equation, which is defined using $y^-_{QL}$, this equation is very hard to solve. The trick to approach this problem is to assume that $\delta y^-$ is very big compared to $y^--y^+$ and $\frac{1}{T_1}$. Hence, the biggest contribution to this equation tells us that
\begin{equation}\label{eq:large-delta-y}
    0=-k \tau  \left(3e^{k  y^-_{QL}/2}\right)+8 \pi  \lambda T_1\simeq -3k \tau+8 \pi  \lambda T_1\,.
\end{equation}
which confirms our initial assumption and tells us that
\begin{equation}
    k \tau=\frac{8\pi T1 \lambda}{3e^{k y^-_{QL}/2}}.
\end{equation}
In particular, we can now rewrite the approximation of $x^+_{QL}$ we found in eq.~\eqref{eq: first approx to x^+_Q for late times} to 
\begin{subequations}\label{eq:QESlate}
\begin{equation}
    x^+_{QL}=t_\infty+\frac{2}{3}t_\infty \lambda+\mathcal{O}\left(k\lambda\right).
    \label{eq: solution of x^+_Q for late phases}
\end{equation}

Lastly, we can use the explicit expressions for $\lambda$ and $\tau$ from eq.~\eqref{eq: approximation of f(u-sigma)} and eq.~\eqref{eq: f-tau}, respectively and solve this equation for $y^-_{QL}$
\begin{equation}
    y^-_{QL}=y^+-\frac{\log \left(\frac{8 \pi  T_1 }{3 k}\right)}{2 \pi T_1}+\mathcal{O}\left(k\left(\log\left(\frac{8\pi T_1}{3k}\right)\right)^2\right).
    \label{eq: solution of y^-_Q for late phases}
\end{equation}
\end{subequations}
This solution confirms the assumption that $\delta y^-$ is very big since it is proportional to $\log\left(\frac{8\pi T_1}{3k}\right)$.

\section{Bounds on the interior reconstruction window}
\label{section: Bounds on blind spot}
In this appendix, we give a detailed explanation on how to derive the bounds on a dimensionless parameter $\frac{E_S}{c(T_1+T_0)}=\frac{\pi}{12}\frac{T_1-T_0}{ k}$, in order for the interior reconstruction window to exhibit the temporary interruption described in scenario 3 of section~\ref{ssection: page curve}. We identify two bounds, the first being a lower bound on  $\frac{E_S}{c\left(T_0+T_1\right)}$ and the second being an upper bound. To derive the bounds, we focus on the competing saddles in the interior reconstruction window at intermediate times. We will copy the corresponding generalized entropies here for convenience
\begin{equation}
    \begin{split}
         S^{\rm CC}_{\rm gen}&=\frac{c}{6} \log \left(\frac{12 \pi E_S y^+_2 f(y^+_1)(y^-_1-y^-_2) }{c\sqrt{f'(y^+_1)}}\right)\\
        &+\frac{2 \pi  T_0 \phi_r+\Tilde{\phi}_0}{4 G_N}+\frac{c}{6}\log 2\,,\\
        S^{\rm ST}_{\rm gen}&=\frac{c}{6} \log \left(\frac{24 \pi E_S x^-_{QS}y^-_1(x^+_{QS}-f(y^+_1))}{c (x^-_{QS}-x^+_{QS})\sqrt{f'(y^+_1)}}\right)\\
        &+\frac{c}{6} \log \left(y^-_2-y^+_2\right)+\frac{\Tilde{\phi}_0+\phi(x^+_{QS},x^-_{QS})}{4 G_N},\\
         S^{\rm LT}_{\rm gen}&=\frac{c}{6} \log \left(\frac{2(y^-_1-y^-_{QL})(x^+_{QL}-f(y^+_1))}{(x^-_{QL}-x^+_{QL})}\sqrt{\frac{f'(y^-_{QL})}{f'(y^+_1)}}\right)\\
        &+\frac{c}{6} \log \left(y^-_2-y^+_2\right)+\frac{\phi (x^+_{QL},y^-_{QL})+\Tilde{\phi}_0}{4 G_N}\,.
    \end{split}
    \label{eq: competing phases in reconstruction window at intermediate times}
\end{equation}
Notice that we can use eq.~\eqref{eq: formula for energy of the shock} and eq.~\eqref{eq: eqn for k related to other parameters} in order to eliminate $c$ and $E_S$.
\subsection{Lower bound}
As discussed in section~\ref{ssection: page curve} around eq.~\eqref{eq:triple}, the lower bound on $\frac{E_S}{c(T_1+T_0)}$ can be established from the triple intersection of the competing phases specified in eq.~\eqref{eq: competing phases in reconstruction window at intermediate times}. We do this by first finding the time in which the ScramblingTrivial and CrossConnected generalized entropies intersect, and equate it to the Page time at which the ScramblingTrivial and LateTrivial entropies are equal.
\\
\paragraph{Time of intersection between $S^{\rm CC}_{\rm gen}$ and $S^{\rm ST}_{\rm gen}$}\mbox{}\\
We begin by finding the time $u$ for which
\begin{equation}
    0=S^{\rm CC}_{\rm gen}-S^{\rm ST}_{\rm gen}.
\end{equation}
We can write this equation out explicitly
\begin{equation}
    0=\frac{\phi_r}{2 G_N} \left(k \log \left(\frac{2 \sigma_2 (x^-_{QS} (\sigma_1+u) (x^+_Q-f(u-\sigma_1,\phi_r,k)))}{(x^-_{QS}-x^+_{QS}) ((\sigma_1-\sigma_2) (u-\sigma_2) f(u-\sigma_1,\phi_r,k))}\right)+\frac{\pi ^2 T_0^2 x^-_{QS} x^+_{QS}-1}{x^-_{QS}-x^+_{QS}}-\pi T_0\right)\,.
\end{equation}
We will solve this equation perturbatively. The location of $(x^+_{QS},x^-_{QS})$ is given by eqs.~\eqref{eq:QESscram}. Using the approximations defined in eq.~\eqref{eq: approximation of f(u-sigma)} and eq.~\eqref{eq: approximations of derivatives of f(u-sigma)}, and introducing the rescalings $\lambda\to \mu\nu\Tilde{\lambda}$, $\tau\to \mu\Tilde{\tau}$ and $k\to\nu\Tilde{k}$, the equation up to leading order in $\mu$ and $\nu$ reads
\begin{equation}
    0=\frac{\phi_r}{2G_N}\left(\nu \Tilde{k}\log\left(\frac{\sigma_2 (\pi  T_0 t_\infty-1) (\sigma_1+u)}{\pi  T_0 t_\infty (\sigma_2-\sigma_1) (u-\sigma_2)}\right)+{\cal O}\left(\nu^2\mu^0\right)\right).
\end{equation}
The solution to this equation is
\begin{equation}
    u_{CCST}=-\frac{\sigma_2 (\pi  \sigma_2 T_0 t_\infty -\sigma_1)}{\pi  \sigma_1 T_0 t_\infty -\sigma_2}+\mathcal{O}\left(k\right).
\end{equation}
\\
\paragraph{Time of intersection between $S^{\rm ST}_{\rm gen}$ and $S^{\rm LT}_{\rm gen}$}\mbox{}\\
Next, we find the time $u$ for which
\begin{equation}
    0=S^{\rm ST}_{\rm gen}-S^{\rm LT}_{\rm gen}.
\end{equation}
This time is referred to as the Page time since it marks the transition to the phase in which a large portion of the black hole interior is accessible, and in which the entropy begins to decrease as argued by Page \cite{Page93}. We can write out this equation explicitly using the dilaton solution given in eq.~\eqref{eq: dilaton profile before and after quench}:
\begin{equation}
\begin{split}
    0&=\frac{\phi_r}{4 G_N} \left(2 k \log \left(\frac{\pi ^2 x^-_{QS} \left(T_1^2-T_0^2\right) (\sigma_1+u) \left(\frac{x^+_{QL}-f(y^-_{QL})}{\sqrt{f'(y^-_{QL})}} \right) (f(u-\sigma_1)x^+_{QS})}{k (x^+_{QS}-x^-_{QS}) (\sigma_1+u-y^-_{QL}) (x^+_{QL}-f(u-\sigma_1))}\right)\right.\\
    &\qquad\qquad\left.-\left(\frac{2 f'(y^-_{QL})}{x^+_{QL}-f(y^-_{QL})}+\frac{f''(y^-_{QL})}{f'(y^-_{QL})}\right)+\frac{2 \left(1-(\pi  T_0)^2 x^-_{QS} x^+_{QS}\right)}{x^+_{QS}-x^-_{QS}}\right).
\end{split}
\end{equation}
As before, we will solve this equation perturbatively. The location of $(x^+_{QS},x^-_{QS})$ is given by eqs.~\eqref{eq:QESscram} and the location of $(x^+_{QL},y^-_{QL})$ is given in eqs.~\eqref{eq:QESlate}. Using the approximations defined in eq.~\eqref{eq: approximation of f(u-sigma)} and eq.~\eqref{eq: approximations of derivatives of f(u-sigma)} as well as the rescalings $\lambda\to \mu\nu\Tilde{\lambda}$, $\tau\to \mu\Tilde{\tau}$ and $k\to\nu\Tilde{k}$, the equation up to leading order in $\mu$ and $\nu$ reads
\begin{equation}
  \begin{split}
    0&=\frac{\phi_r}{4 G_N} \left(2 \Tilde{k} \log \left(\frac{3 \left(\pi ^{3/2} \Tilde{\tau}   T_1 \sqrt{e^{k y^-_{QL}/2}} \left(T_1^2-T_0^2\right) (1-\pi  T_0 t_\infty) (\sigma_1+u)\right)}{8 \nu^2 \mu^{1/2} \Tilde{k} \Tilde{\lambda}   T_0 \sqrt{\Tilde{\tau}  T_1 t_\infty} \left(\log \left(\frac{8 \pi  T_1}{3 \Tilde{k}\nu  }\right)+4 \pi  \sigma_1 T_1\right)}\right)\right.\\
    &\qquad\qquad\left.+ \tilde{k} \nu +2 \pi  \left(T_0-T_1 e^{-\frac{k y^-_{QL}}{2}}\right)+\mathcal{O}\left(\nu^2\mu^0\right)\right).
\end{split}  
\end{equation}
We can use the explicit definition of $\lambda$ and $\tau$ from eq.~\eqref{eq: approximation of f(u-sigma)} and eq.~\eqref{eq: f-tau} as well as the fact that even though $k y^-_{QL}$ and $k y^+$ are fixed, both are still small, such that the equation becomes
\begin{equation}
  \begin{split}
    0&=\frac{\phi_r}{4 G_N} \left(\Tilde{k}  \nu \left(2 \log \left(\frac{\pi^3 T_1 \left(T_1-T_0\right)^2 (T_1 +T_0) (\sigma_1+u)}{\Tilde{k}^2 \nu^2 T_0  (\log \left(\frac{8\pi T_1}{3 \Tilde{k}\nu}\right)+4 \pi  \sigma_1 T_1)}\right)-\log \left(\frac{8\pi T_1}{3 \Tilde{k}\nu}\right)-2 \pi  \sigma_1 T_1+2 \pi  T_1 u\right)\right.\\
    &\qquad\qquad\left.+\frac{\nu}{2}   \left(-\Tilde{k} \log \left(\frac{8 \pi  T_1}{3 \Tilde{k}\nu  }\right)-2 \pi  \Tilde{k} \sigma_1 T_1+2 \pi  \Tilde{k} T_1 u+2 \Tilde{k}\right)+2 \pi  (T_0-T_1)+\mathcal{O}\left(\nu^2\mu^0\right)\right).
\end{split}  
\end{equation}
This equation can be solved recursively by first solving
\begin{equation}
  \begin{split}
    0&=\frac{\phi_r}{4 G_N} \left(k  \left(-\log \left(\frac{8\pi T_1}{3 k}\right)-2 \pi  \sigma_1 T_1+2 \pi  T_1 u\right)\right.\\
    &\qquad\qquad\left.+\frac{1}{2}   \left(-k \log \left(\frac{8 \pi  T_1}{3 k  }\right)-2 \pi  k \sigma_1 T_1+2 \pi  k T_1 u+2 k\right)+2 \pi  (T_0-T_1)\right),
\end{split}  
\end{equation}
which gives
\begin{equation}
    u^0_{STLT}=\frac{2}{3}\frac{T_1-T_0}{k T_1}+\frac{\log \left(\frac{8 \pi  T_1}{3 k }\right)}{2 \pi T_1}+\sigma_1-\frac{1}{3\pi T_1}.
\end{equation}
The first recursive correction to this solution can be found by solving
\begin{equation}
  \begin{split}
    0&=\frac{\phi_r}{4 G_N} \left(k  \left(2 \log \left(\frac{\pi^3 T_1 \left(T_1-T_0\right)^2 (T_1 +T_0) (\sigma_1+u^0_{STLT})}{k^2 \nu^2 T_0  (\log \left(\frac{8\pi T_1}{3 k}\right)+4 \pi  \sigma_1 T_1)}\right)-\log \left(\frac{8\pi T_1}{3 k}\right)-2 \pi  \sigma_1 T_1+2 \pi  T_1 u\right)\right.\\
    &\qquad\qquad\left.+\frac{1}{2}   \left(-k \log \left(\frac{8 \pi  T_1}{3 k  }\right)-2 \pi  k \sigma_1 T_1+2 \pi  k T_1 u+2 k\right)+2 \pi  (T_0-T_1)\right),
\end{split}  
\end{equation}
and is given by
\begin{equation}
\begin{split}\label{eq:Pagetime}
    u^1_{STLT}&=\frac{2}{3}\frac{T_1-T_0}{k T_1}+\frac{\log \left(\frac{8 \pi  T_1}{3 k }\right)}{2 \pi T_1}+\sigma_1-\frac{1}{3\pi T_1}\\
    &\quad-\frac{2}{3\pi T_1}\log \left(\frac{\pi^3 T_1 \left(T_1-T_0\right)^2 (T_1 +T_0) (\sigma_1+u^0_{STLT})}{k^2 \nu^2 T_0  (\log \left(\frac{8\pi T_1}{3 k}\right)+4 \pi  \sigma_1 T_1)}\right)+\mathcal{O}\left(k\right).
    \end{split}
\end{equation}
\\
\paragraph{Finding a lower bound}\mbox{}\\	
In order to find the first bound, we now equate $u_{CCST}$ and $u^1_{STLT}$ and solve for the backreaction parameter $k$. More specifically, we solve
\begin{equation}
\begin{split}
    \frac{\sigma_2 (\pi  \sigma_2 T_0 t_\infty -\sigma_1)}{\sigma_2-\pi  \sigma_1 T_0 t_\infty}&=\frac{2}{3\pi T_1}\log \left(\frac{3 \pi ^{3/2} T_1\left(T_1^2-T_0^2\right) (1-\pi T_0 t_\infty) (\sigma_1+u^0_{STLT})}{8 k \nu T_0\sqrt{T_1 t_\infty} \left(\log \left(\frac{8 \pi  T_1}{3 k}\right)+4 \pi  \sigma_1 T_1\right)}\right)\\ &\quad+\frac{2 (T_1-T_0)}{3 k T_1}-\frac{\log \left(\frac{8 \pi  T_1}{3 k }\right)}{6 \pi T_1}+\sigma_1-\frac{1}{3\pi T_1}+\mathcal{O}\left(k\right)
   .
    \end{split}   
\end{equation}
As before, we approach this equation recursively and find the zeroth order approximation by solving
\begin{equation}
\begin{split}
    \frac{\sigma_2 (\pi  \sigma_2 T_0 t_\infty -\sigma_1)}{\sigma_2-\pi  \sigma_1 T_0 t_\infty}&=\frac{2 (T_1-T_0)}{3 k T_1}+\sigma_1-\frac{1}{3\pi T_1}\,,
    \end{split}   
\end{equation}
which leads to
\begin{equation}
\begin{split}
    k^0_{FB}&=\frac{2 \pi  (T_1-T_0) (\sigma_2-\pi  \sigma_1 T_0 t_\infty)}{ \left(\sigma_2+3 \pi ^2 T_0 T_1 t_\infty \left(\sigma_1^2+\sigma_2^2\right)-\pi  \sigma_1 (T_0 t_\infty+6 \sigma_2 T_1)\right)}\,.
    \end{split}   
\end{equation}
The first recursive correction can then be found by solving
\begin{equation}
\begin{split}
    \frac{\sigma_2 (\pi  \sigma_2 T_0 t_\infty -\sigma_1)}{\sigma_2-\pi  \sigma_1 T_0 t_\infty}&=\frac{2 (T_1-T_0)}{3 k T_1}-\frac{\log \left(\frac{8 \pi  T_1}{3 k^0_{FB} }\right)}{6 \pi T_1}+\sigma_1-\frac{1}{3\pi T_1}\\
    &\quad+\frac{2}{3\pi T_1}\log \left(\frac{3 \pi ^{3/2} T_1\left(T_1^2-T_0^2\right) (1-\pi T_0 t_\infty) (\sigma_1+u^0_{STLT}|_{k=k^0_{FB}})}{8 k^0_{FB} T_0\sqrt{T_1 t_\infty} \left(\log \left(\frac{8 \pi  T_1}{3 k^0_{FB}}\right)+4 \pi  \sigma_1 T_1\right)}\right).
    \end{split}   
    \label{eqn:first bound}
\end{equation}
We know from figure~\ref{figure: finite segments intermediate times} that this triple intersection of $S^{\rm CC}_{\rm gen}=S^{\rm ST}_{\rm gen}=S^{\rm LT}_{\rm gen}$ is a limiting behaviour of the temporary blindness phenomenon and happens for large values of $k$. Hence, in order to have the temporary blindness occur, the value of $k $ should be below the bound we just derived. In other words, we established a lower bound on the dimensionless parameter $\frac{E_S}{c(T_1+T_0)}=\frac{\pi}{12}\frac{T_1-T_0}{ k}$ for temporary blindness to occur, which is given by
\begin{equation}
\begin{split}
   \frac{E_S}{c\left(T_1+T_0\right)}&=\frac{\pi T_1}{8}\left[\frac{\sigma_2 (\pi  \sigma_2 T_0 t_\infty -\sigma_1)}{\sigma_2-\pi  \sigma_1 T_0 t_\infty} + \frac{\log \left(\frac{8 \pi  T_1}{3 k^0_{FB} }\right)}{6 \pi T_1}-\sigma_1+\frac{1}{3\pi T_1}\right.\\
    &\left.\quad-\frac{2}{3\pi T_1}\log \left(\frac{3 \pi ^{3/2} T_1\left(T_1^2-T_0^2\right) (1-\pi T_0 t_\infty) (\sigma_1+u^0_{STLT}|_{k=k^0_{FB}})}{8 k^0_{FB} T_0\sqrt{T_1 t_\infty} \left(\log \left(\frac{8 \pi  T_1}{3 k^0_{FB}}\right)+4 \pi  \sigma_1 T_1\right)}\right)\right].
    \end{split}   
    \label{eqn:first bound on param}
\end{equation}

\subsection{Upper bound}	
As discussed in section~\ref{ssection: page curve} around eq.~\eqref{eq:upper bound}, the upper bound on $\frac{E_S}{c(T_1+T_0)}$ can be established from the intersection of $S^{\rm CC}_{\rm gen}$ and $S^{\rm LT}_{\rm gen}$ specified in eq.~\eqref{eq: competing phases in reconstruction window at intermediate times} and the intersection of their derivatives.	
\\
\paragraph{Time of intersection between $S^{\rm CC}_{\rm gen}$ and $S^{\rm LT}_{\rm gen}$}\mbox{}\\
First, we find the time $u$ for which
\begin{equation}
    0=S^{\rm LT}_{\rm gen}-S^{\rm CC}_{\rm gen}.
\end{equation}
We can write this equation out explicitly using the dilaton profile given in eq.~\eqref{eq: dilaton profile before and after quench}. The equation simplifies to
\begin{equation}
    \begin{split}
        0&=\frac{\phi_r}{4 G_N} \left(2 k \log \left(\frac{2 k \sigma_2 (\sigma_1+u-y^-_{QL}) \sqrt{f'(y^-_{QL})} (x^+_{QL}-f(u-\sigma_1))}{(x^+_{QL}-f(y^-_{QL})) \pi ^2 (\sigma_1-\sigma_2) \left(T_1^2-T_0^2\right) (u-\sigma_2) f(u-\sigma_1)}\right)\right.\\
        &\qquad\left.+\frac{2 f'(y^-_{QL})}{x^+_{QL}-f(y^-_{QL})}+\frac{f''(y^-_{QL})}{f'(y^-_{QL})}-2 \pi  T_0\right).
        \label{eq: Slatetrivial-Scrossconnected}
    \end{split}
\end{equation}
As usual, we will solve this equation perturbatively. The perturbative location of $(x^+_{QL},y^-_{QL})$ is given in eqs.~\eqref{eq:QESlate}. Moreover, we can use the approximations defined in eq.~\eqref{eq: approximation of f(u-sigma)} and eq.~\eqref{eq: approximations of derivatives of f(u-sigma)}. As before, we rescale the parameters $\lambda\to \mu\nu\Tilde{\lambda}$, $\tau\to \mu\Tilde{\tau}$ and $k\to\nu\Tilde{k}$. We can use the explicit definition of $\tau$ and $\lambda$ given in eq.~\eqref{eq: approximation of f(u-sigma)} and eq.~\eqref{eq: f-tau} as well as the fact that even though $k y^-_{QL}$ and $k y^+$ are fixed, both are still small, such that the equation up to leading order in $\mu$ and $\nu$ becomes
\begin{equation}
    \begin{split}
        0&=8 \pi ^2 T_1 (T_1-T_0)+2 \pi  \Tilde{k} \nu  T_1 \left(-\log \left(\frac{8 \pi  T_1}{3 \Tilde{k} \nu }\right)+6 \log \left(\frac{8 \pi  T_1}{3 \Tilde{k}\nu}\right)+6 \pi  T_1 (u-\sigma_1)+2\right.\\
        &\qquad\left.-4 \log \left(-\frac{64 \sqrt{\mu } \nu^{3/2} \sigma_2 \sqrt{T_1} \left(\log \left(\frac{8 \pi  T_1}{3 \Tilde{k} \nu }\right)+4 \pi \sigma_1 T_1\right)}{9 \pi ^{3/2} (\sigma_2-\sigma_1) \left(T_1^2-T_0^2\right) (u-\sigma_2) \sqrt{ t_\infty}}\right)\right)+\mathcal{O}\left(\nu^2\mu^0\right).
    \end{split}
\end{equation}

This equation can be solved recursively by first solving
\begin{equation}
  \begin{split}
        0&=8 \pi ^2 T_1 (T_1-T_0)+2 \pi  k   T_1 \left(-\log \left(\frac{8 \pi  T_1}{3 k  }\right)+6 \log \left(\frac{8 \pi  T_1}{3 k}\right)+6 \pi  T_1 (u-\sigma_1)+2\right.\\
        &\qquad\left.-4 \log \left(\frac{64 \sigma_2  T_1^{3/2} \left(\log \left(\frac{8 \pi  T_1}{3 k }\right)+4 \pi \sigma_1 T_1\right)}{9 \sqrt{\pi} (\sigma_2-\sigma_1) \left(T_1^2-T_0^2\right)  \sqrt{ t_\infty}}\right)\right),
\end{split}  
\end{equation}
which gives
\begin{equation}
\begin{split}
    u^0_{LTCC}&=\frac{2 (T_1-T_0)}{3 k  T_1}+\frac{4 \log \left(\frac{64  \sigma_2   T_1^{3/2} \left(\log \left(\frac{8 \pi  T_1}{3 k}\right)+4 \pi  \sigma_1 T_1\right)}{9 \sqrt{\pi } \sqrt{t_\infty} (\sigma_2-\sigma_1) \left(T_1^2-T_0^2\right)}\right)-5 \log \left(\frac{8 \pi  T_1}{3 k}\right)+6 \pi  \sigma_1 T_1-2}{6 \pi  T_1}\\
    &\quad+\mathcal{O}\left(k\right).
    \end{split}
\end{equation}
The first recursive correction to this solution can be found by solving
\begin{equation}
  \begin{split}
        0&=8 \pi ^2 T_1 (T_1-T_0)+2 \pi  k   T_1 \left(-\log \left(\frac{8 \pi  T_1}{3 k  }\right)+6 \log \left(\frac{8 \pi  T_1}{3 k}\right)+6 \pi  T_1 (u-\sigma_1)+2\right.\\
        &\qquad\left.-4 \log \left(-\frac{64  \sigma_2 \sqrt{T_1} \left(\log \left(\frac{8 \pi  T_1}{3k }\right)+4 \pi \sigma_1 T_1\right)}{9 \pi ^{3/2} (\sigma_2-\sigma_1) \left(T_1^2-T_0^2\right) (u^0_{LTCC}-\sigma_2) \sqrt{ t_\infty}}\right)\right),
\end{split}  
\end{equation}
and is given by
\begin{equation}
\begin{split}
    u^1_{LTCC}&=\frac{2 (T_1-T_0)}{3 k  T_1}+\frac{4 \log \left(\frac{-64  \sigma_2 \sqrt{T_1} \left(\log \left(\frac{8 \pi  T_1}{3 k}\right)+4 \pi  \sigma_1 T_1\right)}{9 \pi^{3/2} \sqrt{t_\infty} (\sigma_2-\sigma_1) \left(T_1^2-T_0^2\right)(u^0_{LTCC}-\sigma_2)}\right)-5 \log \left(\frac{8 \pi  T_1}{3 k}\right)+6 \pi  \sigma_1 T_1-2}{6 \pi  T_1}\\
    &\quad+\mathcal{O}\left(k\right).
    \end{split}
\end{equation}
\\
\paragraph{Time of intersection between $S^{'\rm CC}_{\rm gen}$ and $S^{'\rm LT}_{\rm gen}$}\mbox{}\\
In this section we intend to find the time $u$ for which
\begin{equation}
    0=S^{'\rm LT}_{\rm gen}-S^{'\rm CC}_{\rm gen}.
\end{equation}
We can simply take the derivative with respect to $u$ of eq.~\eqref{eq: Slatetrivial-Scrossconnected} and find
\begin{equation}
\begin{split}
    0&=\frac{k \phi_r}{2 G_N (u-\sigma_2) (\sigma_1+u-y^-_{QL}) f(u-\sigma_1) (x^+_{QL}-f(u-\sigma_1))} \times\\
    &\left(-x^+_{QL} (u-\sigma_2) (\sigma_1+u-y^-_{QL}) f'(u-\sigma_1)\right.\\
    &\qquad\qquad\left.+x^+_{QL} (-\sigma_1-\sigma_2+y^-_{QL}) f(u-\sigma_1)+(\sigma_1+\sigma_2-y^-_{QL}) f(u-\sigma_1)^2\right).
\end{split}
\end{equation}
As before, we will solve this equation perturbatively
\begin{equation}
    \begin{split}
        0=\pi  (u-\sigma_2)T_1  \log \left(\frac{8 \pi  T_1 }{3 k}\right)+2 \log \left(\frac{8 \pi  T_1 }{3 k}\right)+12 \pi ^2 T_1 ^2\sigma_1\left( u- \sigma_2 \right) +4 \pi  T_1 \left( \sigma_2- u \right)+\mathcal{O}\left(k \lambda\right).
    \end{split}
\end{equation}
This equation is perfectly solvable and the solution is given by
\begin{equation}
    u_{LT'CC'}=\frac{(3 \pi  \sigma_2 T_1-2) \log \left(\frac{8 \pi  T_1}{3 k}\right)+4 \pi  T_1 (\sigma_1 (3 \pi  \sigma_2 T_1-2)-\sigma_2)}{\pi  T_1 \left(3 \log \left(\frac{8 \pi  T_1}{3 k}\right)+12 \pi  \sigma_1 T_1-4\right)}+\mathcal{O}\left(k\right).
\end{equation}
\\
\paragraph{Finding an upper bound}\mbox{}\\
In order to find the second bound, we now equate $u_{LT'CC'}$ and $u^1_{LTCC}$ and solve for the backreaction parameter $k$. More specifically, we solve
\begin{equation}
\begin{split}
    \frac{(3 \pi  \sigma_2 T_1-2) \log \left(\frac{8 \pi  T_1}{3 k}\right)+4 \pi  T_1 (\sigma_1 (3 \pi  \sigma_2 T_1-2)-\sigma_2)}{\pi  T_1 \left(3 \log \left(\frac{8 \pi  T_1}{3 k}\right)+12 \pi  \sigma_1 T_1-4\right)}&=\frac{2 (T_1-T_0)}{3 k  T_1}
    \\
    \quad+\frac{4 \log \left(\frac{-64  \sigma_2 \sqrt{T_1} \left(\log \left(\frac{8 \pi  T_1}{3 k}\right)+4 \pi  \sigma_1 T_1\right)}{9 \pi^{3/2} \sqrt{t_\infty} (\sigma_2-\sigma_1) \left(T_1^2-T_0^2\right)(u^0_{LTCC}-\sigma_2)}\right)-5 \log \left(\frac{8 \pi  T_1}{3 k}\right)+6 \pi  \sigma_1 T_1-2}{6 \pi  T_1}
    &+\mathcal{O}\left(k\right).
    \end{split}   
\end{equation}	

As before, we can solve this equation recursively by first solving
\begin{equation}
  \begin{split}
        \frac{3 \pi  \sigma_2 T_1-2}{3 \pi  T_1}&=\frac{2 (T_1-T_0)}{3 k  T_1}+\frac{6 \pi  \sigma_1 T_1-2}{6 \pi  T_1},
\end{split}  
\end{equation}
which gives
\begin{equation}
    k^0_{SB}=\frac{2 \pi  (T_1-T_0)}{3 \pi  \text{T1} (\sigma_2-\sigma_1)+1}.
\end{equation}
The first recursive correction to this solution can be found by solving 
\begin{equation}
  \begin{split}
        &\frac{(3 \pi  \sigma_2 T_1-2) \log \left(\frac{8 \pi  T_1}{3 k^0_{SB}}\right)+4 \pi  T_1 (\sigma_1 (3 \pi  \sigma_2 T_1-2)-\sigma_2)}{\pi  T_1 \left(3 \log \left(\frac{8 \pi  T_1}{3 k^0_{SB}}\right)+12 \pi  \sigma_1 T_1-4\right)}=\frac{2 (T_1-T_0)}{3 k  T_1}
    \\
    &\quad+\frac{4 \log \left(\frac{-64  \sigma_2 \sqrt{T_1} \left(\log \left(\frac{8 \pi  T_1}{3 k^0_{SB}}\right)+4 \pi  \sigma_1 T_1\right)}{9 \pi^{3/2} \sqrt{t_\infty} (\sigma_2-\sigma_1) \left(T_1^2-T_0^2\right)(u^0_{LTCC}\Big|_{k=k^0_{SB}}-\sigma_2)}\right)-5 \log \left(\frac{8 \pi  T_1}{3 k^0_{SB}}\right)+6 \pi  \sigma_1 T_1-2}{6 \pi  T_1}.
\end{split}  
\end{equation}
As shown in figure~\ref{figure: finite segments intermediate times}, this intersection of $S^{\rm CC}_{\rm gen}=S^{\rm LT}_{\rm gen}$ and $S^{'\rm CC}_{\rm gen}=S^{'\rm LT}_{\rm gen}$ is a limiting behaviour of the temporary blindness phenomenon and happens for small values of $k$. Hence, in order to have the temporary blindness occur, the value of $k $ should be above the bound we just derived. In other words, we established an upper bound on the dimensionless parameter $\frac{E_S}{c(T_1+T_0)}=\frac{\pi}{12}\frac{T_1-T_0}{ k}$ for temporary blindness to occur.
\begin{equation}
  \begin{split}
        \frac{E_S}{c\left(T_1+T_0\right)}&=\frac{(3 \pi  \sigma_2 T_1-2) \log \left(\frac{8 \pi  T_1}{3 k^0_{SB}}\right)+4 \pi  T_1 (\sigma_1 (3 \pi  \sigma_2 T_1-2)-\sigma_2)}{8 \left(3 \log \left(\frac{8 \pi  T_1}{3 k^0_{SB}}\right)+12 \pi  \sigma_1 T_1-4\right)}
    \\
    &\quad-\frac{1}{12}\log \left(\frac{-64  \sigma_2 \sqrt{T_1} \left(\log \left(\frac{8 \pi  T_1}{3 k^0_{SB}}\right)+4 \pi  \sigma_1 T_1\right)}{9 \pi^{3/2} \sqrt{t_\infty} (\sigma_2-\sigma_1) \left(T_1^2-T_0^2\right)(u^0_{LTCC}\Big|_{k=k^0_{SB}}-\sigma_2)}\right)\\
    &\quad-\frac{5}{48} \log \left(\frac{8 \pi  T_1}{3 k^0_{SB}}\right)+\frac{1}{8} \pi  \sigma_1 T_1-\frac{1}{24}\,.
\end{split}  
\label{eqn: second bound on param}
\end{equation}

\bibliographystyle{JHEP}
\bibliography{evaporation_paper.bib}

\end{document}